\documentclass[%
 amsmath,amssymb,
 aps, rmp,
 eqsecnum
 ]{revtex4-1}
 
\usepackage[utf8]{inputenc}
\usepackage{graphicx}
\usepackage{dcolumn}
\usepackage{bm}
\usepackage{hyperref}
\usepackage{xcolor}

\def\be{\begin{eqnarray}}
\def\ee{\end{eqnarray}}

\def\d{{\rm d}}

\newcommand{\beq}{\begin{equation}}
\newcommand{\eeq}{\end{equation}}
\newcommand{\beqa}{\begin{eqnarray}}
\newcommand{\eeqa}{\end{eqnarray}}
\newcommand{\beql}{\begin{align}}
\newcommand{\eeql}{\end{align}}

\newcommand{\Or}[1]{\mathcal{O}\left(#1\right)}
\newcommand{\lp}{\left(}
\newcommand{\rp}{\right)}
\newcommand{\vep}{\varepsilon}

\newcommand{\R}{\mathbb{R}}

\newcommand{\sR}{\mathsf{R}}

\newcommand{\cP}{\mathcal{P}}

\begin{document}

\title{The Large D Limit of Einstein's Equations}

\author{Roberto Emparan$^{ab}$}
\email{emparan@ub.edu}
\author{Christopher P.~Herzog$^c$}%
 \email{christopher.herzog@kcl.ac.uk}
\affiliation{%
 ${}^a$ Departament de F$\acute{\imath}$sica Qu\`antica i Astrof$\acute{\imath}$sica and Institut de Ci{\`e}ncies del Cosmos, Universitat de Barcelona, Mart$\acute{\imath}$ i Franqu\`es 1, E-08028 Barcelona, Spain \\
 ${}^b$ Instituci\'o Catalana de Recerca i Estudis Avan\c{c}ats (ICREA),
 Passeig Llu$\acute{\imath}$s Companys 23, E-08010 Barcelona, Spain \\
${}^c$ King's College London, The Strand, London WC2R 2LS, England
}%

\begin{abstract}
We review recent progress in taking the large dimension limit of Einstein's equations.  Most of our analysis is classical in nature and concerns situations where there is a black hole horizon although we briefly discuss various extensions that include quantum gravitational effects.
The review consists of two main parts: the first a discussion of general aspects of black holes and effective membrane theories in this large dimension limit, and the second 
a series of applications of this limit to interesting physical problems.  
The first part includes a discussion of quasinormal modes which leads naturally into a description of effective hydrodynamic-like equations that describe the near horizon geometry.  There are two main approaches to these effective theories 
--  a fully covariant approach and a partially gauge-fixed one -- which we discuss in relation to each other.
In the second part we divide the applications up into three main categories: the Gregory-Laflamme instability, black hole collisions and mergers, and the anti-de Sitter/conformal field theory correspondence (AdS/CFT). 
AdS/CFT posits an equivalence between a gravitational theory and a strongly interacting field theory, allowing us to extend our spectrum of applications to problems in hydrodynamics, condensed matter physics, and nuclear physics.  A final, shorter part of the review describes further promising directions where there have been, as yet, few published research articles.
\end{abstract}

\maketitle

\tableofcontents{}

\newpage

\part{General aspects}
\label{part:general}

\section{Large $N$ and large $D$ expansions}

Our theories of physics frequently have parameters -- continuous ones, such as coupling constants and masses, but also discrete ones, such as numbers of fields and internal degrees of freedom -- that can be varied from their actual values in Nature while still maintaining consistency. An often fruitful strategy is to explore the theories when these parameters approach the boundaries of their allowed range, since in some of these limits the equations become significantly simpler. If a solution can be found, then, by perturbatively correcting it in an expansion around these limits, one hopes to obtain an approximation to the theoretical predictions for real-world parameter values.

A prime example is the theory of quantum electrodynamics. Much of what we know about it begins with the limit where the electron charge is zero and the theory is free, which then provides a convenient starting point for the perturbative analysis. For other theories, such as QCD at low energies, and more generally, for theories of strongly coupled non-linear systems, this strategy is inappropriate, and one should try to identify less obvious but still useful parameters. This search becomes harder the simpler and more elegant the formulation of the theory is. As a case in point, consider non-abelian Yang-Mills theory and General Relativity, in the absence of any quarks or matter content. The Lagrangians that define them,
\beq\label{YMGR}
\mathcal{L}_{YM}=-\frac14 \text{Tr}\,F^2,\qquad \mathcal{L}_{GR}=\sqrt{-g}R\,,
\eeq
are notorious for the lack of any apparent adjustable knobs. However, it has been known for many years that, retaining the same form \eqref{YMGR} of the Yang-Mills Lagrangian, we can sneak in a parameter as the rank of the gauge group $SU(N)$, with $N$ taking any value $\geq 2$. The limit $N\to\infty$ not only results in a simpler description of gluon dynamics, but also in a reorganization of the degrees of freedom in terms of worldsheets  \cite{tHooft:1973alw}, eventually leading to a reformulation as a holographic theory of strings and quantum gravity \cite{Maldacena:1997re}.

The theory of General Relativity in \eqref{YMGR} contains a somewhat similar variable parameter, namely, the number of spacetime dimensions $D$. The classical theory is well defined in any $D\geq 4$, and moreover it retains its most remarkable features: black hole solutions and propagating degrees of freedom (i.e., gravitational waves). One may then hope that the limit $D\to\infty$ results in a convenient simplification of the equations and possibly also a novel reformulation of the dynamics, at least for some phenomena.

In this article we will review the progress made in recent years in realizing these hopes, especially in the study of classical black hole physics and in applications to a variety of problems. We will also describe prospects for further extensions, possibly beyond the realm of black holes and into the quantum regime. 

\subsection{The nature of the large $D$ limit}

What does it mean to take the large $D$ limit of Einstein's gravitational theory \eqref{YMGR}? The comparison with the large $N$ limit of gauge theories is illustrative, both for its similarities and its differences. The $SU(N\to\infty)$ Yang-Mills theory is an instance of a limit of a quantum theory where the number of degrees of freedom at each point in space grows very large. These limits are widespread in theoretical physics, including $O(N\to\infty)$ vector models, Potts models, matrix, Sachdev-Ye-Kitaev (SYK) and tensor models, and large central charge limits of conformal field theories, among many others. The simplifications that these afford can vary considerably; for instance, the SYK model hits the right balance between the overkill simplicity of vector theories and the difficulty of solving matrix quantum mechanics.

A different type of limit is obtained when what grows large is the number of connections to nearby points, or directions out of each point in space. This is familiar in statistical mechanics: when the coordination number of a lattice diverges, one recovers a mean field theory. The intuition here is that spatial quantum fluctuations average themselves out, leading to a semiclassical collective theory -- either the conventional mean field, or a richer dynamical mean field theory \cite{Georges:1996zz}. However, such a limit is seldom considered for continuum quantum field theories, the reason being that, although the effects of long-distance quantum fluctuations are suppressed, the short distance, ultraviolet divergences 
get uncontrollably strong.\footnote{Conformal blocks of conformal field theories are free from UV divergences and can be solved in an expansion in $1/D$ \cite{Fitzpatrick:2013sya}. Other properties of CFTs in the limit of large $D$ are investigated in \cite{Gadde:2020nwg}.} Nevertheless, the idea can be fruitfully applied as long as one appropriately focuses on infrared physics, and in particular when the theory is a classical one.\footnote{In mathematics, geometry in the limit of infinite dimensions has often been studied, for instance in Perelman's work on Ricci flow; see sec.~6 of \cite{Perelman:2006un}.}

Interestingly, the large $D$ limit in gravity contains aspects of these two types of limit. On the one hand, the number of graviton polarizations (i.e., the number of gravitational degrees of freedom at each point) grows like $\sim D^2$ at large $D$, similar to the growth $\sim N^2$ of gluon polarizations in $SU(N)$ gauge theories. One may then hope that the simplifications that occur in the latter have a counterpart in the semiclassical expansion of Einstein's theory around flat space. Do the Feynman diagrams of perturbative quantum General Relativity -- itself a gauge theory -- arrange themselves into something similar to the topological expansion of $SU(N)$ Yang-Mills theory? Unfortunately, the answers found so far are not encouraging. The classes of diagrams enhanced at large $D$ by the combinatorics of polarization indices in gravity do not bring in anything as suggestive as the worldsheets of large $N$ gauge theory \cite{Strominger:1981jg,BjerrumBohr:2003zd}. Other approaches that, with varying degrees of success, also utilize the large number of graviton degrees of freedom include \cite{Hamber:2005vc,Canfora:2009dx,Sloan:2016kbc}.

On the other hand, $D$ also appears in General Relativity in its basic role as the ``directions out of a point'', into which the gravitational lines of force get more and more diluted as $D$ grows large. This effect is manifested in the radial decay of the potential
\beq\label{PhiD}
\Phi\sim \frac1{r^{D-3}}
\eeq
that solves the Laplace equation in $D-1$ spatial dimensions. The larger $D$ is, the faster the interaction potential decays at long distance, and the more steeply it increases at small scales. That is, as $D$ grows the gravitational field is increasingly strongly localized near its source, while at larger distances it is suppressed in a manner that is non-perturbative in $1/D$. This suggests that this ``mean field-like'' aspect of large $D$ gravity -- and not its ``large $N$-like'' quality -- may be fruitfully exploited for the study of gravitating massive objects, and in particular black holes.

The first works to make successful use of this idea were \cite{Kol:2004pn,Asnin:2007rw}, which performed a $1/D$ expansion to analytically solve the Euclidean negative mode of the Schwarzschild solution, or equivalently, the zero mode at the threshold of the Gregory-Laflamme (GL) instability of black strings \cite{Gregory:1993vy}.\footnote{Ref.~\cite{Asnin:2007rw} also introduced an expansion around $D=3$, but this idea has not been followed up.}  Although the study was not taken beyond this problem, \cite{Asnin:2007rw} correctly identified and exploited two important generic features of the large $D$ limit of black holes: the parametric separation between a near-horizon region and a far region, and the localization near the horizon of the dynamics of interest.

Despite this initial success, and besides incidental discussion of features of black holes as $D$ is made large \cite{Caldarelli:2008mv,Hod:2011zzb}, a systematic study did not begin until the work of \cite{Emparan:2013moa}, which prompted the development of the subject in earnest. By now much has been understood that can be coherently reviewed. A new perspective with novel insights on black hole dynamics has emerged, which has also found application to current topical problems. For instance, efficient analytical calculations of black hole quasinormal modes, and simplified numerical simulations of black hole collisions, may eventually be relevant to gravitational wave astronomy (within limitations that we will discuss later). Another natural niche for applications is the AdS/CFT correspondence, with studies of dual superconductors, the dynamics of quark-gluon plasmas, the structure of entanglement, and others. In addition, the large $D$ methods can, and already have, shed light on thorny problems in black hole theory for which conventional numerical methods are unwieldy, or at least can well do with efficient guidance from simpler, more intuitive techniques. These problems include basic issues about instability of black holes and their evolution, cosmic censorship, and possibly critical collapse.

\subsection{Organization of this review}

Part~\ref{part:general} follows the path that led from the initial explorations of black holes at large $D$ to the development of non-linear effective theories,
with some benefit of hindsight and discussion of issues not often found in the literature. Part~\ref{part:applications} focuses on applications, mostly (but not only) of the effective theories. In this part the different sections can be read quite independently of each other, although they require background developed in Part~\ref{part:general}. 
In the briefer Part~\ref{part:ramblings} we discuss open ends and a sample of ideas that may admit development.

\subsubsection*{Outline}

We begin in sec.~\ref{sec:nearfar} with an explanation of how the large $D$ limit concentrates the gravitational effects of a black hole close to the horizon and makes the near-horizon geometry universal. Sec.~\ref{sec:qnm} shows that the effect can be regarded as a `decoupling limit' that separates the gravitational perturbations of the black hole into two distinct sectors. Sec.~\ref{sec:eft} introduces the non-linear effective theories of black holes. Of these, we have selected two versions: sec.~\ref{subsec:effmemb} is devoted to the theory in \cite{Bhattacharyya:2015fdk}, as the one with broadest generality, and sec.~\ref{subsec:eftbb} explains the effective theory for black branes \cite{Emparan:2015gva}, which is the basis of many applications. 

Part~\ref{part:applications} of the review discusses various applications of the large $D$ limit.
In  sec.~\ref{sec:GL}, we explore how the Gregory-Laflamme instability of black strings depends on $D$. 
Then in sec.~\ref{sec:collisions}, we review how black hole collisions and mergers behave as $D$ becomes large, revealing a novel violation of cosmic censorship, and we examine the prospects for shedding light on black hole mergers in four dimensions.
Sec.~\ref{sec:holography} explores how large $D$ techniques have been applied to the AdS/CFT correspondence.
By taking a large $D$ limit of AdS/CFT, one may gain further leverage from Einstein's equations in explaining various phenomena in strongly interacting field theories.  In this AdS/CFT context, we will look at applications to hydrodynamics, condensed matter physics, and nuclear physics.

Part~\ref{part:ramblings} surveys other directions of the large $D$ programme, combining subjects that are already being investigated and other more speculative lines that may hold promise for the future. Sec.~\ref{sec:hawkingetal} examines the prospects of a large $D$ approach to quantum and stringy physics. Sec.~\ref{sec:hider} is a brief overview of work done in the context of higher-derivative theories of gravity. Sec.~\ref{sec:beyond} explores open-ended opportunities to study phenomena in large $D$ gravity beyond the scope of the effective theories. Sec.~\ref{sec:gwaves} ends the review on a speculative note about gravitational radiation.

\paragraph*{Note.}
We often use, instead of $D$, a number $n$ defined as the power of the radial decay of the gravitational potential. Generally,
\beq\label{nDp}
n=D-p-3\,,
\eeq
where $p$, which remains finite as $D\to\infty$, stands for the spatial dimensionality of a $p$-brane worldvolume (so, e.g., $p=0$ for the Schwarzschild-Tangherlini black holes). Throughout this review, $D$ and $n$ are interchangeably used as large expansion parameters.

\section{The large $D$ geometry of stationary black holes}
\label{sec:nearfar}

\subsection{Near and far zones}

Let us begin with the most elementary black hole in $D$ dimensions: the Schwarzschild-Tangherlini solution \cite{Schwarzschild:1916uq,Tangherlini:1963bw},
\beq\label{SchD}
ds^2=-f(r) \d t^2+\frac{\d r^2}{f(r)}+r^2 \d \Omega_{D-2}\,,\qquad f(r)=1-\lp\frac{r_0}{r}\rp^{D-3}\,.
\eeq
The main difference with respect to the four-dimensional Schwarzschild solution, besides the addition of angles to the sphere $S^{D-2}$, is the replacement of the Newtonian potential $\sim 1/r$ with its $D$ dimensional counterpart \eqref{PhiD}. Here we normalize it with a scale set by the horizon radius $r_0$,
\beq\label{Phir0D}
\Phi\sim \lp\frac{r_0}{r}\rp^{D-3}\,.
\eeq

When $D$ is considered a fixed number, $r_0$ is the only length that sets the scale for all phenomena in this geometry. However, when $D$ is regarded as a variable parameter that we take to be large, we find a different scale that characterizes the gravitational force near the horizon,
\beq
\nabla\Phi\bigr|_{r_0}\sim \frac{D}{r_0}\,.
\eeq
That is, the large $D$ limit introduces a new, parametrically separated length scale in the system, namely,
\beq\label{sepscales}
\frac{r_0}{D}\ll r_0\,.
\eeq
There may appear other scales, too; we will see that for black branes $r_0/\sqrt{D}$ plays an important role.

The length $r_0/D$ not only characterizes the slope of $\Phi$, but also the radial distance over which the gravitational field is appreciably non-zero. That is, as $D\to\infty$ we have
\beq\label{farzone}
r>r_0\quad \Rightarrow\quad \Phi(r)\to 0\,,
\eeq
but also
\beq\label{nearzone}
r-r_0\lesssim \frac{r_0}{D}\quad\Rightarrow\quad \Phi(r) =\Or{1}\,.
\eeq
These behaviors define two regions in the geometry: the ``far zone'' \eqref{farzone} and the ``near(-horizon) zone'' \eqref{nearzone}.\footnote{We will not discuss the black hole interior, but similar considerations apply there \cite{Emparan:2013moa}.} In the far zone, the Schwarzschild-Tangherlini geometry \eqref{SchD} becomes flat Minkowski space. The near-horizon geometry is more interesting. 
As explained in \eqref{nDp} it is convenient to use 
\beq\label{nD3}
n=D-3
\eeq
instead of $D$ as the perturbation parameter, and introduce
\beq\label{sR}
\sR=\lp \frac{r}{r_0}\rp^n
\eeq
as the finite radial variable appropriate for the near-horizon region, which is defined by $\ln \sR\ll n$. Introduce also a near-horizon time
\beq
\bar t=\frac{n}{2 r_0} t\,.
\eeq
Then, for large $n$ one finds \cite{Soda:1993xc,Emparan:2013xia}
\beq\label{nhbh}
\d s^2_{nh}\to \frac{4r_0^2}{n^2}\lp-\lp 1-\frac1{\sR}\rp \d \bar t^2+\frac{\d \sR^2}{\sR(\sR-1)}\rp +r_0^2 \sR^{2/n} \d \Omega_{n+1}\,.
\eeq
Note that in the sphere radius $r_0 \sR^{1/n}\simeq r_0\lp 1+\frac{\ln \sR}{n}\rp$ we are retaining an apparently subleading-order term, since it is promoted to leading order whenever the area element $r_0^n \sR^2$ is involved.

Let us focus now on the $(\bar t,\sR)$ part of the metric. Its prefactor reflects that the short scale $r_0/n$ is a measure of the small proper radial extent of this zone, and of the short proper time required to cross it, measured in units of the distances and times of the far zone. The two-dimensional metric in \eqref{nhbh} is actually well known. Defining the proper radius $\rho$ by
\beq
\cosh^2\rho=\sR\,,
\eeq
this part of the geometry takes the form
\beq\label{2Dbh}
\d s^2_2=\frac{4r_0^2}{n^2}\lp-\tanh^2\rho\, \d \bar t^2+\d \rho^2\rp\,,
\eeq
which is the black hole solution of the dilaton-gravity theory that appears in the low energy limit of two-dimensional string theory \cite{Mandal:1991tz,Elitzur:1991cb,Witten:1991yr}, with dilaton profile
\beq\label{lindil}
\phi=-\ln \cosh\rho\,.
\eeq
This follows from an observation in \cite{Soda:1993xc,Grumiller:2002nm} about the spherical reduction of Einstein gravity in the limit $D\to\infty$. For a metric of the form
\beq\label{redn}
\d s^2=g_{\mu\nu}\d x^\mu \d x^\nu + r_0^2 e^{-4\phi/(n+1)}\d \Omega_{n+1}^2
\eeq
where $g_{\mu\nu}(x^\lambda)$ is a 2D metric and $\phi(x^\lambda)$ a 2D scalar field, the Einstein-Hilbert action reduces to\footnote{The apparently wrong sign of the kinetic term for $\phi$ is not a problem since this theory does not have any local dynamics. From the higher dimensional point of view, this statement is Birkhoff's theorem, or the absence of any s-wave dynamics in vacuum gravity.}
\beq
I=\frac{\Omega_{n+1}r_0^{n+1}}{16\pi G}\int d^2 x\sqrt{-g}\, e^{-2\phi}
\lp R+\frac{4n}{n+1}(\nabla\phi)^2 + \frac{n(n+1)}{r_0^2} e^{4\phi/(n+1)}\rp\,,
\eeq
which in the limit $n\to\infty$ becomes the 2D string effective action
\beq
I=\frac{1}{16\pi G_2}\,\int d^2 x\sqrt{-g}\, e^{-2\phi}
\lp R+4(\nabla\phi)^2 + 4\Lambda_{2}^2\rp
\label{eq:larged3}
\eeq
with
\beq
G_2 = \lim_{n\to\infty} \frac{G}{\Omega_{n+1}r_0^{n+1}}\,, \qquad \Lambda_{2}=\frac{n}{2r_0}\,.
\eeq
Notice that keeping $r_0/n$ finite amounts to keeping finite the Hawking--temperature, 
\beq\label{TH}
T_H=\frac{\Lambda_{2}}{2\pi}\,
\eeq
of the large $D$ Schwarzschild-Tangherlini black hole, while for the 2D string black hole, the temperature conjugate to the time $\bar t$ is $1/{2\pi}$, i.e., $\sim 1/n$ times smaller. In this identification, Minkowski space at $D\to\infty$ corresponds to the linear dilaton vacuum, which appears as the asymptotic geometry at $\rho\to\infty$ of the black hole \eqref{2Dbh},  \eqref{lindil}. The string length would be $\sim r_0/D$.

\subsection{Universality}
\label{subsec:universal}

The appearance of the 2D string black hole \eqref{2Dbh} in the context of a pure gravity theory is intriguing, but perhaps more surprising is that it appears universally as the large $D$ near-horizon geometry of all neutral, non-extremal black holes of Einstein's theory, including solutions with rotation and a cosmological constant \cite{Emparan:2013xia}. This universality, together with the fact that this geometry has enhanced symmetry (it is the quotient space $SL(2,\R)/U(1)$), entails very significant simplifications in the study of large $D$ black hole physics.

In the case of static (A)dS black holes, it is easy to verify that when the limit $n\to\infty$ is taken keeping  finite $\sR$ and
\beq
\lambda=\frac{2 r_0^2}{n^2}\Lambda, 
\eeq
which parametrizes the cosmological constant $\Lambda$ in units of $r_0$, then to leading order the geometry in the $(t,\sR)$ directions is the same as \eqref{nhbh}, only up to a rescaling of the coordinates that changes the overall size of the near-horizon region relative to the far zone scales: a negative cosmological constant shrinks it, a positive one enlarges it.

The effect of rotation is more interesting. Considering, for instance, the Myers-Perry solutions with a single angular momentum turned on \cite{Myers:1986un}, and focusing on the region around a given polar angle $\theta=\theta_0$, the near-horizon geometry takes the form of the 2D string black hole, with an extra dimension (for the rotation direction) added to it, and then boosted along this direction with a velocity equal to the local velocity of the horizon at latitude $\theta_0$. Therefore the rotation is accounted for by a local boost in a spatial direction parallel to the horizon. This feature will be of great importance when developing a general effective theory of large $D$ black hole dynamics.

By considering the addition of charge and possibly dilatonic scalar fields to the black hole one obtains different universality classes of near-horizon geometries \cite{Emparan:2013xia} (see also \cite{Guo:2015swu}). These geometries can all be regarded as solutions of theories where matter is added to \eqref{eq:larged3}. Just like in the asymptotically flat case, the gravitational effects of this matter around the black hole decay at large values of $\sR$ and are relevant only close to the horizon. So the asymptotic geometry of the near horizon zone in all cases has a 2D sector
\beq
ds^2_2 \to \frac{r_0^2}{n^2}\lp -d\bar t^2+\frac{d\sR^2}{\sR^2}\rp
\eeq
(up to an overall $n$-independent constant) and a dilaton $\phi \to -\ln\sqrt{R}$. That is, the linear dilaton vacuum plays the same role near the horizon at large $D$ as Minkowski space plays in asymptotically flat spacetimes -- but more universally, since it is also the asymptotics of near-horizon zones when there is a $D$-dimensional cosmological constant.

Interestingly, the same 2D black hole and physics similar to the large $D$ limit appears in a holographic plasma with a five-dimensional dilaton gravity bulk dual, whose conformal critical point of the plasma is in correspondence with $D\to\infty$ \cite{Betzios:2018kwn}.

\subsection{Strings near the horizon?}
\label{sec:crazy}

This `stringy nature' of the black hole in the limit $D\to\infty$ is also manifested in its entropy. For the Schwarzschild-Tangherlini black hole, it behaves as
\beq
S(M)\sim M^{1+\frac{1}{D-3}}\,.
\eeq
The fact that the exponent is $>1$ means that the specific heat of the black hole is negative, and that the merger of two black holes is highly irreversible. However, when $D\to\infty$ we find
\beq\label{hagedorn}
S(M)\to M\,,
\eeq
which is the same behavior as that of a gas of strings with a Hagedorn spectrum (whose thermodynamic stability depends on the details of logarithmic corrections to the leading order result \eqref{hagedorn}). To leading order at large $D$, there is no entropy production in a merger, a feature that we will encounter again in sec.~\ref{sssection:stress}. Although this effect might be attributed to the vanishingly short interaction range in a large number of dimensions, one should bear in mind that it does not hold for charged black holes. 

The behavior \eqref{hagedorn} is a consequence of the properties of the two-dimensional dilaton black hole \eqref{2Dbh}, but it is a tantalizing possibility that it also indicates that large $D$ black holes may be described in terms of effective strings.\footnote{This would be similar to the evidence for `little strings' in the throat of NS fivebranes \cite{Aharony:1998ub}.} Curiously, the entropy of a conformal gas with energy $E$ in $D$ dimensions,
\beq
S(E)\sim E^{1-\frac{1}{D}}
\eeq
also approaches the Hagedorn behavior $S\sim E$ when $D\to\infty$, now from the side of positive specific heat.
The meaning, if any, of this coincidence in the large $D$ limit of the degeneracy of the high-energy spectra of quantum field theories, string theories, and black holes, is unclear.

\section{Decoupled and non-decoupled dynamics of large $D$ black holes}
\label{sec:qnm}

The previous discussion referred to stationary black hole states. In order to obtain an idea of their dynamical behavior at large $D$, it is convenient to start with small fluctuations.

\subsection{Quasinormal spectrum}

Slightly perturbed black holes oscillate with a set of characteristic frequencies known as the quasinormal mode spectrum. In many respects, these are analogous to normal modes, but they have a dissipative part (imaginary frequency) due to the absorptive nature of the horizon. These modes dominate the late-time ringdown phase of a black hole formed in a merger, and also the return to equilibrium when a black hole has been perturbed, e.g., by a particle falling into it. An instability of a black hole is often manifested in the presence of a quasinormal mode with a positive imaginary part of the frequency, whose amplitude grows exponentially in time.

Like normal mode spectra, the quasinormal frequencies depend on the specific features of the system, such as shape and composition or field content. Since the quasinormal oscillations of a black hole propagate away as gravitational radiation, they are a primary source of information in gravitational wave astronomy \cite{Berti:2009kk}. 
Quasinormal modes also play an important role in the AdS/CFT correspondence: the longest lived, least damped modes of a black brane characterize the late time approach to equilibrium of the dual thermal state \cite{Horowitz:1999jd}.

The study of quasinormal modes of large $D$ black holes has revealed one of their most basic properties: the existence of two different dynamical regimes, distinguished by frequencies parametrically separated in $1/D$. These are:
\begin{itemize}
\item \textit{Fast, non-decoupled modes}, with frequencies $\omega\sim D/r_0$. These modes straddle between the near horizon zone and the far zone. Most quasinormal modes are in this category, and their spectrum falls into large universality classes with no information (to leading order at large $D$) about the black hole other than its horizon radius or shape.
\item \textit{Slow, decoupled modes}, with $\omega\sim 1/r_0$. These oscillations are localized within the near-horizon zone, and they vanish in the far zone to all perturbative orders in $1/D$. There are only a few of them, and they are specific to each black hole. Hydrodynamic behavior and horizon instabilities appear in this sector of the dynamics. 
\end{itemize}
The existence of the first set of modes is expected: generically, the characteristic oscillation frequency of the black hole is set by the surface gravity $\kappa \sim D/r_0$, which is also the inverse of the light-crossing time for the near-horizon region.

More surprising is the presence of the second set of much slower modes, indeed static ones relative to the characteristic time of the near zone. They were first found numerically in \cite{Dias:2014eua}, and the further understanding of their nature in \cite{Emparan:2014aba} prompted the development of non-linear effective theories for the slow dynamics of black hole fluctuations.

\subsection{Qualitative features from the radial potential}

The main qualitative features of the quasinormal spectrum at large $D$ can be understood from the form of the effective radial potential for the perturbations of the Schwarzschild-Tangherlini black hole \cite{Kodama:2003jz}. In the large $D$ limit we will find a universality of quasinormal spectra that is larger than that implied by the universality classes of near-horizon geometries. Namely, the potential that non-decoupled modes perceive will be insensitive to anything but the shape of the horizon -- not to its charges or couplings to scalars. On the other hand, the universality of near-horizon geometries will imply that a given universality class will share the same set of decoupled modes (e.g., one vector and two scalars for all neutral black holes), even though their actual frequencies will be different since they depend on the near-horizon geometry at the next order.

Let us then consider linearized gravitational perturbations $\delta g_{\mu\nu}=e^{-i\omega t}h_{\mu\nu}(r,\Omega)$ around the Schwarzschild-Tangherlini black hole solution \eqref{SchD}. The angular dependence can be separated and the perturbations classified according to their algebraic transformation properties under the $SO(n+2)$ symmetry of the sphere $S^{n+1}$: scalar-type ($S$), vector-type ($V$) and tensor-type ($T$). Tensor perturbations exists only in five or more spacetime dimensions ($n\geq 2$). 

\begin{figure}[t]
 \begin{center}
  \includegraphics[width=.5\textwidth,angle=0]{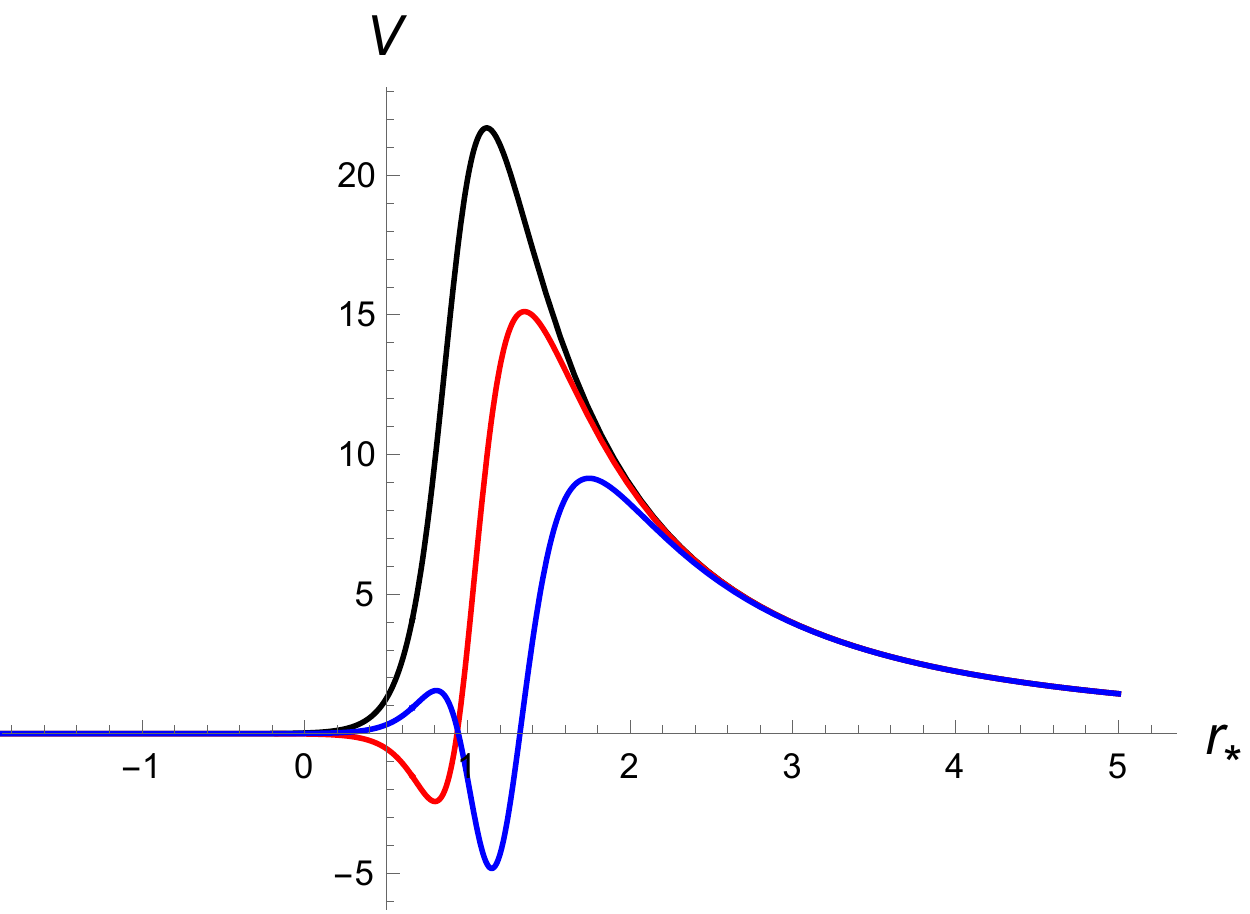}
   \end{center}
 \vspace{-5mm}
 \caption{\small Radial potentials $V_s(r_*)$ for perturbations of the Schwarzschild black hole for $n=8$ and $\ell=2$. The horizon is at $r_*\to-\infty$. We use the coding black/red/blue $=$ tensor/vector/scalar in this and in the next two figures. Units are $r_0=1$.}
 \label{fig:n8l2}
\end{figure}

Ref.~\cite{Kodama:2003jz} obtained decoupled master variables $\Psi_s(r_*)$, with $r_*=\int dr/f$, for each of these perturbations, which satisfy equations of the form
\beq\label{master}
\lp \frac{d^2}{d r_*^2}+\omega^2-V_s\rp \Psi_s=0\qquad\quad  s=S,V,T\,.
\eeq
Explicit expressions for $V_s$ can be found in \cite{Kodama:2003jz} (and in the present context, in \cite{Emparan:2014aba}). Fig.~\ref{fig:n8l2} illustrates $V_s(r_*)$ for moderate values of $n$ and $\ell$. There is a barrier, which grows with $\ell$, corresponding to radial gradients and centrifugal energy. For small enough $\ell/n$, the scalar and vector potentials possess additional minima and maxima closer to the horizon, which are absent for the tensor perturbations.
\begin{figure}[t]
 \begin{center}
  \includegraphics[width=.46\textwidth,angle=0]{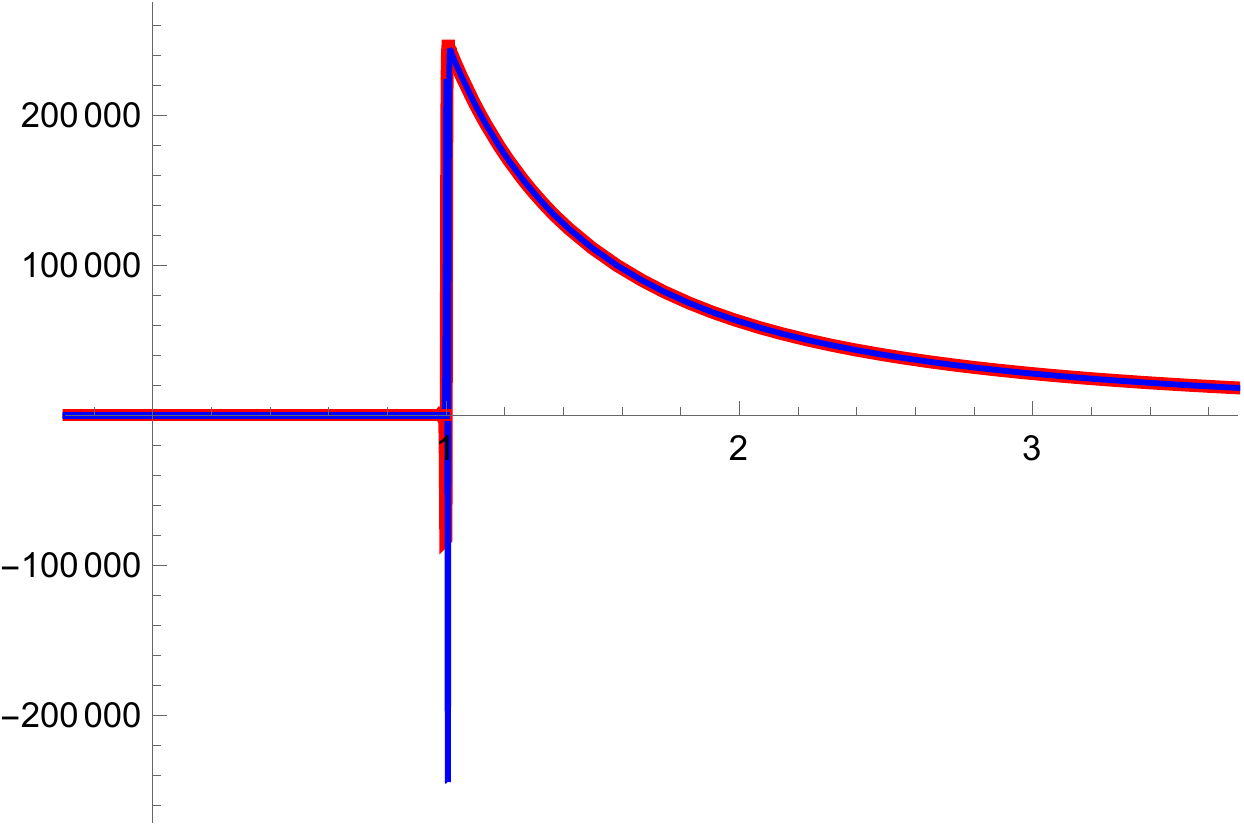}
  \hspace{7mm}
  \includegraphics[width=.46\textwidth,angle=0]{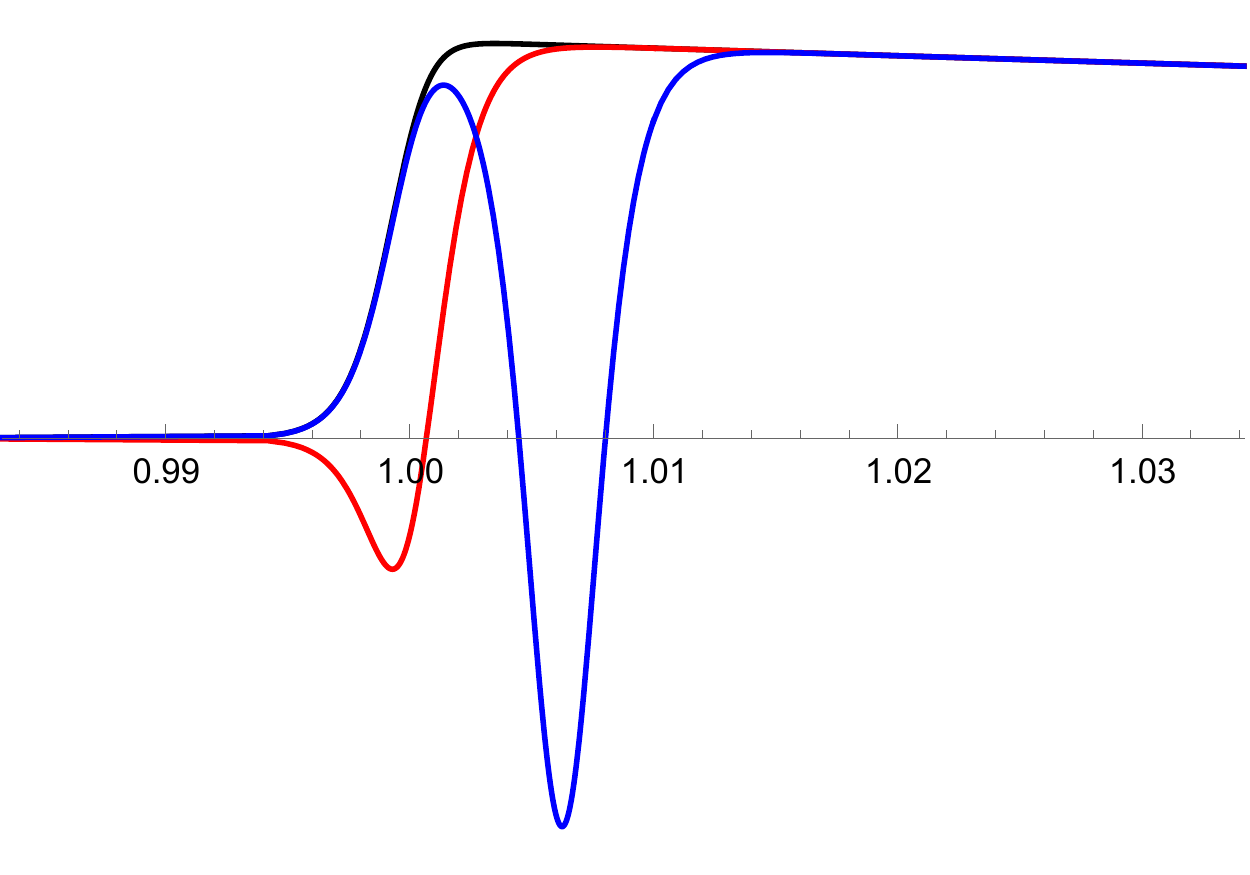}
   \end{center}
 \vspace{-5mm}
 \caption{\small Radial potentials $V_s(r_*)$ for $n=997$ and $\ell=2$. On the right is a blow-up of the potential near the peak at $r_*\simeq 1$.}
 \label{fig:n997l2}
\end{figure}
When $n$ is very large these features of the potential become more marked, as illustrated in fig.~\ref{fig:n997l2}. The left and right figure represent, respectively, the view of the far zone and a blow up of the near-horizon zone. The former is dominated by the centrifugal barrier $\sim 1/r^2$, which reaches a maximum at the photon sphere near $r_*\simeq 1$, approaching
\beq\label{Vmax}
V_s^\text{max}\to n^2\omega_c^2
\eeq
where
\beq
\omega_c=\frac1{2r_0}\lp 1+\frac{2\ell}{n}\rp\,.
\eeq
As a consequence, waves with frequency $\omega=\Or{1/r_0}\ll n\omega_c$ cannot penetrate the potential: they stay either outside or inside the barrier, since their tunneling probability is infinitely suppressed as $n\to \infty$. Thus, large $D$ induces a decoupling of low-frequency dynamics, and quasinormal modes will be discussed separately according to whether their frequency is $\omega \sim n/r_0$ or instead $\omega \sim 1/r_0$.

\begin{figure}[t]
 \begin{center}
  \includegraphics[width=.6\textwidth,angle=0]{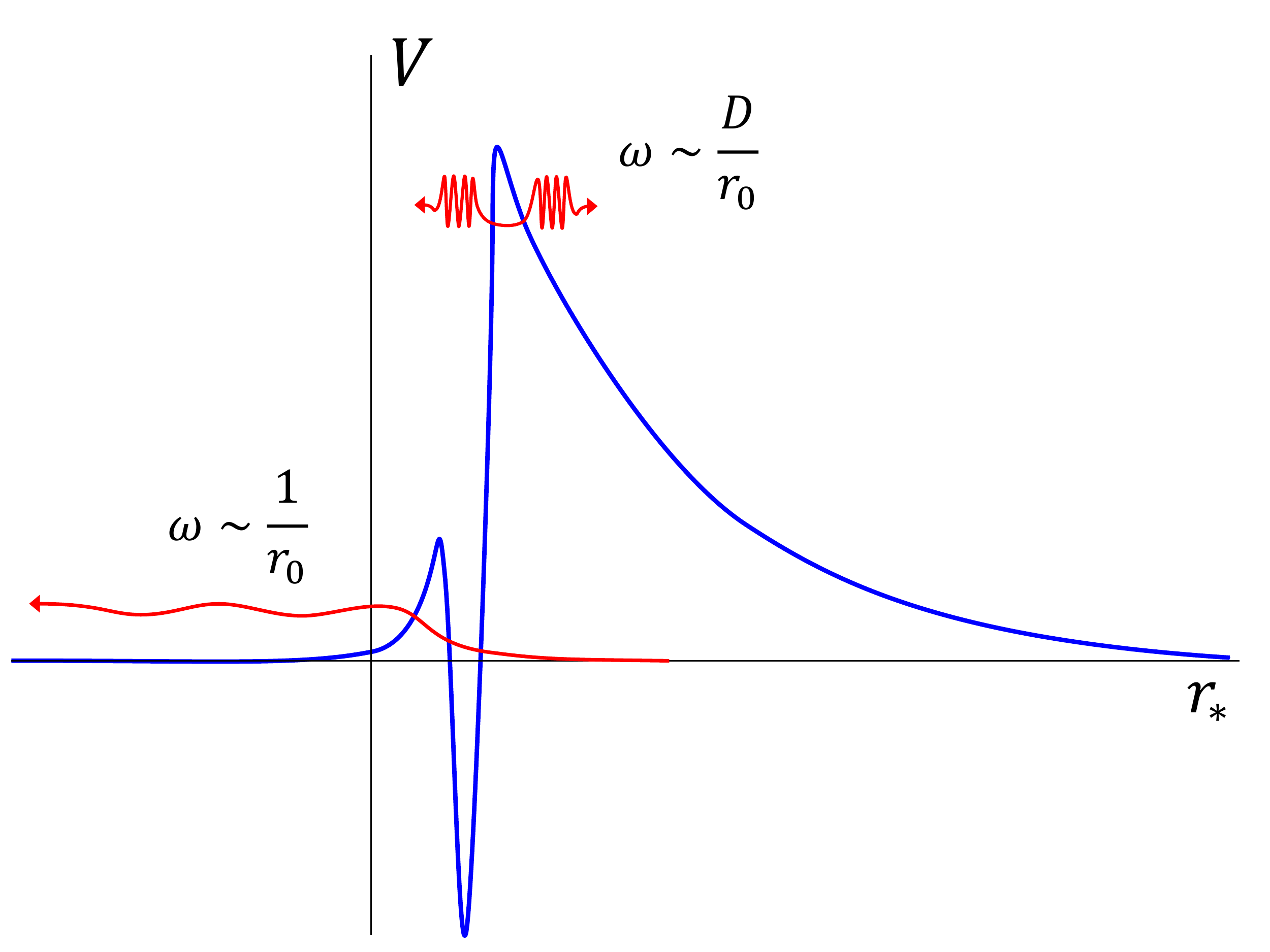}
   \end{center}
 \vspace{-5mm}
 \caption{\small Illustration of quasinormal modes at $D\gg 1$: non-decoupled modes near the peak of the potential have high frequencies and straddle between the near and far zones. Decoupled modes have low frequencies and are trapped in the near-horizon geometry.}
 \label{fig:qnms}
\end{figure}

Quasinormal modes are solutions of \eqref{master} characterized by the absence of any amplitudes coming in from infinity or coming out of the horizon, see fig.~\ref{fig:qnms}.
Using the coordinate in \eqref{sR},
the ingoing boundary condition at the future horizon at $\sR=1$  is 
\beq\label{horbc}
\Psi_s(\sR) \propto (\sR -1)^{-i\omega r_0/n}\,.
\eeq

We can now expect to find quasinormal modes as solutions that connect outgoing and ingoing waves by joining them below the peak of the potential like in fig.~\ref{fig:qnms}. 
The potentials that these modes `see' are the ones on the left in fig.~\ref{fig:n997l2}. The frequencies of these modes are of the order of the height of the potential \eqref{Vmax},\footnote{By $\omega$ we refer to the real part of the quasinormal frequency. The imaginary part turns out to also be large, but, as we will see, less so.}
\beq\label{largefreq}
\omega \simeq n \omega_c \sim \frac{D}{r_0}\,,
\eeq
i.e., in the range of the fast time scale $r_0/D$. These oscillations travel ballistically between the near and far zones, and are insensitive to any structure of the potential other than its peak. This is almost featureless as it depends only on the black hole radius $r_0$, making this non-decoupled spectrum a universal feature of static black holes.

But the potential for scalar and vector modes also possesses structure to the left of the peak, deeper in the near-horizon zone, as seen on the right in fig.~\ref{fig:n997l2}.  We can expect quasinormal oscillations that are ingoing at the horizon and are trapped inside the barrier, with wavefunctions that vanish like $\sim e^{-1/n}$ outside the barrier, hence without any incoming component at any perturbation order.\footnote{We elaborate on this in sec.~\ref{subsec:nonconv}.} Their frequency will be slow,
\beq
\omega\sim \frac1{r_0}
\eeq
and will depend on the specific properties of the potential wells. These modes constitute the decoupled spectrum, and they dominate the late time decay of black hole perturbations.

\subsection{Non-decoupled modes}
\label{subsec:nondecqnm}

Following the previous discussion, we first focus on waves with high frequency \eqref{largefreq} near the peak of the potential.

The tensor potential for all $\ell$, and the vector and scalar potentials for $\ell=\Or{n}$, all approach the form in fig.~\ref{fig:n997l2}, left, that is,
\beq\label{Vslimit}
V_s\to \frac{n^2\omega_c^2r_0^2}{r_*^2}\Theta(r_*-r_0)\,.
\eeq
Therefore the three kinds of perturbations will be isospectral. For each $\ell$ there is a sequence of modes, called  `overtones', whose wavefunctions have $k-1$ nodes, $k=1,2,\dots$. The least damped overtones, with $k\ll n$, are sensitive only to the structure near the tip of the potential, which approaches a triangular shape that makes it easy to obtain their frequencies by employing a WKB approximation \cite{Emparan:2014cia,Emparan:2014aba}. The result is
\beq\label{onomega}
\omega r_0 = \frac{n}2+\ell -a_{k}\left(\frac{e^{i\pi}}{2}\lp \frac{n}2+\ell\rp\right)^{1/3}\,,
\eeq
where $a_k$ are the zeros of the Airy function ($k=1,2,\ldots$), very accurately approximated by
\beq
a_k\simeq \lp\frac{3\pi}{8}(4k-1)\rp^{2/3}\,.
\eeq
The damping ratio of these modes
\beq
\frac{\text{Im}\,\omega}{\text{Re}\,\omega}\sim n^{-2/3}
\eeq
vanishes as $n\to\infty$, so they are long-lived in their characteristic time scale. They limit to undamped normal modes.

The comparison of \eqref{onomega} with a numerical calculation of the frequencies for a large number of values of $n$ \cite{Dias:2014eua} gives agreement of the real part with an accuracy well matched by $\approx 1/(2n)$, slightly better than the naive expectation of errors $\approx 1/n$. The imaginary part shows poorer agreement even at $D=100$, but this can be explained from features of the potential peak \cite{Emparan:2014aba}.

This calculation applies to all static, spherically symmetric black holes of the same radius in the limit $D\to\infty$, possibly with charge, dilatonic couplings, and cosmological constant \cite{Emparan:2014cia}. Non-decoupled modes are the oscillations of a field in a flat spacetime with a hole of radius $r_0$ in it. 

\subsection{Decoupled modes}
\label{subsec:decmodes}

Quasinormal modes with frequency $\omega/n\to 0$ correspond to static, zero-energy states, which can only exist if the potential has a negative minimum, as is the case for vectors and scalars with $\ell=\Or{1}$. These states are unique for a given $\ell$, with no other overtones close to them. 

The analysis of these modes starts with the wave equation
\beq
\lp\mathcal{L}+U_s\rp\Psi_s(\sR)=-\frac{\sR-1}{\sR^{1/n}}\frac{d}{d\sR}\lp\frac{\sR-1}{\sR^{1/n}}\frac{d}{d\sR}\Psi_s\rp+ \frac1{n^2}\lp V_s(\sR)-\omega^2\rp\Psi_s=0
\eeq
in the near-horizon zone. At the boundary of this zone, $\sR\gg 1$, we require that the divergent behavior $\Psi\sim \sqrt{\sR}$ is absent and only allow wavefunctions that are normalizable within this region,
\beq\label{asybc}
\Psi(\sR\to\infty)\to \frac1{\sqrt{\sR}}\,.
\eeq
At the future horizon we impose \eqref{horbc}. 

The procedure to solve the equation in a perturbative expansion in $1/n$ is straightforward \cite{Emparan:2014aba}. Here we only show that decoupled modes, with the previous boundary conditions, exist as solutions that to leading order are static. Their non-zero frequencies arise at the next order. (For the remainder of this section we fix $r_0=1$.)

It is easy to see that there are no decoupled tensor modes. To leading order in $1/n$ the potential is
\beq
U_{T}^{(0)}=\frac{\sR^2-1}{4\sR^{2}}\,,
\eeq
and there is no combination of the two independent solutions
\beq
u_0=\sqrt{\sR}\,,\qquad v_0=\sqrt{\sR}\,\ln\lp 1-\sR^{-1}\rp
\eeq
that satisfies the two boundary conditions \eqref{horbc} and \eqref{asybc}.

The vector potential is
\beq
U_{V}=\frac{\sR-1}{4\sR^{1+2/n}}\left[\lp 1+\frac{2\ell}{n}\rp^2-\frac1{n^2}-\frac{3}{\sR}\lp 1+\frac1{n}\rp^2 \right]-\frac{\omega^2}{n^2}\,,
\eeq
and now there is a leading-order solution with the required behavior at the boundaries, namely,
\beq
\Psi_{V}^{(0)}=\frac1{\sqrt{\sR}}\,.
\eeq
So there does exist a vector quasinormal mode.
In order to determine its frequency we need to go to the next order. The solution is
\beq
\Psi_{V}^{(1)}=-\frac{(\ell-1)\ln(\sR-1)+\ln\sqrt{\sR}}{\sqrt{\sR}}\,,
\eeq
and now the boundary condition at the horizon \eqref{horbc} fixes
\beq
\omega=-i(\ell-1)\,.
\eeq
These mode frequencies are purely imaginary.

For the scalar modes the calculation is more involved, but one can again show the existence at leading order of a static solution. The next order solution yields two frequencies, related by $\omega_-=-\omega_+^*$, 
\beq\label{omsc0}
\omega_{\pm}=\pm\sqrt{\ell-1}-i (\ell-1)\,.
\eeq
These modes have both a real and an imaginary part. There are no other overtones nearby these vector and scalar modes in the complex $\omega$ plane.

It is straightforward to continue the expansion to higher orders and obtain corrections to the frequencies \cite{Emparan:2014aba,Emparan:2015rva}. The accuracy of the large $D$ expansion can then be tested by comparing with direct numerical calculations at finite $D$ \cite{Dias:2014eua}. Vector mode frequencies are very accurately captured by the large $D$ expansion, with agreement of up to eight digits at $n=100$. Perhaps more remarkably, even down to $D=4$ (i.e., $n=1$!) the frequency of the `algebraically special mode' has been reproduced with  $6\%$ accuracy from a calculation including up to terms $1/n^3$. This level of agreement suggests again that the expansion parameter is better thought to be $1/(2n)$ rather than $1/n$.

\subsection{Cosmological constant and rotation}

The universality of the near-horizon geometry becomes extremely useful for the study of perturbations of black holes with rotation and a cosmological constant. The presence of decoupled quasinormal modes is automatic, since it follows from the existence of static solutions in the universal near-horizon geometry, proven in sec.~\ref{subsec:decmodes}. The values of their frequencies $\omega\sim 1/r_0$  depend on the $1/D$ corrections to the metrics, and these are different for each black hole. Nevertheless, including a cosmological constant is straightforward \cite{Emparan:2015rva}. In AdS, when the black hole becomes much larger than the cosmological radius and approaches a black brane, the scalar and vector modes become hydrodynamic sound and shear modes. We will encounter them again in secs.~\ref{subsec:eftbb} and \ref{subsec:holohydro}.

Adding rotation makes the calculation less simple but still feasible analytically. The latter is particularly interesting, since higher-dimensional rotating black holes possess complex patterns of behavior \cite{Myers:1986un,Emparan:2003sy}. For now we shall only mention a highlight of the analysis in \cite{Suzuki:2015iha}.

Unlike the Kerr black hole, the spin of singly-rotating Myers-Perry black holes in dimension $D\geq 6$ is not limited by an extremality bound \cite{Myers:1986un}. It was noted in \cite{Emparan:2003sy} that, as the spin increases, the horizon spreads along the plane of rotation. Since black membranes are known to be unstable, it was proposed that Myers-Perry black holes with large enough angular momenta should also become unstable. The onset of these instabilities would be marked by the appearance of zero-mode perturbations of the black holes as their spin is increased. This was confirmed by numerical solution of the perturbation equations in \cite{Dias:2009iu,Dias:2010maa} in $D=6,\dots,11$. For instance, in $D=8$ the zero modes appear when the rotation parameter $a$, in units of the mass-radius $r_m$, is
\beq
\frac{a}{r_m}=1.77,\,2.27,\,2.72\,\dots
\eeq
Ref.~\cite{Suzuki:2015iha} solved the problem analytically in the large $D$ limit, and found zero modes for\footnote{We describe a simpler approach to this in sec.~\ref{sssec:blobs}.}
\beq\label{MPultraspin}
\frac{a}{r_m}=\sqrt{3},\,\sqrt{5},\,\sqrt{7}\,\dots
\eeq
The two results agree with an error smaller than $2.7\%$.

\subsection{Asymptotic non-convergence of the expansion}
\label{subsec:nonconv}

The startling accuracy of the calculations of decoupled quasinormal frequencies prompts the question of whether successive expansion orders will continue to improve the results for finite values of $D$. In other words, is the $1/D$ expansion in the decoupled sector convergent, or is it instead only asymptotic?

A first hint comes from the fact that the potential $(r_0/r)^{D-3}$ is non-analytic in $1/D$. 
One then expects that non-perturbative effects from the far zone must be present in the near-horizon zone which spoil the convergence of the expansion.

We can see this by solving the scalar field equation in the far zone of the Schwarzschild black hole (or indeed, any spherical black hole) and studying its behavior in the `overlap' region $r_0/n\ll r-r_0\ll r_0$, i.e., $1\ll\sR\ll e^n$, where it must match the field in the near zone \cite{Emparan:2015rva}. Outgoing quasinormal waves are Hankel functions,
\beq
\Psi=\sqrt{r}H^{(1)}_{n\omega_c r_0}(\omega r)\,.
\eeq
In the overlap zone, and for frequencies $\omega\sim \Or{1}$, these behave, schematically, like
\beq\label{overlap}
\Psi(r)\sim \frac1{\sqrt{\sR}}\sum_{i\geq 0}\lp \frac{\ell}{n}\rp^i \sR^{i/n}\,.
\eeq
Matching to a decoupled wave in this region is possible if \eqref{asybc} is satisfied, which requires that $\ell\ll n$. However, large orders in the expansion in \eqref{overlap}, with $i=n+\Or{1}$, give a behavior $\Psi\sim \sqrt{\sR}$ which violates the decoupling condition \eqref{asybc}. The breakdown is non-perturbative in $1/n$, and for waves with $\ell=\Or{1}$ one obtains
\beq
\text{non-perturbative corrections}=\Or{n^{-n}}\,,
\eeq
confirming the general expectation.

The non-decoupled quasinormal modes can be thought of as non-perturbative, or trans-series, 
corrections to the decoupled sector \cite{Casalderrey-Solana:2018uag}.
While the decoupled modes have $\Or{1}$  imaginary part in the large $D$ expansion, they are still relatively long lived compared to the non-decoupled modes, which damp out and oscillate on time scales $\sim n^{-1/3}$ and $\sim n^{-1}$ respectively.  The long time behavior of the black holes is thus governed by the decoupled modes as a sort of attractor to which   
the non-decoupled modes serve as highly damped nonperturbative corrections.

\section{Effective theories of black holes at large $D$}
\label{sec:eft}

Can the previous study of black hole fluctuations be taken beyond the linearized approximation, into a fully non-linear theory of large $D$ black hole dynamics?  In order to understand how this is possible and what it implies, it is convenient to broadly set the framework for the discussion.

\subsection{Aspects of effective theories of black hole dynamics}

The efficient study of a physical system often relies on the ability to parametrically separate dynamical regimes at different scales: high energy vs.\ low energy, short distance vs.\ long distance, early time vs.\ late time. Effective theories encode in a few parameters the long distance effects of short distance physics.

From this perspective, the main reason that the dynamics of a black hole is generally hard to solve is that there is only one scale in the system: the horizon radius, or equivalently the black hole mass. Nevertheless, there are many situations with another length scale that permits successful effective descriptions. The point-particle limit, in which a black hole moves in gravitational fields of curvature radii much larger than its horizon size, is a well known example. But if we are interested in the dynamics of the horizon -- the range of its possible shapes and fluctuations -- we must keep finite and non-zero the horizon size. New length scales can be introduced with the addition of rotation and charge: close to the extremality bounds, a throat in the radial direction appears which is much longer than the horizon radius.  This is the basis of the decoupling between the dynamics in the throat and outside it that pervades correspondences of the AdS/CFT type. In other instances, the horizon along some directions is much longer than the scale of variation in directions transverse to it (which is typically set by the surface gravity or horizon temperature). This happens for black branes in asymptotically flat and AdS spacetimes, and in ultraspinning black holes in higher dimensions. It is then possible to isolate the horizon fluctuations of wavelength much longer than the radial scale, and formulate effective theories organized in an expansion in horizon gradients. These are the Fluid/Gravity correspondence \cite{Bhattacharyya:2008jc} and the blackfold approach \cite{Emparan:2009at,Camps:2012hw}.

The large $D$ limit is useful since it provides another generic length scale without recourse to any large charge, rotation, nor long directions along the horizon, i.e., a scale that is present even for neutral, slowly rotating black holes. As we saw earlier, in this limit the horizon radius $r_0$ and its surface gravity $\kappa$ define two separate scales, such that $\kappa^{-1}/r_0\sim 1/D$. Moreover, we have found that quasinormal oscillations of frequency $\omega\ll \kappa$ are decoupled from the physics far from the horizon. Then it seems possible to develop a fully non-linear effective theory of the slow fluctuations of black holes.

\subsection{Hydro Review}
\label{sec:hydroreview}

In view of the close connections between on the one hand the large $D$ limit of Einstein's equations, the Fluid/Gravity correspondence, and the AdS/CFT correspondence, and on the other hand hydrodynamics, 
we would like to begin with a brief review of hydrodynamics as an effective description of physical systems.  More details
can be found in the excellent lectures \cite{Kovtun:2012rj}.

There is lore that most interacting many body systems   -- at nonzero temperature, close to equilibrium, and at long enough time and length scales --  are well described by hydrodynamics.  Hydrodynamics is intended here in a generalized sense that encompasses the Navier-Stokes equations as a particular example.  
One identifies the conserved (or quasi-conserved) currents that follow
from the symmetries of the theory.  In the Lorentzian setting, there is at least a stress-energy tensor $T^{\mu\nu}$, but possibly additional conserved currents $J_I^\mu$ as well, $I = 1, 2, \ldots$.  The conservation rules for these currents $\partial_\mu T^{\mu\nu}=0=\partial_\mu J_I^\mu$ govern the behavior of the slowest modes in the system, and thus determine the physics at the longest time and length scales.

For each of these conservation equations $\partial_\mu T^{\mu\nu} = 0$ and $\partial_\mu J_I^\mu = 0$, 
we can introduce one functional quantity to describe the behavior of the fluid.
For the stress tensor, one conventionally chooses local temperature $T$ and the fluid four velocity $u^\mu$, $u^2 = -1$.  When $J^\mu_I$ exist, we can introduce associated charge densities $\rho_I$ as well. 
From the near equilibrium assumption, $T^{\mu\nu}$ must have an expansion in gradients of $T$ and $u^\mu$ which can then be further constrained by symmetry.  This expansion is often called the stress tensor constitutive relation. 

Let us focus on the particular example of a Lorentz invariant system which has a stress tensor $T^{\mu\nu}$ but no other conserved currents. We can write down the constitutive relation for the stress tensor quite generally as a gradient expansion in the fluid four-velocity $u^\mu$ and the temperature $T$:
\beq\label{hystress}
T_{\mu\nu}=\vep(T) u_\mu u_\nu + P(T) \Pi_{\mu\nu}-2\eta\, \sigma_{\mu\nu} - \zeta\, \Pi_{\mu\nu} \partial_\lambda u^\lambda
+ \Or{\partial^2}
\eeq
where 
\be
\Pi^{\mu\nu} &=& \eta^{\mu\nu} + u^\mu u^\nu \ , \\
\sigma_{\mu\nu} &=& \frac{1}{2} {\Pi_\mu}^\lambda {\Pi_\nu}^\sigma (\partial_\lambda u_\sigma + \partial_\sigma u_\lambda) - \frac{1}{d-1} \Pi^{\mu\nu} \partial_\lambda u^\lambda \ .
\ee
Here $P(T)$ is the pressure and $\varepsilon(T)$ is the energy density, both of which depend on $T$ through equations of state.  The coefficients $\eta$ and $\zeta$ are conventionally called the shear and bulk viscosities respectively.  

The careful reader may be confused at this point by the claim that (\ref{hystress}) is the general result, given that it lacks several possible first order gradient corrections.  For example, no terms involving gradients of $T$ have been included.  A confusing aspect of hydrodynamic descriptions is that $T$ and $u^\mu$ are only well defined concepts when they do not vary from point to point.  We can redefine $T$ and $u$ by gradients, e.g.\ $T \to T + u^\mu \partial_\mu T$, and get a new constitutive relation for $T^{\mu\nu}$ with a correspondingly altered set of hydrodynamic equations $\partial_\mu T^{\mu\nu} = 0$.   We have used this freedom to eliminate several possible gradient corrections from (\ref{hystress}).  The particular choice (\ref{hystress}) is called Landau frame, which defines $u$ as the unit-normalized timelike eigenvector of the stress tensor, i.e., $T^{\mu\nu} u_\nu = \varepsilon u^\mu$. 

The second law of thermodynamics puts constraints on the hydrodynamic expansion.   We expect that the entropy will increase during hydrodynamic flows.  This increase is encoded in a positive divergence of the entropy current.  If we introduce an entropy density $s$ that satisfies the thermodynamic relation
\beq\label{thermo}
\vep +P=Ts \ ,
\eeq
the entropy current at leading order in the gradient expansion is 
\beq\label{hyent}
J_S^\mu=s u^\mu + \Or{\partial^2} \ .
\eeq
From conservation of the stress tensor, it then follows 
\beq\label{hyentprod}
\partial_\mu J_S^\mu=\frac{2\eta}{T}\sigma_{\mu\nu}\sigma^{\mu\nu}+\frac{\zeta}{T}(\partial_\mu u^\mu)^2 + \ldots \ ,
\eeq
which means the viscosities $\eta$ and $\zeta$ must be non-negative for the fluid to satisfy the second law of thermodynamics.

The generality of this discussion, which did not specify the nature of the system being considered, suggests that the hydrodynamic description should also apply to the long wavelength fluctuations of extended black hole horizons. As we mentioned above, for black branes in AdS spacetimes this idea was developed into the subject of the Fluid/Gravity correspondence \cite{Bhattacharyya:2008jc,Hubeny:2011hd}.
Motivated by this context, we will often be interested in conformal fluids. For these fluids, the stress tensor is traceless, which puts many additional constraints on the behavior of the system.  In particular, one finds $\varepsilon = (d-1) P$ and the bulk viscosity vanishes $\zeta = 0$.  Scale invariance further implies the energy density and pressure dependence on the temperature is power law, 
\beq
\varepsilon  = (d-1) P \sim T^d\,.
\eeq 

Other applications will be concerned with the fluids dual to asymptotically flat black branes, which are not conformal \cite{Emparan:2009at,Camps:2010br}. Nevertheless they can be linked to the conformal fluids for AdS black branes, inheriting from them contraints on the constitutive relation \cite{Kanitscheider:2009as,Caldarelli:2012hy}. Specifically, they imply that for an asymptotically flat black $p$-brane in $D=n+p+3$ the fluid has 
\beq\label{AFrels}
\vep = -(n+1) P\,,\qquad \zeta=2\eta\lp \frac1{p}+\frac1{n+1}\rp\,.
\eeq
 
\subsection{Large $D$ vs. AdS/CFT decoupling and Fluid/Gravity}

The large $D$ approach shares features with the decoupling limit of AdS/CFT and with the Fluid/Gravity correspondence, but with important differences between them. In contrast to the long AdS throats of (near-)extremal black holes, where highly-redshifted modes are trapped, the near-horizon zone of a large $D$ black hole is very short. Still, it contains dynamics of its own: the small set of almost static gravitational modes that are decoupled from the far zone, which were described in sec.~\ref{subsec:decmodes}.

The connections are even closer between large $D$ effective theories and the Fluid/Gravity correspondences. In both cases one integrates the radial dependence orthogonal to the horizon,
leaving an effective theory for the fluctuations parallel to the horizon, with frequencies and wavenumbers $\omega, k\ll \kappa, T_H$. The two methods differ in how the expansion parameter $k/T_H$ is made small. In the Fluid/Gravity correspondence, $T_H$ remains finite while $k$ is taken to be infinitesimal, so the theory is organized in powers of $k$, i.e., it is a gradient expansion of hydrodynamic type. In the large $D$ effective theory, instead, $k$ remains finite while $T_H$ diverges like $D$. The smallness of $k/T_H$ does not require a gradient expansion, so the large $D$ effective theory can consistently consider fluctuations with finite (non-infinitesimal) wavenumbers $k$ -- at the price of introducing large dimensionality. For this same reason, it also applies to black holes of finite extent, including the Schwarzschild-Tangherlini solution, and not only to black branes or ultraspinning black holes.
In general, the expansions in $1/D$ and in gradients (powers of $k$) have overlapping but non-coincident ranges of applicability; see \cite{Bhattacharyya:2018iwt,Bhattacharyya:2019mbz,Patra:2019hlq} for a detailed comparison.

\subsection{The large $D$ effective membrane theory}
\label{subsec:effmemb}

The large $D$ effective theory may be conceived of in a very pictorial way. Recall that the effect of taking $D\gg 1$ is to concentrate the gravitational dynamics of the black hole within a thin sliver outside the horizon, leaving a hole in an otherwise undistorted background geometry (e.g., Minkowski or AdS spacetime). We can then envisage the surface of the hole as a membrane in that background, with properties obtained by integrating the Einstein equations near the horizon of the black hole. This is what we describe next.

Currently there exist several formulations of effective theories of large $D$ black holes -- roughly divided into the Japan-Barcelona variety \cite{Emparan:2015hwa,Emparan:2015gva,Suzuki:2015iha,Emparan:2016sjk} and the India (Mumbai-Kanpur-Pune) variety \cite{Bhattacharyya:2015dva,Bhattacharyya:2015fdk,Dandekar:2016jrp,Bhattacharyya:2016nhn,Bhattacharyya:2017hpj}.\footnote{Contributions from Beijing and other places will be mentioned later.} They employ different ans\"atze, gauges and variables, and result in equations whose complete equivalence has not been established in general. Nevertheless, they all share the same basic concepts, and whenever precise comparisons have been made, e.g., in computations of quasinormal frequencies, they agree perfectly. Each formulation has its advantages, but it seems fair to say that the Indian formulation, which is fully covariant, is more elegant and encompasses in a single, compact set of equations what the Japan-Barcelona methods achieve with several partially gauge-fixed theories. Thus, in the following we explain the Indian `hardcover edition' of the effective theory -- the so-called ``large $D$ membrane paradigm''. Afterwards, in sec.~\ref{subsec:eftbb} we will devote attention to a remarkable subcase -- a very usable `pocket edition' of the equations -- initially found in \cite{Emparan:2015gva}.

\subsubsection{The membrane equations}
\label{subsec:membeqs}

We start again from the Schwarzschild-Tangherlini black hole, now written in Kerr-Schild coordinates,
\beq
\d s^2=-\d t^2+\d r^2+ r^2 \d \Omega_{n+1}+\lp\frac{r_0}{r}\rp^n (\d t-\d r)^2\,.
\eeq
This is convenient since the deviations from flat space appear in a linear-looking form, and  when $n$ is large, the separation is neat between the flat space at $r>r_0$ and the thin, membrane-like, near-horizon region $r\approx r_0$. Observe that the one-form $\d t -\d r$ is null both in the complete geometry and in the flat spacetime metric. 

A covariant, boosted version is obtained if we replace $\d t$ with a timelike one-form $u$ with unit norm (a velocity vector) in the Minkowski spacetime, so the metric is
\beq
g_{MN}=\eta_{MN}+\lp\frac{r_0}{r}\rp^n \lp \partial_M r -u_M\rp \lp \partial_N r -u_N\rp
\eeq
with the radius function
\beq
r^2(x) =x^M x^N \lp \eta_{MN}+u_M u_N\rp\,.
\eeq
Written like this, the solution contains a constant parameter, the horizon radius at $r=r_0$ with normal $\d r=\partial_M r\, \d x^M$,  and a constant one-form, the velocity $u$. We will turn them into collective fields by letting them vary over the horizon much more slowly than the scale $\sim D/r_0$ of radial gradients. The horizon can then have non-uniform, varying shapes, and move with varying velocity in tangential directions. We saw in sec.~\ref{subsec:universal} that, when $D$ is large, the rotation of the black hole acts like a local boost along the horizon. Then, with this formalism we expect to be able to describe large-$D$ rotating black holes and their fluctuations.

When the constant parameters are allowed to vary along the horizon, the metric will not solve anymore the Einstein equations $R_{MN}=0$. To remedy this, one introduces small corrections to the metric, of the order $1/D$ of the ratio between horizon gradients and radial gradients, such that the Einstein equations can be satisfied consistently in an expansion in $1/D$. In the spirit of a `radial ADM decomposition', one separates the radial evolution equations from the constraint equations on surfaces at constant radius. The hierarchy between radial gradients and gradients along the horizon allows to solve explicitly the radial dependence of the evolution equations. Then, the constraint equations, with dependence only along horizon directions, become effective equations for the collective fields. This philosophy, which is common to all the large-$D$ effective theories, was first employed in the Fluid/Gravity correspondence \cite{Bhattacharyya:2008jc}, and later used to derive the blackfold equations \cite{Camps:2010br,Camps:2012hw}.

According to this, we consider a metric of the form
\beq
g_{MN}=g_{MN}^{(0)} + \frac{g_{MN}^{(1)}}{n}+\dots
\eeq
with
\beq\label{KSansatz}
g_{MN}^{(0)}=\eta_{MN}+\rho^{-n} \lp n_M -u_M\rp \lp n_N -u_N\rp\,.
\eeq
Here $\rho(x)$ is now a function of the coordinates in flat space, $n_M$ is a one-form field normal to surfaces of constant $\rho$,
\beq
n_M=\frac{\partial_M\rho}{\sqrt{\eta^{NP} \partial_N\rho\partial_P\rho}}\,,
\eeq
(not to be confused with the number of dimensions $n$) and $u_M$ is a one-form velocity field tangent to these surfaces, 
\beq
\eta^{MN} u_M u_N=-1\,,\qquad \eta^{MN} u_M n_N=0\,.
\eeq
These one-forms are unit-normalized with the Minkowski metric. In the full spacetime we have
\beq
g^{MN}_{(0)}n_M n_N=1-\rho^{-n}\,,
\eeq
which implies that $\rho=1$ is a null hypersurface generated by $u$. This will be the event horizon. As a submanifold of flat space, $\rho=1$ is a timelike codimension-one surface which is regarded as the membrane worldvolume, and $u$ is a velocity vector in its tangent space.

One can see that this geometry will be a good approximation to the metric of a (boosted) Schwarzschild black hole, within a region of size $\sim 1/D$ of the horizon, if $\nabla^2\lp\rho^{-n}\rp=0$ and
\beq\label{divu}
\nabla^A u_A=0\,.
\eeq
Here and henceforth all indices $A, B\ldots$ and derivatives refer to the metric on the membrane.

The Einstein radial evolution equations for $g_{MN}^{(1)}$ are now decomposed into scalars, vectors, and tensors according to their tensorial character under the group $SO(n+1)$ of rotations orthogonal to $u_M$ and $n_M$ -- like we did in the study of linear perturbations.\footnote{Actually, in the derivation one separates a finite number $p$ of directions where there is non-trivial dependence, and preserves a large $SO(n-p+1)$ symmetry. Remarkably, the final equations can be covariantized without distinguishing worldvolume directions.} The radial dependence in $g_{MN}^{(1)}$ can then be integrated, and the remaining constraint equations only involve the collective fields. They take the form
\beq\label{consteq}
\cP_C^A\lp u^B\lp \nabla_B u_A-K_{BA}\rp+\frac{\nabla_A K-\nabla^2 u_A}{K}\rp=0\,.
\eeq
Here
\beq\label{Pproj}
\cP_B^A=\delta_B^A+u_B u^A\,,
\eeq
is the spatial projector orthogonal to the velocity on the membrane worldvolume, $K_{AB}$ is the extrinsic curvature of the membrane in Minkowski space, and $K$ its trace.

Together, \eqref{divu} and \eqref{consteq} constitute a set of $n+2$ equations for $n+2$ variables: the $n+1$ independent components of the unit velocity field along the membrane, plus the function that specifies the shape of the membrane. The derivation ensures that solving these equations provides a solution to the vacuum Einstein equations to first order in $1/D$. As a check, it has been verified that the decoupled vector and scalar quasinormal frequencies of the Schwarzschild black hole that we studied in sec.~\ref{subsec:decmodes} are correctly reproduced from the linearized fluctuations of a spherical membrane \cite{Bhattacharyya:2015dva,Bhattacharyya:2015fdk}.

\subsubsection{Playing with soap bubbles}

The elastic aspects of the membrane are more manifest for stationary configurations where the velocity is proportional to a Killing vector on the membrane, 
\beq
u^A=\gamma k^A\,,
\eeq
with $\gamma=|k|^{-1}$ the Lorentz redshift factor relative to Killing time. Then the equations simplify and admit a first integral, reducing to \cite{Dandekar:2017aiv,Mandlik:2018wnw}
\beq\label{soapbubble}
\gamma^{-1}K= 2\kappa\,.
\eeq
The integration constant $\kappa$ corresponds to the surface gravity of the black hole, normalized relative to Killing time.

This equation was initially derived following a different approach in \cite{Emparan:2015hwa,Suzuki:2015iha}, who referred to it as the `soap bubble equation'. When the membrane is static, $\gamma=1$, we recover the Laplace-Young equation that governs the shape of fluid interfaces (such as soap films) as surfaces of constant mean curvature. Equation \eqref{soapbubble} is a relativistic version of the Laplace-Young equation. It can be extended to soap films in curved backgrounds by including the gravitational time dilation on the same footing as the Lorentz dilation factor $\gamma$ \cite{Emparan:2015hwa,Suzuki:2015iha}. One can solve this equation to find bubbles for rotating Myers-Perry black holes \cite{Suzuki:2015iha,Mandlik:2018wnw} and for `black droplets' in AdS \cite{Emparan:2015hwa}.

We are thus led to a satisfyingly suggestive picture: large $D$ black holes are spherical and ellipsoidal soap bubbles in flat spacetime, and their tension is given by the surface gravity. Black strings and black branes are also soap films at large $D$, but if you ask any child, they will tell you that these bubbles must be unstable. And of course they are right.

\subsubsection{Fluid stress-energy, entropy, action, and coupling to radiation}
\label{sssection:stress}

Since \eqref{consteq} arises as momentum-type constraint equations on the surface of the membrane, one naturally expects that they can be interpreted as the conservation equations of a quasilocal stress-energy tensor for the membrane,
\beq\label{memstresseq}
\nabla^A T_{AB}=0\,.
\eeq
This stress-energy tensor was found in \cite{Bhattacharyya:2016nhn} to be
\beq\label{memstress}
16 \pi T_{AB}= K u_A u_B+K_{AB} -2 \sigma_{AB}\,,
\eeq
where $\sigma_{AB}$ is the shear tensor of the velocity field $u$. Moreover, an entropy current associated to the horizon area was also identified as
\beq\label{JS}
J_S=\frac{u}{4}\,.
\eeq

We can now attempt to interpret equations \eqref{memstresseq} and \eqref{JS} as describing an effective fluid on a dynamical, curved membrane.\footnote{These are similar in many respects to the equations of the blackfold approach \cite{Emparan:2009at}, but, being obtained in different expansions, the effective fluids are not the same except in common regimes of validity. The blackfold fluid of \cite{Emparan:2009at} has non-zero energy density and is compressible.}
Following \cite{Dandekar:2017aiv} we first subtract an automatically conserved Brown-York term, $16 \pi T_{AB}^\text{BY}=K_{AB}-K g_{AB}$, which leaves
\beq\label{fluidstress}
16 \pi T_{AB}^\text{fluid}= K P_{AB} -2 \sigma_{AB}\,.
\eeq
Comparing this and \eqref{JS} with the general expressions of first-order hydrodynamics \eqref{hystress}, \eqref{hyent}, \eqref{thermo}, yields the effective energy density, pressure, shear viscosity, entropy density and temperature as
\beq\label{memfluid}
\vep=0\,,\qquad P=\frac{K}{16\pi}\,,\qquad \eta=\frac{1}{16\pi}\,,\qquad s=\frac14\,,\qquad T=\frac{K}{4\pi}\,.
\eeq
Observe that \eqref{divu} implies that the fluid is incompressible, so $\zeta=0$.  It is nevertheless a peculiar fluid since its energy density vanishes. We will see in sec.~\ref{subsec:eftbb} that a more conventional fluid is found when the velocity is non-relativistic and the membrane geometry is flat.

Additionally, \eqref{divu} means that entropy is conserved to leading order at large $D$. At the next order this equation is modified to \cite{Dandekar:2016fvw}
\beq\label{divu2}
\nabla_A u^A =\frac1{8K}\sigma_{AB}\sigma^{AB}\,,
\eeq
which, plugging in \eqref{memfluid}, reproduces precisely the standard hydrodynamic result for viscous entropy production \eqref{hyentprod}.

Within this same context, ref.~\cite{Dandekar:2017aiv} made a remarkable observation. The stress-energy tensor \eqref{memstress} satisfies the equation
\beq\label{KTeq}
K_{AB}T^{AB}=0
\eeq 
to leading order at large $D$. This equation is known to generically govern the elastic dynamics of relativistic branes \cite{Carter:2000wv,Emparan:2009at,Camps:2012hw} (it generalizes the geodesic equation for particles). Here it is satisfied as an identity, although only in the first non-trivial order in the $1/D$ expansion. Then \cite{Dandekar:2017aiv} noted that it is possible to `improve' the membrane stress-energy tensor such that \eqref{KTeq} is an exact algebraic identity at finite values of $D$. The improved tensor is
\beq\label{impstress}
16 \pi \tilde T_{AB}= \tilde K \cP_{AB} -2 \sigma_{AB}+ K_{AB}-K g_{AB}\,,
\eeq
where $g_{AB}$ is the metric induced on the membrane, and the improvement term is
\beq
\tilde K =\frac{K^2-K^{AB}K_{AB}+2K^{AB}\sigma_{AB}}{K+K_{AB}u^A u^B}\,.
\eeq
Using the appropriate rules for counting powers of $D$, one can verify that $\tilde K\to K$ when $D\to \infty$, so $\tilde T_{AB}\to T_{AB}$ is recovered. Since \eqref{KTeq} is satisfied, it defines consistent membrane dynamics at finite $D$. Moreover, the improved stress tensor also reproduces the equation for entropy production \eqref{hyentprod}.

For stationary membranes the conservation of $\tilde T_{AB}$ leads to the improved soap-bubble equation $\gamma^{-1}\tilde K= 2\kappa$. \cite{Dandekar:2017aiv} proved that this equation follows from the extremization of the worldvolume action
\beq
I=\frac1{16\pi}\int_M \sqrt{-g}\lp K-2\gamma\kappa\rp\,.
\eeq
Using this action one defines thermodynamic quantities for stationary membranes: energy, entropy, and temperature. Surprisingly, for static spherical membranes these reproduce the properties of Schwarzschild black holes exactly at finite values of $D$. This agreement also holds if one includes a background cosmological constant, but it does not when applied to rotating black holes.

Finally, since we are viewing the black hole as a physical membrane moving in a background geometry, we can envisage computing the gravitational radiation that this motion creates. In the same spirit as the quadrupole coupling of a source to the radiation field, we can couple the membrane stress tensor to linearized gravitational perturbations of the background. It is clear from our earlier discussion that such a coupling between near and far zones is non-perturbative in $1/D$: the emission of radiation from the slowly oscillating black hole is suppressed exponentially, $\sim e^{-D}$, or even factorially, $D^{-D}$ \cite{Emparan:2013moa,Bhattacharyya:2016nhn,Andrade:2019edf}.

The theory of the radiation emission from the coupling to the membrane has been worked out in detail in \cite{Bhattacharyya:2016nhn}. It will be very interesting to put it to use to calculate the gravitational waves emitted from, e.g., a non-linearly relaxing black hole, or, much more ambitiously, the collision and merger of two black holes. As we will see in sec.~\ref{sec:collisions}, these phenomena have already been studied in a large $D$ membrane approach, but so far gravitational wave emission is estimated only heuristically \cite{Andrade:2019edf}.

\subsubsection{Extensions}
\label{sssec:generalize}

The membrane equations \eqref{consteq} have been extended in many ways:
\begin{itemize}
\item The membrane can carry charge as a conserved current in its worldvolume, and much of the previous discussion extends to this case \cite{Bhattacharyya:2015fdk,Bhattacharyya:2016nhn,Mandlik:2018wnw,Kundu:2018dvx}.

\item Subleading corrections in $1/D$, as well as the inclusion of a cosmological constant, have been successfully incorporated into the equations \cite{Dandekar:2016fvw,Bhattacharyya:2017hpj,Kundu:2018dvx,Bhattacharyya:2018szu,Biswas:2019xip}.

\item Membrane equations have been derived for gravitational theories with higher-curvature terms in the action, namely Einstein-Gauss-Bonnet gravity \cite{Saha:2018elg} and a general fourth-derivative gravity theory \cite{Kar:2019kyz}. The analysis is only possible perturbatively in the couplings to the higher-derivative terms, a point that we will return to later in sec.~\ref{sec:hider}.

\end{itemize}

In all these extensions, whenever there is a previous calculation of quasinormal mode frequencies at large $D$, one finds agreement.

\subsubsection{Limitations}
\label{sssec:limits}

It is clear from the derivation of the effective theory that the equations will not be valid when temporal gradients or spatial gradients along the horizon are $\Or{D/r_0}$ (where $r_0$ is a characteristic length of the horizon). Since the effective equations (of sec.~\ref{subsec:membeqs}, and also those of \cite{Suzuki:2015iha,Tanabe:2015hda}) capture correctly the properties of Myers-Perry black holes and black rings, it is possible to use these theories for stationary configurations where the horizon modifies its shape away from sphericity by amounts of order $\Or{r_0}$. The effective theories can also deal with non-uniform black strings with inhomogeneity of size $\Or{r_0/\sqrt{D}}$ (see sec.~\ref{sec:GL}), but not on much shorter scales $\Or{r_0/D}$.\footnote{Non-uniformity on black strings on scales $\Or{r_0/D}$ \emph{can} be described at large $D$ if one leaves the remit of these effective theories \cite{Emparan:2019obu}.}

However, when time evolution is involved, the effective theories in the Japan-Barcelona framework \cite{Emparan:2015hwa,Suzuki:2015iha,Emparan:2015gva,Tanabe:2015hda,Tanabe:2016pjr,Emparan:unp} encounter a limitation. In these formulations, one has to first specify a stationary membrane shape by solving the soap bubble equation \eqref{soapbubble}; then, one lets this shape fluctuate dynamically with an amplitude that is $\Or{r_0/D}$, so it remains within the near-horizon region of the stationary solution. Attempting to make this amplitude larger, of size $\Or{r_0}$, renders the fluctuation non-normalizable in the near zone, thus entering the regime of non-decoupled dynamics. As a consequence, these theories can only deal with time-dependent variations of the radius of the horizon that are small, with amplitude not more than $\Or{r_0/D}$ away from the stationary shape.\footnote{One might envisage a membrane that changes slowly away from the stationary shape, with time gradients of $\Or{1/r_0}$, building up over time to a large, $\Or{r_0}$ deformation in shape. The Jp-Bcn framework cannot accommodate for this, and it is not clear either if it is compatible with the constraint in the Indian theories that the velocity be tangential to the horizon.} 

This limitation is not apparent in the India formulation, so in principle \eqref{divu} and \eqref{consteq} could describe large, $\Or{r_0}$ time-dependent fluctuations of an amplitude comparable to the horizon size.  However, these equations have not been applied yet to investigate non-linear time-dependent processes. So far, all such studies employ the effective theory of black branes that we will see in sec.~\ref{subsec:eftbb}, for which, as eq.~\eqref{r1pluslog} shows, the fluctuations in $r_0$ are of size $\Or{1/D}$. In the derivation of these equations in \cite{Dandekar:2016jrp}, this condition follows necessarily from \eqref{divu}.

A complete understanding of the scope of the large $D$ effective theories would require to verify explicitly whether the equations \eqref{divu} and \eqref{consteq} can overcome this limitation.\footnote{The Indian framework could not be extended to cases of large temporal gradients if such fluctuations would always belong to non-decoupled dynamics and proceed on fast time scales, of order $1/D$. Heuristically, recall that non-decoupled frequencies are $\omega \sim D/r_0$, and they may be excited when the black hole fluctuates. A deformation away from a stationary solution of the same order (in $D$) as $r_0$ might be expected to excite frequencies $\Or{D/r_0}$, so the evolution will happen on fast timescales $\Or{r_0/D}$.} As a concrete and interesting test, one may try to derive from them a set of two non-linear $1+1$ PDEs for black strings with large dynamical fluctuations, with an amplitude that is an $\Or{1}$ fraction of the black string thickness, and which reduce to \eqref{consm} and \eqref{consmom} when the amplitudes are $\Or{r_0/D}$. 

Another interesting phenomenon where this question is relevant is the evolution of the merger of two black holes. The method employed in \cite{Andrade:2018yqu,Andrade:2019edf} to investigate these collisions, which we will review in sec.~\ref{subsec:ccv}, requires that the horizon be within an amplitude $\Or{r_0/D}$ of the stationary shape.  If the large $D$ techniques could go beyond this regime, a more complete description of the merger would be possible.

\subsubsection{Membrane paradigms}

Let us comment on the relation between the large $D$ effective theories (of any variety) and the older membrane paradigm of \cite{Damour:1982,Price:1986yy}. In spite of broad similarities, the large $D$ theories differ from it in crucial ways. Like the large $D$ equations for the collective coordinates, the membrane paradigm consists of constraint equations on a surface on, or just outside, a horizon. However, unlike the large $D$ approach, these constraints are not imposed after having integrated the dynamics near the horizon. In fact, they apply generally to any null hypersurface in Einstein's theory in any dimension, without requiring any separation of scales. They only employ arbitrarily large boosts close to a Rindler horizon. 

The membrane paradigm is a very suggestive way of writing the boundary conditions on a null hypersurface, but not more than that. The radial dependence is not integrated, so it is not an effective theory in the sense that we discussed above, with dynamical information about the possible shapes and fluctuations of the horizon. As a consequence, the membrane paradigm is more general, but also more limited than the large $D$ equations for solving the physics of black holes.

\subsection{Effective theory for black branes}
\label{subsec:eftbb}

\subsubsection{Black brane equations}
\label{sssec:bbeqns}

Let us specialize now to the fluctuations of a black brane extended along a finite number $p$ of spatial directions. It is known from the work of \cite{Kol:2004pn,Asnin:2007rw,Camps:2010br,Emparan:2013moa} that the most interesting dynamics of a black brane -- the Gregory-Laflamme instability -- occurs over length scales of order $\sim 1/\sqrt{D}$ along the horizon. Therefore we appropriately rescale the spatial directions $\sigma^i\to\sigma^i/\sqrt{n}$, so that the background metric where the membrane moves is 
\beq\label{flatbrane}
\d s^2=-\d t^2 + \frac1{n}\delta_{ij}\d\sigma^i \d\sigma^j+\d \rho^2 +\rho^2\d\Omega_{n+1}\,.
\eeq
Here $i = 1, \ldots, p$, so now 
\beq
n=D-p-3\,.
\eeq
Time is not rescaled since the frequency of Gregory-Laflamme modes is of order one. Then we will consider small velocities
\beq
u_M=\lp -1+\Or{n^{-1}},\frac{v_i(t,\sigma)}{\sqrt{n}}\rp\,.
\eeq
If the fluctuating radius of the membrane surface is at
\beq
\rho^n=m(t,\sigma)
\eeq
then, when $n$ is large,
\beq\label{r1pluslog}
\rho=1+\frac{\ln m(t,\sigma)}{n}\,.
\eeq
Plugging all of this into the membrane equations \eqref{divu} and \eqref{consteq}, they become \cite{Dandekar:2016jrp}
\beq\label{consm}
\partial_t m+\nabla_i(m v^i)=0
\eeq
and
\beq\label{consmom}
\partial_t(m v_i)+\nabla^j (m v_i v_j +\tau_{ij})=0
\eeq
with
\beq\label{stresstau}
\tau_{ij}=-\epsilon m\delta_{ij}-2m\nabla_{(i}v_{j)}-m\nabla_j\nabla_i\ln m\,.
\eeq
We have introduced here a sign parameter $\epsilon$ to accommodate the rather remarkable fact that the  same effective equations apply for a black brane in AdS, only changing\footnote{In AdS the worldvolume is infinite-dimensional but we consider that there is non-trivial dependence in only a finite number $p$ of directions.}
\beq\label{AFAdS}
\epsilon=\begin{cases}+1
&\text{for asymptotically flat (AF)}\,,\\ -1 &\text{for AdS}\,.
\end{cases}
\eeq
The easiest way to derive this connection uses the AdS/Ricci flat correspondence \cite{Caldarelli:2012hy,Caldarelli:2013aaa}, which implies a close relationship between brane solutions in AdS and asymptotically flat space. In the limit $D\to\infty$, it yields the result above.

The equations are invariant under constant Galilean boosts,
\beq\label{galiboost}
\sigma^i\to\sigma^i-w^i t\,,\qquad v_i\to v_i+ w_i\,,
\eeq
which is expected since we have taken small velocities.

There is a definite hydrodynamic flavor to these equations: \eqref{consm} is the continuity equation for a fluid with mass density $m$ and velocity field $v_ i$, and \eqref{consmom} is the conservation equation for the momentum density $m v_i$ with stress tensor $\tau_{ij}$. This is a non-relativistic limit of the fluid of the previous section, now moving in the flat geometry \eqref{flatbrane}. 
This fluid is compressible, with pressure
\beq
P=-\epsilon m\,,
\eeq
and shear and bulk viscosities (the latter is non-zero for an AF black $p$-brane, see \eqref{AFrels})
\beq
\eta=m\,,\qquad \zeta =\frac{1+\epsilon}{p}\eta\,.
\eeq
The entropy density is $s=4\pi m$, and since the mass is conserved, so is the entropy too.
Again, viscous entropy production enters only at the next order in $1/n$ \cite{Herzog:2016hob}. However, entropy can be generated at leading order in charged fluids when there is charge diffusion, i.e., via Joule heating \cite{Emparan:2016sjk}.

These properties agree with the large $D$ limit of the effective fluids for AF and AdS black branes in the hydrodynamic theories of \cite{Bhattacharyya:2008mz,Camps:2010br}. But the differences in the theories are instructive: the constitutive relation for the stress-energy tensor \eqref{stresstau} contains terms with one and two derivatives, and, unlike in hydrodynamics, here these are leading order terms (in $1/D$) and hence allowed to become as large as the zero-derivative, perfect fluid pressure term. The reason why the derivative expansion terminates in \eqref{stresstau} is not that higher gradients are assumed small and neglected, but that an infinite number of higher-order transport coefficients of the black brane vanish when $D\to\infty$. We will return to this point in sec.~\ref{subsec:holohydro}, but for now let us mention that the elastic interpretation of the equations (where $m$ plays the role of the membrane radius) makes this feature less mysterious, as these two-derivative terms are needed to complete the expression for the extrinsic curvature of the membrane to leading order at large $D$ \cite{Emparan:2016sjk}.

A conveniently simple form of the equations is obtained by introducing
\beq
p_i=m v_i +\nabla_i m\,.
\eeq
Then \eqref{consm} and \eqref{consmom} become
\be
\label{conshigherdone}
\partial_t m - \nabla^2 m &=& -\epsilon\, \nabla_i p^i \ , \\
\label{conshigherdtwo}
\partial_t p_i - \nabla^2 p_i &=& \nabla_i m - \nabla^j \left( \frac{p_i p_j}{m} \right) \ .
\ee

These equations were first derived in \cite{Emparan:2015gva} using the metric ans\"atze
\be\label{EFbrane}
\d s^2 = 2 \d t \, \d \rho - A \, \d t^2 - \frac{2}{n} C_i \d \sigma^i \, \d t + \frac{1}{n} G_{ij} \d \sigma^i \, \d \sigma^j + \rho^2 \d \Omega_{n+1} 
\ee
in the AF case, and
\be\label{AdSbrane}
\d s^2 = 2 \d t \, \d \rho+ \rho^2 \lp - A \, \d t^2 - \frac{2}{n} C_i \d \sigma^i \, \d t + \frac{1}{n} G_{ij} \d \sigma^i \, \d \sigma^j\rp 
\ee
in AdS. Requiring that these metrics solve the Einstein equations in an expansion in $1/n$ yields 
\be\label{ACG}
A &=& 1 - \frac{m(t, \sigma)}{\rho^n} \ , \; \; \;
C_i = \frac{p_i(t, \sigma)}{\rho^n} \ , \\
G_{ij} &=& \delta_{ij} + \frac{1}{n} \frac{p_i(t, \sigma) p_j(t, \sigma)}{m(t, \sigma)\rho^n} \, ,
\label{ACGtwo}
\ee
with $m$ and $p_i$ constrained to satisfy (\ref{conshigherdone}) and (\ref{conshigherdtwo}). In order to transform the Eddington-Finkelstein metric \eqref{EFbrane} into the Kerr-Schild form \eqref{KSansatz}, change $t+\rho \to t$ and identify $m(t,\sigma)/\rho^n\to \rho^{-n}(t,\sigma)$. 

In this form, (\ref{conshigherdone}) and (\ref{conshigherdtwo}) resemble (inhomogeneous) heat equations, whose dissipative properties make them well behaved and suited for numerical evolution. Intriguingly, all the non-linearity of the gravitational Einstein equations has been reduced to the last term in \eqref{conshigherdtwo}.

 The sign difference $\epsilon$ between AdS and AF black branes is not innocuous. Consider small fluctuations and long wavelengths, such that we can neglect the non-linear term and the two-derivative terms in \eqref{conshigherdone} and \eqref{conshigherdtwo}. Then $m$ and $p_i$ satisfy wave equations
\beq
(\partial_t^2-\epsilon \nabla^2) (m, p_i) \approx 0\,.
\eeq
For AdS this describes sound waves with speed one, which, in physical length units, is the correct sound velocity $c_s=1/\sqrt{n}$ of a conformally invariant fluid at large $n$. For asymptotically flat black branes, the sound speed is instead imaginary: density fluctuations are unstable and tend to clump. This is the Gregory-Laflamme instability of the black brane. We will return to it at length in sec.~\ref{sec:GL}.

\subsubsection{Extensions}
\label{sssec:bbextend}

The black brane effective equations have been taken further in several different directions:\footnote{These have been obtained independently of the extensions mentioned in sec.~\ref{sssec:generalize} of the `parent equations' \eqref{divu}, \eqref{consteq}.}
\begin{itemize}

\item branes with electric charge and $p$-brane charge, and external electric field \cite{Emparan:2016sjk}. 

\item higher orders in $1/D$ \cite{Herzog:2016hob,Rozali:2016yhw,Emparan:2018bmi}.

\item curved background geometries (deformed boundary metrics) \cite{Andrade:2018zeb}.

\item higher-curvature theories \cite{Chen:2017rxa,Chen:2018nbh,Chen:2018vbv}.

\item other classes of deformations of AdS black branes \cite{Iizuka:2018zgt}.

\item similar effective theories for  finite black holes \cite{Tanabe:2015hda,Tanabe:2015isb,Tanabe:2016pjr,Chen:2017wpf,Herzog:2017qwp,Chen:2018vbv}.

\end{itemize}

The simplicity that these equations bring to the non-linear dynamics of black branes has made them a successful tool for applications of the large $D$ expansion. Some of these follow from an unexpected feature of the equations that we explain next.

\subsubsection{Black holes as blobs on a brane}
\label{sssec:blobs}

Although the effective equations \eqref{conshigherdone} and \eqref{conshigherdtwo} have been derived for fluctuating black branes with an infinite extent, they turn out to also capture a surprising amount of the basic physics of localized black holes, affording a simpler approach to them than the general membrane equations.

\cite{Andrade:2018nsz} observed that \eqref{conshigherdone} and \eqref{conshigherdtwo} support a class of time-independent solutions localized with Gaussian profiles.  Restricting to 2+1 dimensions with $\delta_{ij} \d \sigma^i \d \sigma^j = \d r^2 + r^2 \d \phi^2$, the simplest of these solutions is 
\be\label{Schblob}
 m(r) &=& m_0\, e^{- r^2/2} \,, \quad p_r(r) = \partial_r m \,,
 \ee
(so $v_i=0$) with constant $m_0$. Then \cite{Andrade:2018nsz} (following \cite{Suzuki:2015axa}) claimed that this Gaussian blob on a membrane (whose density does not asymptote to a constant but vanishes exponentially) approximates well the geometry and properties of the large $D$ Schwarzschild black hole.

In order to see how this can be so, and to make contact with the ansatz \eqref{EFbrane} for a 2-brane, we go (once again) to the mother of all black hole metrics \eqref{SchD} and write it in $D=n+2+3$ dimensions as
\beq\label{Schn5}
\d s^2=-\lp 1-\hat r^{-(n+2)}\rp \d\hat t^2+\frac{\d\hat r^2}{1-\hat r^{-(n+2)}}+\hat r^2\lp \d\theta^2+\sin^2\theta \d\phi^2+\cos^2\theta \d\Omega_{n+1}\rp
\eeq
(we set $r_0=1$ for simplicity). Define new coordinates $\sR$ and $r$ as
\beq\label{toblob}
\sR^{1/n}=\hat r\cos\theta\,,\qquad \frac{r}{\sqrt{n}}=\hat r\sin\theta
\eeq
(such that $\hat r^n =\sR\, e^{r^2/2}$ as $n\to\infty$) and also an Eddington-Finkelstein time coordinate
\beq
t=\hat t -\frac1{n}\ln\lp \sR-e^{-r^2/2}\rp\,.
\eeq
Now take the large $n$ limit of \eqref{Schn5} keeping $t$, $r$ and $\sR$ finite; the metric takes the form of a black membrane \eqref{EFbrane} with the Gaussian profile of \eqref{Schblob}, normalized to $m_0=1$. Observe that \eqref{toblob} implies that the blob extends over a region of small angular size $\Delta\theta\sim 1/\sqrt{n}$ around $\theta=0$. Indeed,  most of the area of the sphere $S^{n+3}$ comes from this region, and for this same reason the total mass and horizon area of the black hole are obtained with exponential accuracy by integrating the mass and area densities of the membrane. Heuristically, these blobs can be regarded as endpoints of the instability of a black brane \cite{Emparan:2015gva}.\footnote{To properly regularize it, put the unstable membrane in a finite box and take the limit where the box becomes infinite keeping the total energy fixed. Bear in mind that when $D\to \infty$ the black brane instability forms the blobs and ends without singular pinches to zero thickness \cite{Emparan:2015gva,Emparan:2018bmi}.}   

This analysis can be extended to Myers-Perry rotating black holes. For a single spin, the profiles are 
 \be
 m &=& m_0 \exp \left( - \frac{r^2}{2(1+a^2)} \right) \,, \quad  p_r = \partial_r m \,, \quad  p_\phi =  m r^2 \frac{a}{1+a^2}  \,,
 \ee
so the effect of the rotation is to spread the blob along the rotation plane. Note the ratio $p^\phi / m = \frac{a}{1+a^2} = \Omega$ is the angular velocity. Integrating the relevant stress-energy components over the Gaussian profile produces a total mass $M$ and angular momentum $J$ for the spinning disks, from which one finds $J/M = 2 a$. 
 \begin{figure}
\begin{center}
a) \includegraphics[width=.4\textwidth]{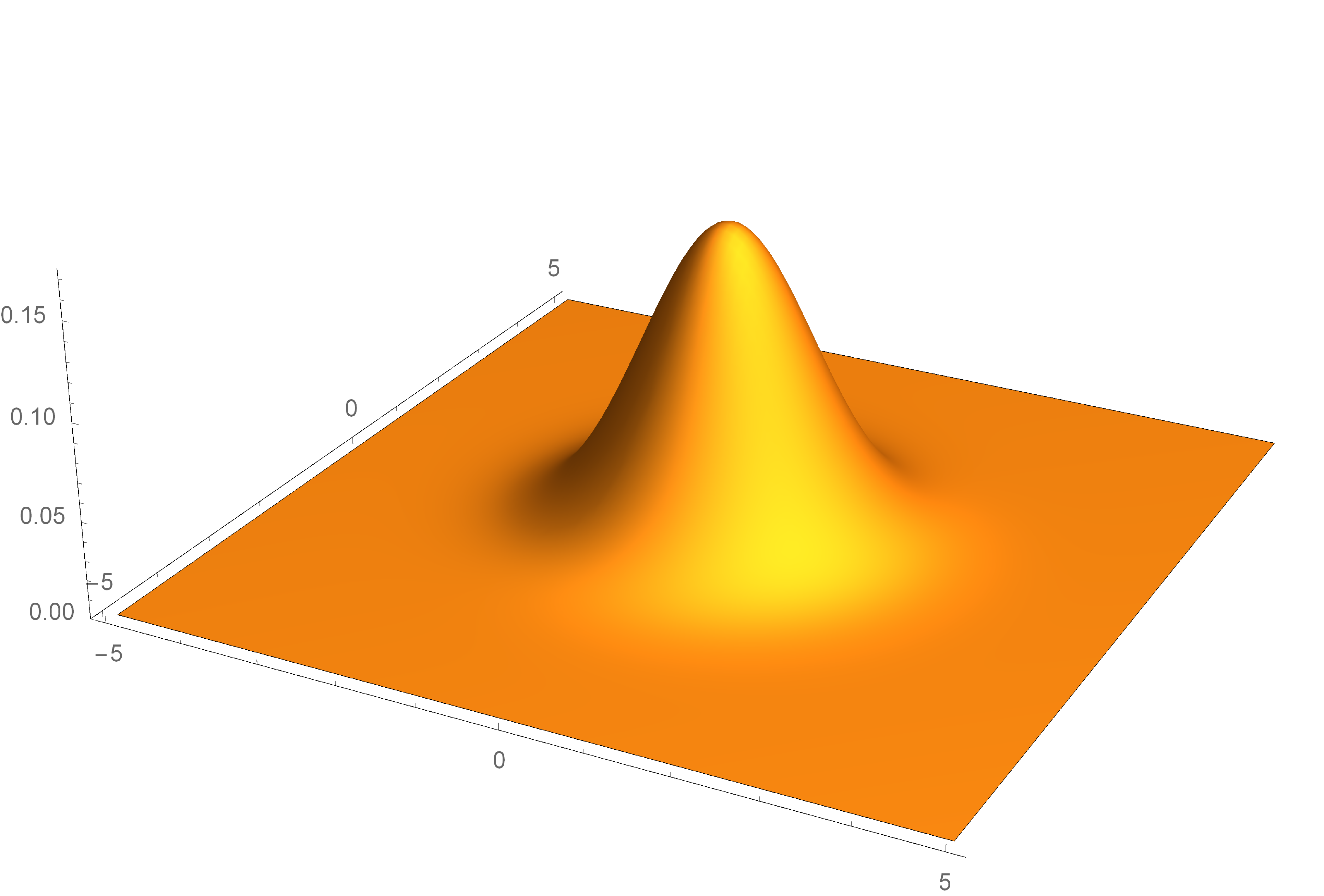}
b) \includegraphics[width=.4\textwidth]{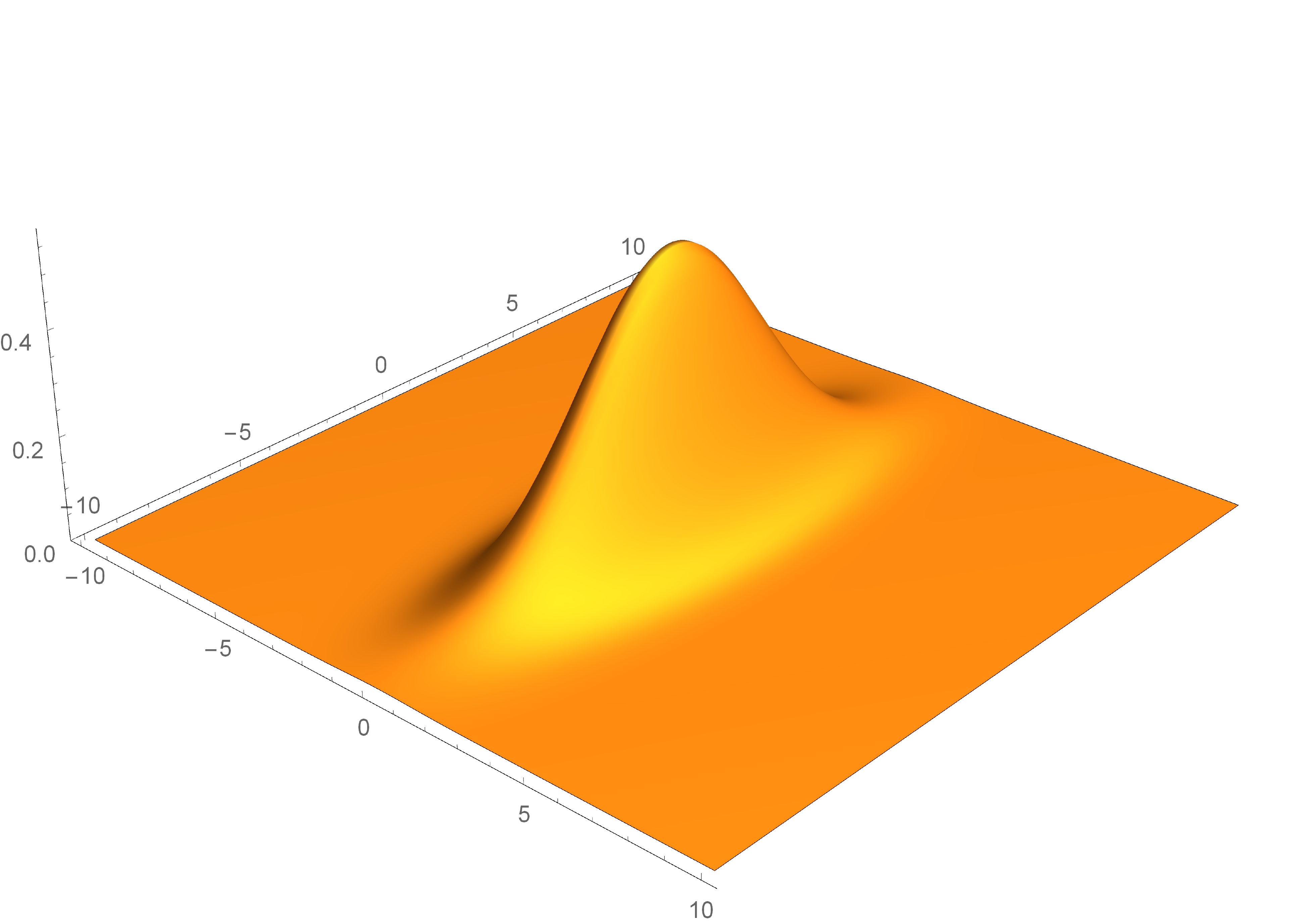}
\end{center}
\caption{The profile $m(r)$  for a) the spinning black disk; b) the spinning black bar.
\label{fig:Gaussianblobs} [[ \cite{Andrade:2018nsz} figs.\ 1 and 3 ]]
}
\end{figure}
%
 The angular velocity is restricted to the range $\Omega \in [0,1/2]$, reaching a maximum at $a=1$.  For velocities $\Omega < 1/2$, $a$ can take one of two values; the angular velocity vanishes both in the static $a \to 0$ and ``ultra-spinning'' $a \to \infty$ limit.  
 
\begin{figure}
 \begin{center}
 \includegraphics[width=.65\textwidth]{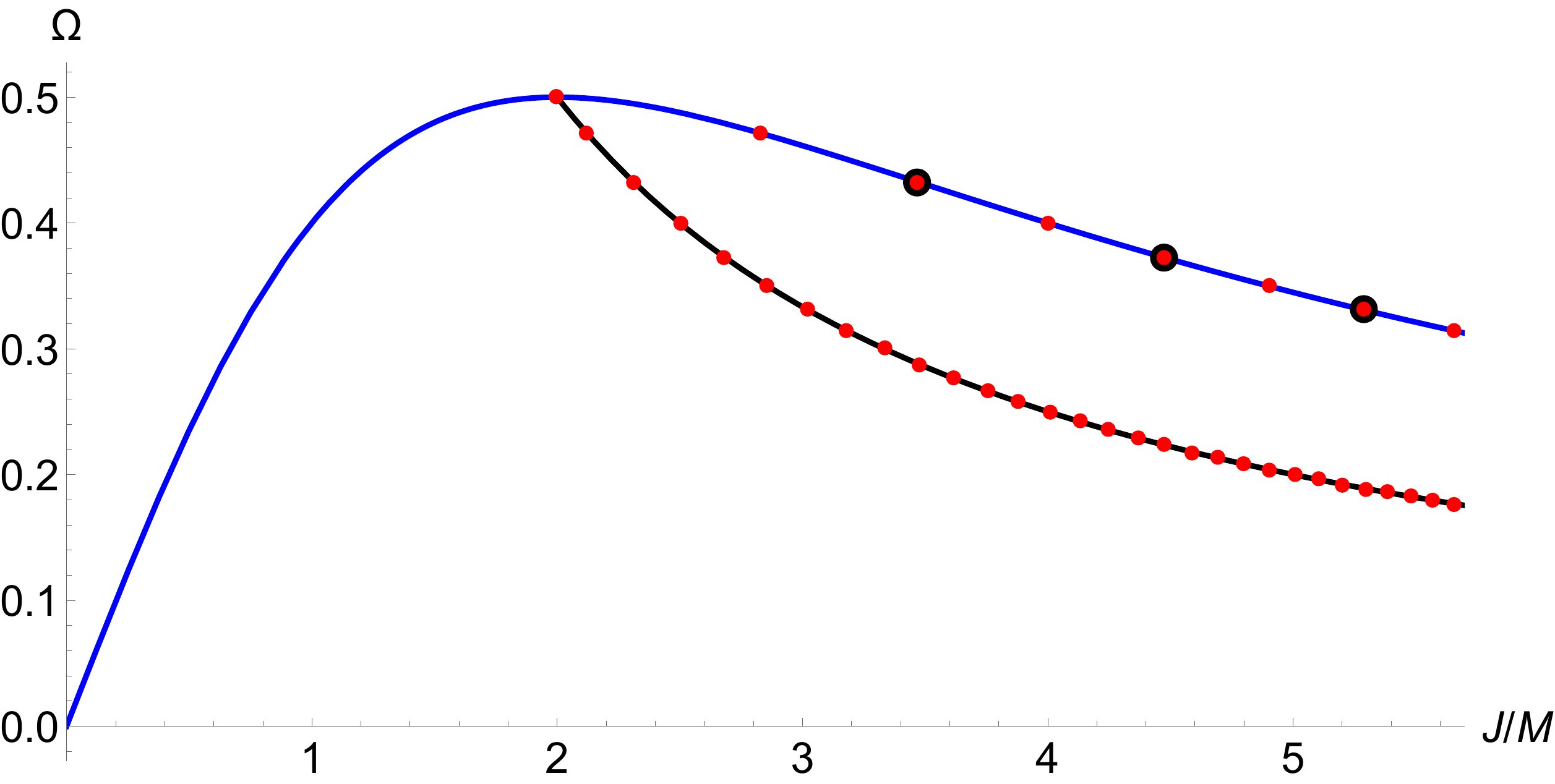}
 \end{center}
 \caption{
 Phases of spinning disks large-$D$, i.e., Myers-Perry black holes (blue line) and spinning black bars (black line) in the plane $(J/M, \Omega)$.
 The red dots indicate the presence of non-axisymmetric corotating zero modes for the spinning disks and bars. 
 The red dots in the disk branch encircled by a black dot indicate also axisymmetric zero modes.  New families of solutions branch from all of these zero modes, as explored in greater detail in \cite{Licht:2020odx}.  [[ \cite{Andrade:2018nsz} fig.\ 4 ]]
 \label{fig:blobspace}
 }
 \end{figure}

\cite{Andrade:2018nsz} also found spinning black bar solutions, which obey 
 \be
 m(t,r,\phi) = \exp \left( 1 - \frac{r^2}{4} \left( 1 + \sqrt{1-4 \Omega^2} \cos(2 (\phi - \Omega t) ) \right) \right) \ .
 \ee
These are oblong Gaussians rotating with angular velocity $\Omega$ (see fig.\ \ref{fig:Gaussianblobs}b). For these profiles, $J/M = 1/\Omega$.    The reality constraint on $m$ restricts the angular velocity to the range $\Omega \in [0,1/2]$.  At $\Omega = 1/2$, the spinning black bars become axisymmetric and join onto the spinning disk solutions, as shown in fig.\ \ref{fig:blobspace}.

These rotating bars break the axial symmetry in the $\phi$ direction. Generically, such an object should emit gravitational waves and could not exist in a stationary configuration. However, when $D\to\infty$ the emission of gravitational radiation is strongly suppressed as $D^{-D}$. Black bars are then expected to exist as long-lived configurations when $D$ is finite but large as we will discuss in greater detail in sec. \ref{sec:collisions}.

With these stationary solutions in hand, one can then perturb them slightly. Requiring that the perturbations are regular and that they decay at infinity, they reproduce the quasinormal modes of the large $D$ Schwarzschild and Myers-Perry black holes. For the latter, when the rotation parameter is $a=\sqrt{3},\,\sqrt{5},\,\sqrt{7}\dots$ there are zero modes that mark the onset of ultraspinning instabilities (recall eq.~\eqref{MPultraspin}).  The ``phase space'' of the spinning disk and bar solutions is shown in fig.\ \ref{fig:blobspace}, along with zero modes that indicate such instabilities.  The qualitative picture that emerges is that for small $J/M$, the spinning disks are stable.  However, once $J/M$ exceeds two, the spinning bar becomes a preferred solution, at least for a little while, before new zero modes appear.  The first (parity odd) mode appears at $J/M = 3 / \sqrt{2}$, followed by a parity even mode at $J/M = 4/\sqrt{3}$.  The work of \cite{Licht:2020odx} suggests that the parity odd modes of the bar may not  lead to instabilities,\footnote{The parity odd zero modes are associated with solutions which have larger angular frequency than the spinning black bar.  As the size of the deformation increases, these solution branches quickly become singular and terminate \cite{Licht:2020odx}. In the case of axisymmetric perturbations of the spinning disk, the parity odd and even modes lead to black Saturns and black rings respectively.\label{foot:odd}} but after the first parity even mode appears, spinning bars decay  in instabilities that are analogues of the GL instability that we consider in sec.~\ref{sec:GL} for the black string.

Besides MP black holes and black bars, the equations \eqref{conshigherdone} and \eqref{conshigherdtwo} have revealed the existence and properties of many other new large $D$ black holes as blobs with a variety of deformation patterns \cite{Andrade:2019edf,Licht:2020odx}. We will comment on some aspects and consequences of the resulting enriched phase diagram in sec.~\ref{sec:collisions}.

Furthermore, using \eqref{galiboost} a blob can be boosted to move with constant velocity. \cite{Andrade:2018yqu} took the further step of throwing two such solutions at each other and observing the time evolution. The results of this `experiment' will also be discussed in sec.~\ref{sec:collisions}.

\newpage

\part{Applications}
\label{part:applications}

\section{The Gregory-Laflamme instability at large $D$}
\label{sec:GL}

About 25 years ago, \cite{Gregory:1993vy,Gregory:1994bj} found that a uniform black string extended along a periodic direction develops a dynamical instability when the length of that direction grows too large.  The endpoint of this instability has been a topic of lengthy and fruitful discussion ever since.  One possibility, advocated by Gregory and Laflamme, is that the endpoint should be a localized black hole.  However, such a transition requires that the horizon pinch off, creating arbitrarily large curvature outside it  and violating cosmic censorship.  \cite{Horowitz:2001cz} argued instead that the endpoint may simply be a non-uniform black string (NUBS).  In seeming support of their argument, 
using the Raychaudhuri equation, they were in fact able to prove that horizon pinch off could only occur at infinite affine parameter, measured with respect to null generators of the horizon.  
However, as pointed out a few years later \cite{Garfinkle:2004em,Marolf:2005vn}, affine time is not the same thing as time measured by an observer far from the black string, and with reasonable assumptions -- for example that the space-time is similar to that of a static Schwarzschild black hole --
 infinite affine time may correspond to finite time for the observer.
Independently, 
several researchers were constructing NUBS solutions in perturbation theory, close to the point of dynamical instability.   \cite{Gubser:2001ac} and \cite{Wiseman:2002ti, Kudoh:2004hs} found in five and respectively six  dimensions that these NUBS solutions had larger entropy and were thermodynamically disfavored, seeming to support the original localized black hole scenario of Gregory and Laflamme.

The situation in low dimension was later clarified by a tour de force time dependent numerical simulation in $D=5$ by \cite{Lehner:2010pn}.  The black string horizon evolves in a self-similar fractal cascade.  At any given time, the solution consists of spherical black holes connected by thin black strings.  The thin black strings between the black holes then evolve further into smaller black holes connected by yet thinner strings.  The behavior is a gravitational analog of the Rayleigh-Plateau instability exhibited by thin streams of water \cite{Cardoso:2006ks}. The numerical simulation crashes when the curvature in the thinner segments becomes very large, but extrapolating the evolution indicates that a singularity will be reached in a finite asymptotic time \cite{Lehner:2010pn,Lehner:2011wc}.\footnote{%
 See also \cite{Figueras:2017zwa} for a related numerical simulation of spinning black holes.
}

Thus, cosmic censorship is violated.\footnote{%
 There is active debate in the literature on these issues.  See e.g.\ \cite{Iizuka:2018zgt} for a large $D$ analysis of ``mushroom'' solutions in AdS that comes to the conclusion that in that instance cosmic censorship is not violated.  They study black  branes ``polarized'' under an electromagnetic field sourced at the boundary, for which violations of cosmic censorship have been argued (in low dimensions) in \cite{Horowitz:2016ezu}. However, the physics of these violations appears to be different than in the case of the GL instability.
}
  The further evolution requires new laws of physics beyond Einstein's classical theory, arguably a quantum theory of gravity. Nevertheless, a plausible evolution across the singularity has been proposed in \cite{Andrade:2018yqu,Andrade:2019edf} such that the input and consequences from quantum gravity are minimal.
The neck that forms in the horizon has very high curvature, and may be regarded as a small, `Planck-size black hole', with very high effective temperature. Such an object, without any conserved charges that could stabilize it, is expected to decay quantum-mechanically by emitting a few Planck-energy quanta in a few Planck times. Then, the neck evaporates in much the same manner as the neck in an unstable fluid-jet evaporates (literally) and breaks the jet into a number of droplets; then, classical hydrodynamics quickly resumes control of droplet evolution after the brief episode of evaporation.
If this picture also applies to black strings, the horizon pinch will quickly evaporate through quantum-gravity effects, splitting the horizon into separate black holes, and classical evolution will take over again. Predictivity of the entire evolution using General Relativity will be maintained to great accuracy, with uncertainties proportional to at most a power of $(M_\text{Planck}/M)$ where $M$ is the total mass of the system.\footnote{%
The loss of predictivity may even be exponentially small: the evolution of a fluid jet, both right before and right after breaking into droplets, is controlled by attractor solutions of the hydrodynamic equations \cite{Eggers:1997zz}.
}

This picture assumes the evolution observed in numerical simulations, where the horizon shrinks without stopping until a singularity forms.
Importantly for our large $D$ perspective, 
 \cite{Sorkin:2004qq} made a key observation that the endpoint of the instability should in fact depend sensitively on the space-time dimension.  Generalizing Gubser's perturbative method to arbitrary dimension, he was able to establish that for $D>13$, there is a second order transition to a NUBS, while for $D\leq 13$, the transition can only be first order, i.e.\ the perturbative NUBS solutions will have higher entropy than their uniform cousins of equal mass.\footnote{In this problem $D$ can actually be regarded as a continuous parameter, so one can more accurately determine that the critical dimension is $D\approx 13.6$ \cite{Sorkin:2004qq,Emparan:2018bmi}.}  
It is this $D$ dependence that makes the Gregory-Laflamme instability a killer application, perhaps {\it the} killer application, for a large $D$ approach to Einstein's equations.

We begin by reviewing a heuristic argument made by \cite{Sorkin:2004qq}, comparing the entropy of a black string and a black hole of the same mass in $D$ dimensions.  For simplicity, we work in the microcanonical ensemble at fixed entropy $S$ and energy $E$.  
To this end, we review the thermodynamic properties of black $p$-branes in $D$ dimensions, where $p$ is the number of spatial dimensions of the membrane.
We start with the asymptotically flat, black brane metric in a $D$ dimensional space-time.  This metric satisfies Einstein's equations in vacuum.  The line element in Eddington-Finkelstein coordinates is\footnote{This is the uniform black brane case of \eqref{EFbrane} and \eqref{ACG} with constant $m=r_0^n$ and $p_i=0$ (and $\rho\to r$, $\sigma^i/\sqrt{n}\to x^i$).}
\be
\d s^2 = 2 \d t \, \d r - \left( 1 - \frac{r_0^n}{r^n} \right) \d t^2 + \delta_{ij} \d x^i \d x^j + r^2 \d \Omega_{n+1}^2\,,
\ee
with $i,j=1\ldots p$ and $\d \Omega_{n+1}^2$ the line element on a sphere with unit radius.

From the Bekenstein-Hawking formula for the entropy of a black hole $S = \frac{A}{4 G_N}$,
where $A$ is the area of the event horizon,
we can read off an entropy density,
\be
\label{sbrane}
s = \frac{r_0^{n+1} \Omega_{n+1}}{4 G_N} \ ,
\ee
per unit volume of the black brane.\footnote{%
 The volume of a $D-1$ dimensional unit sphere 
 \[
 \Omega_{D-1} = \frac{2 \pi^{\frac{D}{2}} }{\Gamma \left( \frac{D}{2} \right)} \sim \left(\frac{2\pi}{D} \right)^{\frac{D}{2}} e^{\frac{D}{2}} \left( \sqrt{\frac{D}{\pi}} + \Or{D^{-1/2}}\right) \ 
 \]  
gets exponentially small as $D$ increases.
}  
The Hawking temperature can be determined from the usual trick of analytically continuing to Euclidean time and insisting that there is no conical singularity at the horizon (or equivalently from the surface gravity):
\be
\label{temperature}
T = \frac{n}{4 \pi r_0} \ .
\ee
The first law of black hole thermodynamics tells us that $\delta \varepsilon = T \delta s$, where $\varepsilon$ is an energy density, from which we can deduce that
\be
\label{eblackbrane}
\varepsilon = \frac{(n+1)r_0^{n} \Omega_{n+1} }{16 \pi G_N} \ .
\ee

We now wish to compare the black hole $p=0$ result with the black string $p=1$ result in $D$ dimensions.  To render thermodynamic quantities finite, we compactify the $x$ direction with periodicity $L$.  
For the string, our general thermodynamic formulae above reduce to the special case
\be
S_s = s L = \frac{L r_s^{D-3} \Omega_{D-3}}{4 G_N} \ , \; \; \; E_s = \varepsilon L = \frac{(D-3) L r_s^{D-4} \Omega_{D-3}}{16 \pi G_N} \ ,
\ee
where the horizon of the string is at $r = r_s$.  
To this result, we compare the corresponding values for a black hole in $D$ dimensions.  Of course our black hole solution is no longer a solution if one of the spatial directions is compactified.  Nevertheless, for small black holes, we expect the corrections to $s$ and $\varepsilon$ to be small.  Indeed,
in the large $D$ limit, where the gravitational effects of the black hole die off as $r^{-n}$, we expect this approximation to become better and better.
We find then that
\be
\label{ShEh}
S_h = \frac{r_h^{D-2} \Omega_{D-2}}{4 G_N} \ , \; \; \; E_h = \frac{(D-2) r_h^{D-3} \Omega_{D-2}}{16 \pi G_N} \ ,
\ee
where the horizon of the black hole is at $r = r_h$.

In the stringy case the entropy grows with energy as $\log S_s \sim \frac{D-3}{D-4} \log E$ while in the black hole case, we find instead $\log S_h \sim \frac{D-2}{D-3} \log E$.  As $\frac{D-3}{D-4} > \frac{D-2}{D-3}$ for the cases of interest where $D>4$, in the microcanonical ensemble we expect a low energy phase where the black hole is stable, $S_h > S_s$, and a high energy phase where the black string is stable, $S_s > S_h$.  The transition point, where the two entropies are equal, occurs when the horizon radii are
\be
\label{rsrhrel}
r^*_s = \frac{D-2}{D-3} r_h^* = L \left( \frac{D-3}{D-2} \right)^{D-2} \frac{\Gamma \left( \frac{D-1}{2}\right)}{\sqrt{\pi} \Gamma \left( \frac{D}{2} -1 \right) } = \frac{L \sqrt{D}}{e \sqrt{2\pi}} + \Or{D^{-1/2}} \ .
\ee
In order for this phase transition to be self-consistent, the black hole should have a diameter that is less than or equal to $L$, i.e.\ it should fit in the box.  This constraint leads to a critical value of the dimension, $D = 13.06$.  In other words, for $D \gtrsim 13$, we do not expect to find a stable localized black hole phase near the point in the phase diagram where the entropies would otherwise be equal.\footnote{%
 We could start with a very thin string, well below the critical thickness and also much smaller than $L$.  
 In this case, there is no constraint from the size of the box.  The isolated black holes of the same energy have a radius $r_h \sim r_s \ll L$ in the large $D$ limit.
}  Instead, for sufficiently large $D$, the uniform black string will evolve into a non-uniform black string for thicknesses
slightly below the critical thickness, as we will see.

In sec.~\ref{sssec:bbeqns}, we already gave the first steps for identifying the GL instability in a large $D$ limit and for constructing the NUBS.  A solution for a large $D$ black string with varying profile along its length is given by inserting into \eqref{EFbrane} and \eqref{ACG} any functions $m(t,z)$ and $p(t,z)$ that solve the equations
\be
\label{hydroeq1}
\partial_t m &=& - \partial_z \left( p - \partial_z m \right)  \ , \\
\partial_t (p  - \partial_z m) &=& -  \partial_z \left( \frac{p^2}{m} +\partial_z^2 m - 2\partial_z p -m    \right) \ .
\ee
We have written the pair of partial differential equations in such a way that they can be interpreted as conservation of a stress tensor $\partial_\mu T^{\mu\nu} = 0$, 
with energy density $T^{00} \sim m$ and momentum $T^{tz} \sim p - \partial_z m$.  Eq.~\eqref{conshigherdtwo} gives an equivalent 
purely second order
presentation of the differential equations that is usually used, where we replace the second equation with a linear combination
\be
\label{hydroeq2}
\partial_t p   - \partial_z^2 p &=& \partial_z \left( m - \frac{p^2}{m} \right) \ .
\ee

The uniform black string solution fixes the constant of proportionality that relates $T^{tt}$ to $m$.
From  
(\ref{eblackbrane}) we see that
\be
T^{tt} &=& \frac{(n+1) \Omega_{n+1}}{16 \pi G_N} \left( m + \Or{n^{-1}} \right) \ .
\ee
The conservation equations (\ref{hydroeq1}) and (\ref{hydroeq2}) then suggest
\be
T^{tz} &=&  \frac{(n+1) \Omega_{n+1}}{16 \pi G_N} \left( p - \partial_z m  + \Or{n^{-1}}  \right) \ , \\
T^{zz} &=&  \frac{(n+1) \Omega_{n+1}}{16 \pi G_N} \left( \frac{p^2}{m} +\partial_z^2 m - 2\partial_z p -m  + \Or{n^{-1}} \right) \ . 
\ee 
\cite{Emparan:2015hwa} obtain the same result for a quasi-local stress tensor defined on a constant large $\sR$ slice of the geometry.

The conservation equations (\ref{hydroeq1}) and (\ref{hydroeq2}) can be solved to find static NUBS solutions.  These static solutions satisfy $T^{tz} = 0$ or equivalently at leading order in the $1/n$ expansion $p = \partial_z m$, leading to the following nonlinear ordinary differential equation for $m(z)$
\be
\frac{(m')^2}{m} - m'' - m = -\tau  \ ,
\ee 
where $\tau$ is a constant proportional to the tension of the black string.
This ODE becomes simpler with the substitution $m(z) = e^{-\rho(z)}$: 
\be
\label{todaeq}
\rho''(z) =  1 - \tau e^{\rho(z)} \ .
\ee
This equation
describes periodic motion of a particle in a Toda-like potential $V(\rho) = \tau e^{\rho}  - \rho$, provided $\tau>0$
\cite{Emparan:2015hwa,Oppo1984,Toda1975}.  As such, it has a ``conserved energy'' and can be reduced to first order form
\be
\label{rhoeq}
\frac{1}{2} \rho'(z)^2 + \tau e^{\rho(z)} - \rho(z) = c
\ee
where $c$ is a second integration constant.  This periodic motion in a bounded potential is reinterpreted here as a periodic profile of the NUBS that can be adapted to a choice of periodic boundary condition.

A particularly simple solution comes from (\ref{todaeq}) in  the limit of vanishing tension $\tau= 0$, in which case the potential is not bounded.  In this case, the solution is a Gaussian $m = e^{-z^2/2}$.
Clearly such a solution will not obey periodic boundary conditions consistent with a compactified $z$ direction.  Yet, if the box is large enough, $m$ will be exponentially close to zero at the edges of the box, and the solution is a very good approximation \cite{Suzuki:2015axa}. Arguing as in sec.~\ref{sssec:blobs}, one sees that this Gaussian solution is a one-dimensional version of the blobs that approximate exponentially well the properties of a spherical Schwarzschild black hole.

Time dependent solutions are more difficult to analyze in general.  
However, in sec.~\ref{sssec:bbeqns} we straightforwardly 
identified the large $D$ version of the GL instability, and here we can make it more precise. Linearizing the conservation equations 
 (\ref{hydroeq1}) and (\ref{hydroeq2}) about the uniform black string solution using the ansatz
 \be
 \label{qnm}
 m = m_0 + \delta m \, e^{\Omega t + i k z} \ , \; \; \; p = \delta p \, e^{\Omega t + i k z} \ ,
 \ee
one finds the dispersion relation
\be
\Omega = \pm k (1 \mp k ) \ .
\ee
 For $0 \leq |k| < 1$, there is a growing mode, and the uniform black string is unstable.  The wave number
 $k_{GL} = 1$ is at the threshold of instability and corresponds to $\Omega = 0$.  
 With some work \cite{Asnin:2007rw,Emparan:2015rva,Emparan:2015rva}, 
 the analysis can be carried to higher order in the $1/n$ expansion:
 \be
 \label{kGL}
 k_{GL} = 1 - \frac{1}{2n} + \frac{7}{8 n^2} + \frac{-\frac{25}{16} + 2 \zeta(3)}{n^3} + \frac{ \frac{363}{128} - 5 \zeta(3)}{n^4} + \Or{n^{-5}} \ .
 \ee
Interestingly, this analytic result reproduces numerical calculations of $k_{GL}$ down to $D=6$ with better than $2.4\%$ accuracy.
 
 To analyze stability, we now follow two distinct approaches  --  thermodynamic and dynamical.  
For thermodynamic stability, we analyze static NUBS solutions.
We extend the linear analysis at $\Omega = 0$ and $k= k_{GL}$ to higher, nonlinear order in the amplitude of the perturbation.  Decomposing the NUBS in a Fourier series, one finds for the static string
  \be
  \label{mofz}
 m(z) &=& 1 + 
 \delta m  \cos k z + 
 \frac{\delta m^2}{6} \cos 2k z + 
 \frac{\delta m^3}{96} \cos 3 k z + \Or{\delta m^4} \ , \\
p(z) &=& \partial_z m(z) \ , \; \; \;
k = 1 - \frac{\delta m^2}{24} + \Or{\delta m^4} \ .
 \ee
We have set $m_0 =1$ and let $k$ depend on $\delta m$ in order to reduce the complexity of the expansion.  
These expressions can straightforwardly be extended to higher order in $\delta m$ and $1/n$ \cite{Emparan:2018bmi}.

To see if this solution is thermodynamically stable, we compare the energy and entropy of the NUBS with its uniform cousin.  We define rescaled energy and entropy densities that absorb some of the $n$ dependence:
\be
M(z) &\equiv& - \frac{16 \pi G_N}{(n+1) \Omega_{n+1}} {T^t}_t  = m(z) + \Or{n^{-1}} \ , \\
S(z) &\equiv& \frac{\sqrt{n}}{\Omega_{n+1}} {\rm Area}(z, \sR_h) = \sqrt{G_{11}(z,\sR_h)} \sR_h^{\frac{n+1}{n}}(z) = m(z) + \Or{n^{-1}} \ ,
\ee
where $G_{11}$ is defined in \eqref{ACGtwo}.
Note that the entropy and energy are the {\it same} at leading order in the large $D$ expansion.  Thus a thermodynamic stability analysis will require going to higher order in the $1/n$ expansion.  We average these local densities over the
$z$ coordinate to define thermodynamic quantities
\be
{\bf M} \equiv \left( \frac{2\pi}{k_{GL} L}\right)^n \int_{-L/2}^{L/2} \frac{\d z}{L} M(z) \ , \; \; \;
{\bf S} \equiv \left( \frac{2\pi}{k_{GL} L}\right)^{n+1} \int_{-L/2}^{L/2} \frac{\d z}{L} S(z) \ .
\ee

\begin{figure}
\includegraphics[width=5in]{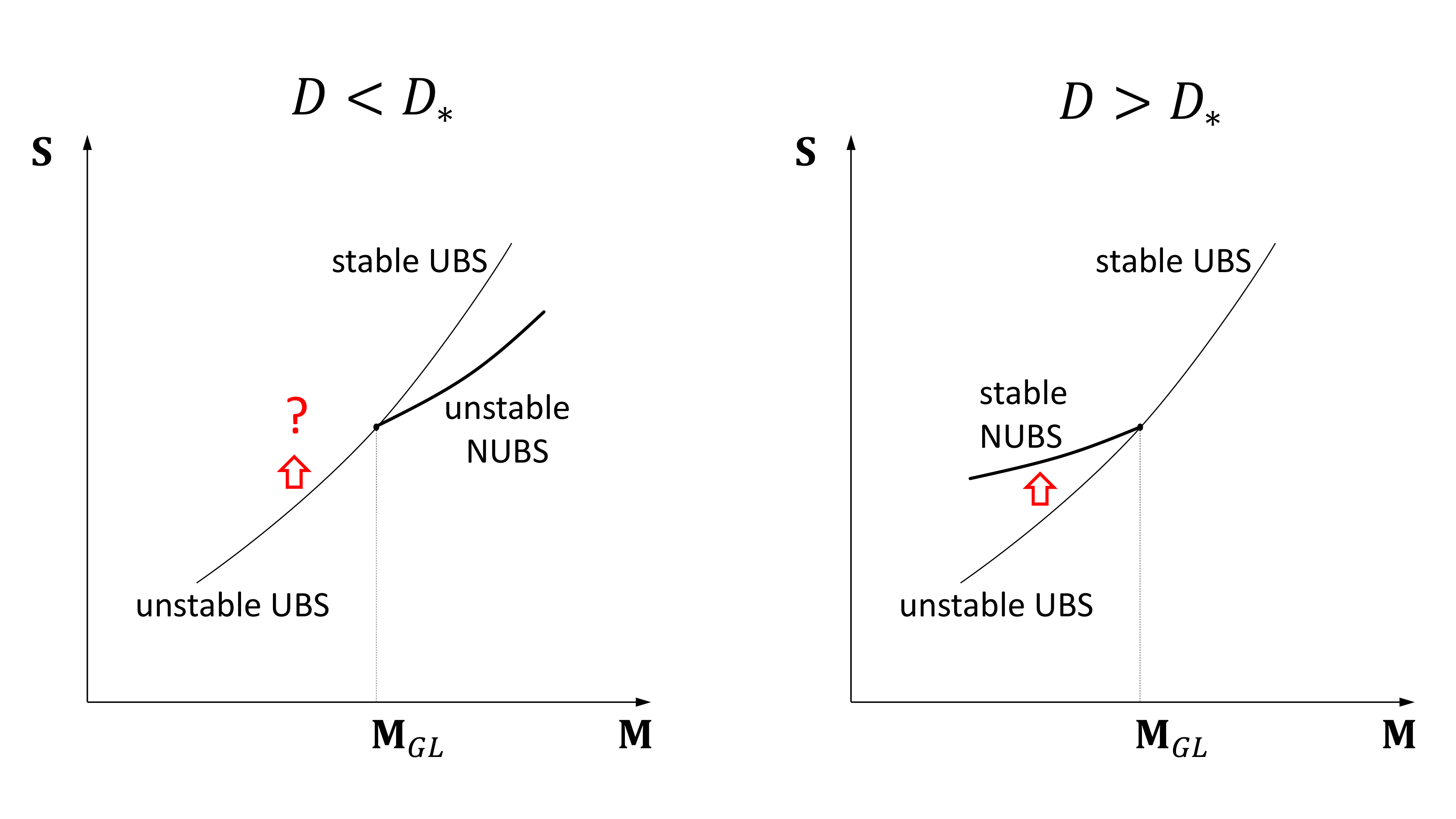}
\caption{For $D<D_*$, there are no weakly non-uniform black strings in the region where the uniform black strings are dynamically unstable.  Therefore, the latter cannot evolve into the former.  The phase transition must happen at some $M_* > M_{GL}$ and be first order in nature.  For $D>D_*$, there do exist weakly non-uniform black strings for the dynamically unstable black strings to decay into, consistent with a second order phase transition at $M = M_{GL}$.  
[[ \cite{Emparan:2018bmi} fig.\ 2 ]]
\label{fig:Dstar}
}
\end{figure}

While the conservation equations (\ref{hydroeq1}) and (\ref{hydroeq2}) only give us the leading order term in the $1/n$ expansion, one can continue the calculation to the next few orders to obtain \cite{Emparan:2018bmi}
\be
{\bf M} &=& 1 + n \, \delta m^2 \left( - \frac{1}{24} + \frac{1}{3n} + \frac{7}{12 n^2} + \Or{n^{-3}} \right) + \Or{\delta m^4} \ , \label{Mnubs}\\
{\bf S} &=& 1 + n \, \delta m^2 \left( - \frac{1}{24} + \frac{11}{12 n} + \frac{7}{24 n^2} + \Or{n^{-3}} \right) + \Or{\delta m^4} \ .
\ee
Crucially, the $\Or{n \, \delta m^2}$ correction to the mass changes sign at $n_* = 9.48$ which corresponds to 
\be
D_* = 13.48
\ee
This critical value of $D_*$ is weakly sensitive to the order in $1/n$ at which one truncates the expansion.  If we were to use only the first order correction, then $D_* = 12$ while keeping further orders yields
$D_* = 13. 65$ and $13.93$ \cite{Emparan:2018bmi}.  As we discussed in sec.~\ref{subsec:nonconv}, the $1/n$ expansion is expected to be asymptotic in nature \cite{Emparan:2015rva}, and so there should be an optimal order at which to truncate the expansion.  Given \cite{Sorkin:2004qq}'s estimate of the critical value $D_* = 13.5$, that order is perhaps $n^{-2}$ or $n^{-3}$.\footnote{The next order term inside the brackets in \eqref{Mnubs} is $\simeq 14/n^3$, and its large coefficient probably reveals that the expansion does not converge.}

Indeed, the fact that the correction to $M(z)$ changes sign means that for $D< D_*$, there is no available weakly non-uniform black string for a dynamically unstable uniform black string to decay into (see fig.\ \ref{fig:Dstar}). The uniform black string must decay before it reaches the GL instability, at some larger $M_* > M_{GL}$, by a first order phase transition instead.  In contrast, for $D > D_*$, there are weakly non-uniform black strings that can be accessed; the uniform black string is expected to decay via a second order phase transition into these candidate solutions.  

To double check this scenario, one should compare also the entropy density of the weakly non-uniform strings with their uniform cousins.  The picture that emerges is shown in fig.\ \ref{fig:Dstar}; the entropy of the weak NUBS is larger than the UBS for $D>D_*$ and less than the UBS for $D<D_*$, as expected on thermodynamic grounds.  The condition on the entropy difference between the uniform and non-uniform black strings leads to similar estimates for the critical dimension $D_* \approx 13.5$ in a $1/n$ expansion.\footnote{%
  This analysis can be repeated in the canonical ensemble, at fixed temperature instead of fixed energy.  The critical value of the dimension is then slightly less, $D_* \approx 12.5$ \cite{Emparan:2018bmi}.
}  
One can further investigate dynamical stability of the weakly non-uniform black strings, repeating the linear fluctuation analysis (\ref{qnm}) but about the non-uniform background.  One finds that above the critical dimension, the weakly non-uniform black strings are dynamically stable \cite{Emparan:2018bmi}.  

For $D<D_*$, even though the transition is first order, the stable phase can still be a NUBS, just with a finite deformation parameter.  In $D=12$ and 13 but not $D=11$, both $1/n$ thermodynamic analysis and a direct numerical approach suggest that the first order phase transition is indeed of such a nature \cite{Figueras:2012xj,Emparan:2018bmi}.  The $1/n$ dynamical fluctuation analysis about the non-uniform black string can be extended beyond linear order to give some support for this scenario as well.  

While we have provided evidence for a phase diagram of the black string, what would really tie everything together is a dynamical closed form solution that evolves a UBS to NUBS for $k < k_{GL}$ or a NUBS to a UBS for $k>k_{GL}$.  Such a solution has not yet been presented in the literature to our knowledge, but is straightforward to work out by modifying (\ref{mofz}) to allow $\delta m$ in (\ref{mofz}) to have the right $t$ and $n$ dependence, $\delta m \to \delta m( t / n) / \sqrt{n}$:
\be
m &=& 1 + \frac{\delta m}{\sqrt{n}} \cos k z + \frac{\delta m^2}{6n} \cos 2 k z + \frac{\delta m^3}{96 n^{3/2}} \cos 3kz + \Or{n^{-2}} \ , \\
p &=& - \frac{(\delta m + \delta \dot m) k }{\sqrt{n}} \sin k z - \frac{ \delta m^2 \, k}{3 n} \sin 2 k z - \frac{ \delta m^3}{32 n^{3/2}} \sin 3 k z + \Or{n^{-2}} \ .
\ee  
We allow the wave number to be freely tuneable $k = 1 - \frac{k_1}{n}$ by an $\Or{n^{-1}}$ amount.  One finds the following differential equation for $\delta m$, 
\be
24 \delta \dot m(t) + \delta m(t)^3 + 12 (1 - 2 k_1) \delta m(t) = 0 \ ,
\ee
which has a solution
\be
\label{interpolatingsoln}
\delta m(t/n) = \sqrt{24 \, \delta k } \, (1 + Ce^{-2\,  t\,  \delta k/n})^{-1/2} \ ,
\ee
where $C$ is an integration constant and $k_1 = \frac{1}{2} + \delta k$.  Comparing with (\ref{kGL}), we see that $\delta k = 0$ corresponds to the $\Or{n^{-1}}$ correction to the critical wave number $k_{GL}$.  For $\delta k > 0$ or correspondingly $k < k_{GL}$, the solution evolves from a UBS in the far past $t \to -\infty$ to a NUBS in the far future $t \to \infty$, assuming $C>0$.  
For $\delta k < 0$ and correspondingly $k > k_{GL}$,
we need to take $C<0$ to find a real solution.  It is a solution that evolves to a UBS from something like a NUBS but which has nonzero time derivatives; 
 it is a solution that does not exist arbitrarily far in the past.  There is a singularity at $e^{2 t_* \delta k/n} = -C$, before
 which the solution ceases to be real. This dependence on the sign of $\delta k$ is consonant with the fact that for $\delta k< 0$, the NUBS is a complex saddle point -- it is not a real solution of the equations of motion. 
 
In conclusion, one can ask what happens if one starts with a uniform thin string in a box of length $L$ in a high number of dimensions, well below the onset of the GL instability, $L \gg L_{\rm GL} \sim n^{-1/2} r_0$.  Is the endpoint still a NUBS?  
\cite{Emparan:2018bmi} indicate that the endpoint of the instability  --  wavy string or beads on a wire  --  will depend both on $n$ and the thickness $r_0/L$ of the initial black string.
At any given finite $n$, thin enough strings will always evolve to beads-on-a-wire; but as $n$ grows larger, this endpoint requires thinner strings.
To gain some insight,
 we analyze $\Or{1/n}$ corrections to the Gaussian solution of \eqref{rhoeq}, which are the large $D$ analogs
 of isolated Schwarzschild black holes.  \cite{Suzuki:2015axa} find that\footnote{%
  We have taken the liberty of correcting some typos in (6.22) of  \cite{Suzuki:2015axa} and
  (5.1) of \cite{Emparan:2018bmi}.
 }
\be
m(z) = e^{-\frac{z^2}{2}} \left(1 - \frac{z^4}{4n} + \frac{z^6(-16+3 z^2)}{96 n^2} + \Or{n^{-3}} \right) \ .
\ee
There is clearly a break down in the $1/n$ approximation once $z \sim n^{1/4}$.  In the original, unscaled spatial coordinate
system, this spatial distance is proportional $n^{-1/4} r_0$.   Inverting the logic, this limit suggests that already
very narrow boxes with a length $L \sim n^{-1/4} r_0$ can support effectively tensionless black hole solutions, at least to leading order in the $1/n$ approximation, that can be interpreted as isolated Schwarzschild black holes.

There is something counter intuitive about this result, that isolated black holes become the favored endpoint in the regime where the box is still very narrow, $L \gtrsim n^{-1/4} r_0$.    From the crude estimate that led to (\ref{rsrhrel}),
one sees that isolated spherical black holes and black strings with the same energy satisfy $r_h = r_s$ to leading order in the $1/n$ approximation, which might suggest that isolated black holes become the favored endpoint of the evolution only once
$L \gtrsim 2 r_0$.  Indeed, a more careful study  \cite{Emparan:2019obu} of the neck formation as the black string pinches off suggests that the isolated black holes fit in the box only once $r_0 = L/2 - \Or{1/n}$.  The gradients involved, however, are beyond what the effective equations (\ref{hydroeq1}) and (\ref{hydroeq2}) can handle.  
Thus in the regime $n^{-1/4} \lesssim L/r_0  \lesssim 2$, we expect the endpoint is in actual fact a highly non-uniform black string which looks tensionless, and hence like an isolated black hole within the approximations of the effective theory. These non-uniform black strings have a neck that is thinner than $\Or{r_0}$, but shifting to an analysis like in \cite{Emparan:2019obu} it is possible to resolve the neck when its thickness is as small as $\sim \Or{r_0/\sqrt{n}}$.

\subsubsection*{Further Extensions}

Generalizations of the black string and black brane effective equations have already been mentioned in sec.~\ref{sssec:bbextend}. We comment here on their application to the GL problem.

The works \cite{Emparan:2016sjk,Rozali:2016yhw} consider charged black branes in a large $D$ limit while 
\cite{Emparan:2016sjk} goes on to consider polarized branes (electric field parallel to the brane) and black branes charged with respect to an abelian $(p+2)$-form.  In all cases, there is a Gregory-Laflamme type instability.  The charge dependence of the solution modifies the dispersion relation for the fluctuations in various interesting ways.  In the charged black brane case, while $k_{GL}=1$ remains the same, for modes with $k < k_{GL}$, the growth rate $\Omega$ is reduced by the presence of the charge. 
In the $(p+2)$-form case, in contrast, the onset of the instability $k_{GL} = \sqrt{1 - q_p^2 / m^2}$ is reduced by the presence of the charge.  Perhaps not surprisingly, in the $p=2$ case at large $D$, the endpoint of the instability appears to be a non-uniform black brane with a triangular (closest packed) lattice symmetry \cite{Rozali:2016yhw}.  

Extensions of the large $D$ analysis of the GL instability to black rings \cite{Tanabe:2015hda}, higher curvature corrections to black rings \cite{Chen:2018vbv}, charged black rings \cite{Chen:2017wpf}, and higher curvature corrections to black strings \cite{Chen:2017rxa} exist.   Note that the black rings will rotate at an angular velocity $\sim D^{-1/2}$ to maintain their shape, somewhat complicating the analysis.  At any rate, when the dust settles, 
the behavior is qualitatively similar to that of the black strings we have already discussed at length.  Thick black rings and strings are stable.  Thin ones exhibit a GL instability.  At small $D$, the
rings and strings tend to break up into isolated black holes.  At large $D$, the end point is a NUBS or non-uniform black ring.  
There are also similarities between the GL instability and the instability of small black holes in $AdS_D \times S^D$ \cite{Herzog:2017qwp} that we will explore in greater depth below in the context of holography.

\section{Black hole collisions and mergers}
\label{sec:collisions}

\subsection{General aspects}

The collision and merger of two black holes is the quintessential phenomenon of General Relativity: it involves its main players -- black holes and gravitational waves, horizons and curvature oscillations -- in a fully dynamical situation with strong gravitational fields, i.e.\ spacetime geometry distorted to the extreme.
One of the most exciting aspects of black hole research in recent years is the observation of gravitational waves from black hole collisions and mergers at the LIGO detectors.   Beyond this astrophysical context, black hole collisions are also of great interest as situations where we can hope to learn more about gravitational dynamics. Do the Einstein equations govern the entire collision process and its outcome, or does the classical theory break down under some circumstances? In other words, does cosmic censorship hold throughout the evolution of all collisions? The lessons we have learned from the GL instability suggest that the answer can depend very sensitively on the dimensionality of spacetime. Besides the fundamental interest of these questions, the study of black hole collisions in dimensions other than four is pertinent for AdS/CFT applications. The collision of two five dimensional black holes describes -- in a dual set up -- the collision of two balls of plasma in four dimensional space-time.

In light of the applications, there is great interest in accurately and efficiently simulating these collision events.
Paper and pencil approaches are typically only useful in the limit where one of the black holes is much more massive than the other, so that the smaller black hole may be treated as a perturbation to the metric created by the larger one.  Full and accurate computer simulations of astrophysical black hole collisions have recently become possible, but still require hours of computing time per event \cite{Lehner:2014asa}.  A key question for us then is whether a large $D$ limit can shed  light on black hole dynamics?

To answer that question, let us begin with some back of the envelope estimates for radiation processes in a large $D$ limit.
Gravitational radiation is sourced by time dependent, quadrupolar (or higher multipolar) mass distributions; the amplitude of metric fluctuations sourced by the quadrupole is linear both in $G_N$ and the quadrupole moment $Q \sim M L^2$ where $M$ and $L$ are characteristic mass and length scales of the source.  The power produced goes as the square of the amplitude.  However, a power of $G_N$ gets absorbed in computing an energy from the metric fluctuations.  
The over all expression is then completed by including an oscillation frequency $\omega$ to give a dimensionally correct result: $P \sim G_N \omega^6 Q^2 \sim G_N \omega^6 M^2 L^4$ in 4d.  In $D$ dimensions, the dependence on $Q$ and $G_N$ remains the same.  However, the dimensionality of $G_N$ changes requiring the power of $\omega$ to be correspondingly adjusted \cite{Cardoso:2002pa}
\be
\frac{dE}{dt} \sim G_N \omega^{D+2} Q^2 \Omega_{D-2}  \sim  G_N \omega^{D+2} M^2 L^4 \Omega_{D-2} \ .
\ee
To capture the $D$ dependence of the result, we include also a sphere volume
 from integrating the power produced per unit angle over the quadrupole radiation pattern.

We can further refine this estimate \cite{Emparan:2013moa} by calculating the energy produced per unit mass per period of oscillation:
\be
\frac{1}{M \omega} \frac{dE}{dt} \sim \Omega_{D-2} G_N \omega^{D+1} M L^4 \sim D^2 \left( \frac{\omega}{T} \right)^D \ ,
\ee
where in the last similarity relation, we have extracted the leading $D$ dependence of the result, employing (\ref{temperature}).  Intriguingly, the temperature of the black hole sets a threshold for radiation effects to kick in.  
For a Schwarzschild black hole or brane in the large $D$ limit, the temperature (\ref{temperature}) scales with $D$.
For frequencies larger than the $T \sim D / r_h$, radiation effects can be quite large while for frequencies smaller than $T$, radiation effects are essentially negligible. 

This scale set by the temperature appears also in quasinormal modes (see section \ref{sec:qnm}).  
The typical quasinormal mode has an imaginary part that produces damping and that scales with the temperature $T$.  Most quasinormal modes will damp out exponentially quickly in a large $D$ limit in a time scale set by $1/T \sim r_h/D$.  Thus in black hole collisions, we expect any energy that is released from these modes to come out rapidly, almost as a shock wave or Dirac delta function shell.  This increase in the speed at which gravitational radiation is released has been observed in numerical simulations of head-on collisions of black holes, as a function of space-time dimension \cite{Cook:2017fec}.

Another important radiation process for black holes is Hawking radiation.  While we expect Hawking radiation to be a very small effect during black hole collisions, as we are discussing radiation processes here, let us make some brief remarks.  Estimating black holes as black bodies, we can derive the large $D$ scaling for power emitted by Hawking radiation from the Stefan-Boltzmann law, that the power $P$ scales as $T^4$ times the surface area in four dimensions.  
In $D$ dimensions, the $T^4$ scaling is replaced by $T^D$.  
For Schwarzschild black holes, from (\ref{sbrane}) and (\ref{temperature}) we find that
\be
P \sim r_h^{-2} \Omega_{D-2} D^D \ .
\ee
In order to keep this effect finite in a large $D$ limit, we must consider very large black holes, with $r_h \sim D^{D/4}$ in units of the Planck length.  
More careful discussions of Hawking radiation 
are provided in refs.\ \cite{Hod:2011zzb,Emparan:2013moa,Holdt-Sorensen:2019tne}, which we will return to in sec.~\ref{subsec:hawking}.

A final back of the envelope estimate uses the second law of black hole thermodynamics to constrain the energy radiated.
If we take two black holes of energy $E_1$ and $E_2$ that merge to form a black hole of energy $E_f$, then the entropy of the final state should exceed the entropy of the initial state.  Moreover, as black holes are highly entropic objects, we will neglect the entropy carried by the radiated energy $E_1 + E_2 - E_f$ in this estimate.  Using the entropy and energy results (\ref{ShEh}) for a $D$-dimensional Schwarzschild black hole, we deduce the inequality 
\be
E_f^\frac{D-2}{D-3} \leq E_1^{\frac{D-2}{D-3}} + E_2^{\frac{D-2}{D-3}} \ .
\ee 
Fixing $E_i = E_1 + E_2$, the energy lost is maximized for equal mass black holes $E_1 = E_2$ and scales with $D$ as
\cite{Witek:2010xi}
\be
\label{radprod}
 E_i -E_f \leq 2^{\frac{1}{D-2}} E_i \ .
\ee
This bound may be violated weakly because of the entropy carried by the emitted radiation, but nevertheless indicates that the radiation is highly suppressed as $D$ increases.

Numerical calculations of head-on collisions \cite{Berti:2010gx,Cook:2017fec} 
produce a more nuanced picture of how gravitational radiation varies with space-time dimension.  The amount of radiation produced depends not just on $D$ but also on the ratio $M_1/M_2$ as well as the initial kinetic energy  associated with the collision.  
In the limit $M_1 / M_2 \to 0$, the smaller mass black hole can be treated as a small perturbation on the metric sourced by the heavier object \cite{Davis:1971gg,Berti:2003si,Berti:2010gx}.  By solving the wave equation in this fixed background, and numerically summing over modes, the radiated momentum and energy are obtained.  
\cite{Berti:2010gx} find that there is a critical value of the space-time dimension where the radiation is minimized.  This critical value depends on the initial kinetic energy of the system.  For black holes that start at rest, the minimum completely disappears and the radiation is a strictly increasing function of $D$.  This monotonically increasing behavior for $D>D_*$ is problematic, as the authors observe; at some point the radiated energy exceeds the mass of the point particle and suggests a break-down in the approximation that the background metric is fixed. 

Some of the same authors \cite{Cook:2017fec} later re-evaluated the $M_1/M_2 \to 0$ limit computations by exactly solving Einstein's equations.  In these later numerical simulations, where the initial kinetic energy is taken to vanish, the monotonically increasing behavior is absent, replaced by a maximum at a relatively small number of dimensions, $D =5$ or 6.  The fact that the radiation decreases in the strict large $D$ limit, seems much more plausible, especially given the estimate (\ref{radprod}).  

These simulations of head-on collisions in $D>4$, while interesting in their own right, are not promising from the point of view of understanding black hole merger events in four dimensions.  The presence of maxima and minima in the energy radiated as a function of dimension seems to exclude any sort of simple extrapolation from a large $D$ limit down to $D=4$.  Additionally, these are head-on collisions, with no angular momentum.  They lack the inspiral phase which produces the characteristic chirp in the gravitational waves observed at LIGO.

Of course it would be nice to look at black hole collisions that involve angular momentum effects in higher $D$.  
Numerically, there are obvious challenges in working with higher dimensional space-time grids to allow for the reduction in symmetry.  Our underlying purpose in taking a large $D$ limit, however, is to obviate the need for numerical simulations, most optimistically to be able to produce simple analytical estimates that can be extrapolated down to $D=4$.
In this context, 
a less obvious challenge is Bertrand's Theorem, which states that closed orbits for central force problems in classical mechanics are only possible for $r^2$ and $1/r$ potentials.  
While the potential experienced by two orbiting black holes in 4d is not strictly speaking $1/r$, when the black holes are far from each other, it is close to $1/r$, allowing for a long period of quasi-stable inspiraling behavior.  In higher dimensions, with a $1/r^{D-3}$ potential, the behavior promises to be more chaotic, and it is not entirely clear what universal lessons applicable to the $D=4$ case can be gleaned from working in the $D \to \infty$ limit. 

To sum up this discussion, collisions performed for large values of $D$ can be expected to resemble four-dimensional ones only when the impact parameters and the total angular momenta are small. For this study, the full scope of the large $D$ techniques, in particular the membrane theory of sec.~\ref{subsec:effmemb}, is still to be elucidated (see the discussion in sec.~\ref{sssec:limits}),
but at the very least one can expect to describe the non-linear relaxation towards the final stationary state using large $D$ effective theories.

Beyond this regime, the value of exploring large $D$ collisions lies in what we can learn about generic gravitational dynamics, and in applications to dual collisions in AdS/CFT. While the latter have received little attention yet, there are detailed analyses of the evolution of black hole collisions in asymptotically flat space which reveal qualitatively new behaviors, in particular, violations of cosmic censorship, which we discuss next.

\subsection{Cosmic censorsip violation in black hole collisions}
\label{subsec:ccv}

The studies in \cite{Andrade:2018yqu,Andrade:2019edf} build on the idea, discussed in sec.~\ref{sssec:blobs}, of treating localized black holes as blobs on a thin black brane -- a picture that can be regarded as borne out of the GL instability. The advantage of this approach is that one can employ  the effective brane equations \eqref{conshigherdone} and \eqref{conshigherdtwo} in order to model the entire collision process, starting from two Gaussian blobs thrown at each other and following their evolution with numerical simulations. These run very stably, and since a collision can be simulated in not more than a few minutes in a conventional computer, it is easy to explore wide ranges of initial conditions.

Because the gravitational potential dies off so quickly with distance, the impact parameter needs to be quite small for the two black holes to have any effect on each other. Another crucial property of these collisions is that no gravitational radiation is emitted, since as we have seen, it is strongly suppressed as $D$ grows. As a consequence, the total mass and angular momentum of the black hole system are conserved. This is of enormous help in obtaining a qualitative picture of the possible outcomes of the merger by examining the available stable configurations with the same mass and spin as the initial state. More precisely, we fix the total mass and add up the orbital angular momentum as well as the intrinsic spin of the black holes, to obtain a value of $J/M$ that characterizes the system.

\begin{figure}
 \begin{center}
 \includegraphics[width=.7\textwidth]{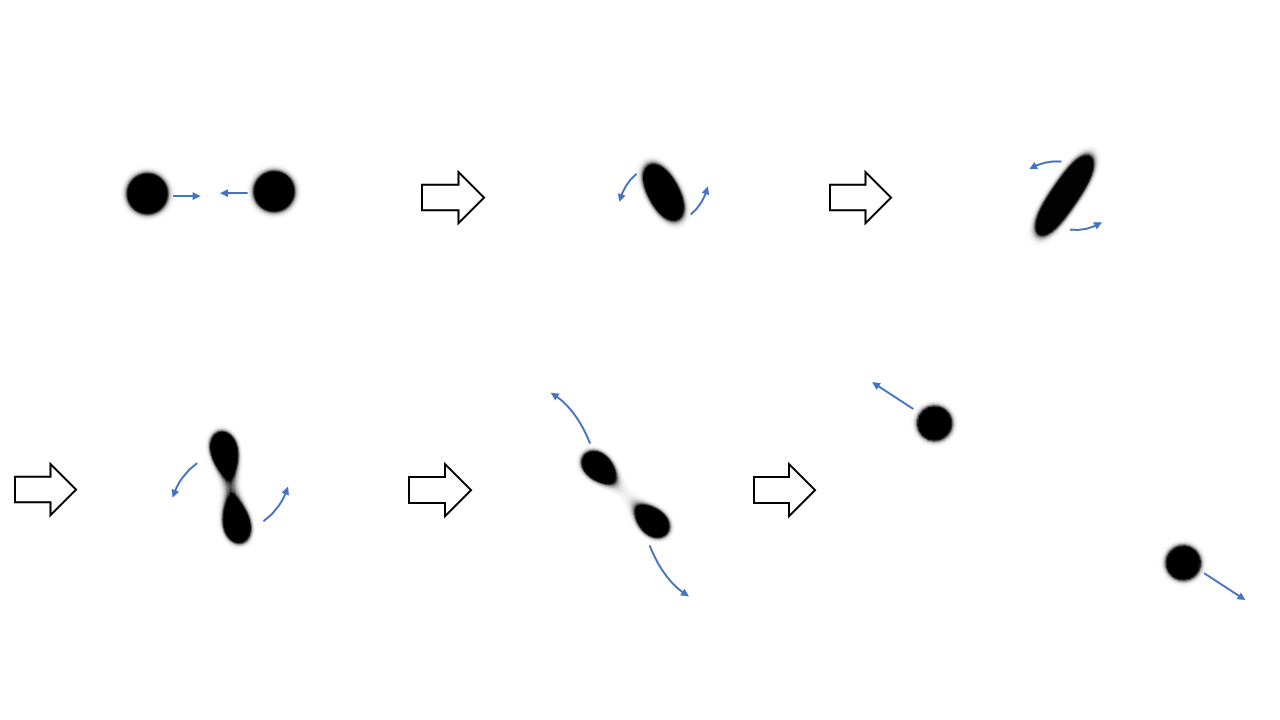}
 \end{center}
 \caption{
 Two spinning black holes collide and form a rotating black bar, which then breaks up into two outgoing black holes different than the initial ones.
 [[ \cite{Andrade:2019edf} figure 1]]
 \label{fig:blobcollision}
 }
 \end{figure}

The basic idea is that, when two blobs merge, if there exists a stable blob with the value of $J/M$ of the initial configuration, then the evolution will end on it. If, instead, no single-blob configuration is stable for that $J/M$, then the system will fragment into separate blobs that fly apart. For values of $J/M$ just beyond a stability threshold, a long-lived, but eventually unstable state -- typically a bar or dumbbell-shaped blob -- can form. These play a role similar to resonances in scattering, and they help understand the physics involved in the fragmentation. Long black bars, with large enough values of $J/M$, exhibit a GL-type of instability, which grows a pinch in their middle and eventually become singular. The violation of cosmic censorship that occurs in the GL instability, and its proposed resolution by fragmenting the horizon, will also appear here. This evolution is borne out by the numerical collision experiments of \cite{Andrade:2019edf}, employing (\ref{conshigherdone}) and (\ref{conshigherdtwo}), as depicted in fig.\ \ref{fig:blobcollision}.

\begin{figure}
 \begin{center}
 \includegraphics[width=.7\textwidth]{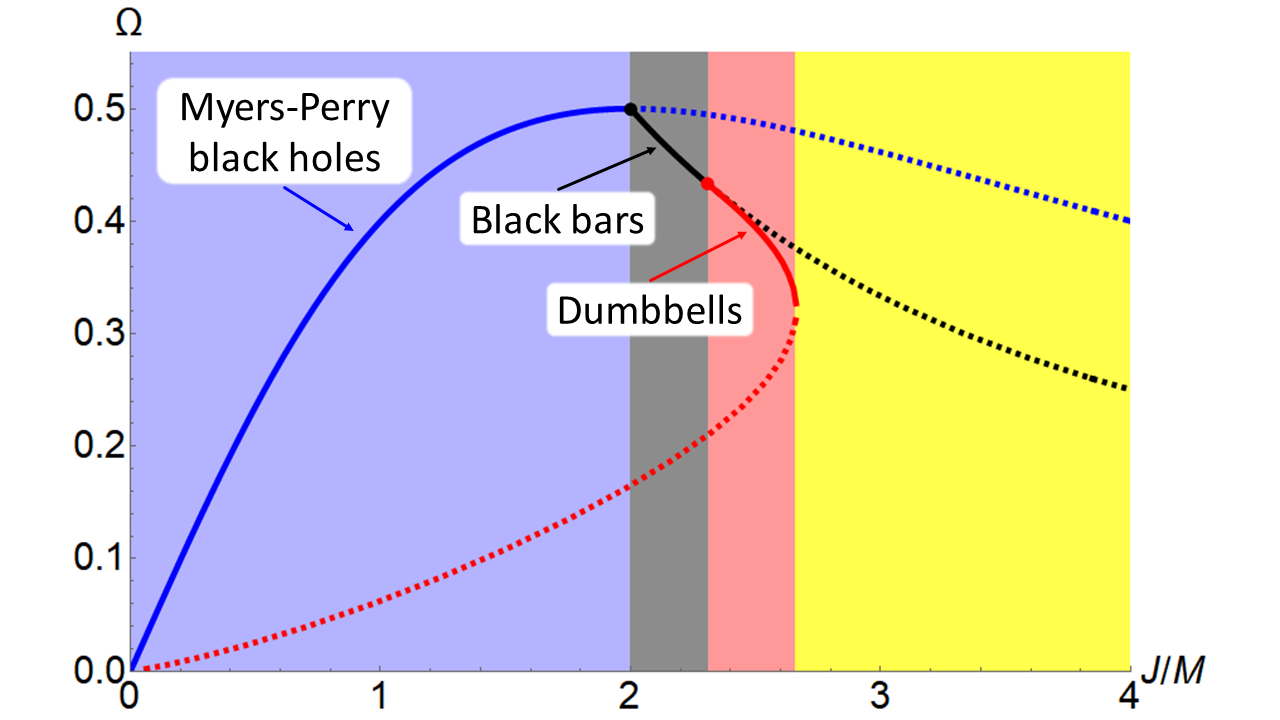}
 \end{center}
 \caption{Stable blobs as the endpoints of merger evolution (cf.~fig.~\ref{fig:blobspace}, and fig.~1 in \cite{Licht:2020odx}). Solid/dashed lines are stable/unstable stationary blobs. Blue: Myers-Perry black holes, stable up to $J/M=2$. Black: black bars, stable up to $J/M=4/\sqrt{3}\approx 2.31$. Red: dumbbells, stable up to $J/M\approx 2.66$ (dumbbells along the dashed line are more like unstable binaries of blobs). The background shading indicates the expected outcome of a merger for an initial value of $J/M$. No stable stationary blobs exist for $J/M\gtrsim 2.66$ (yellow), so, if a merger occurs in this region, it can only evolve to a horizon pinch, violating cosmic censorship (figure from \cite{Andrade:2020unpa}). 
 \label{fig:endpoints}
 }
 \end{figure}

Let us examine this proposed picture in more detail. In fig.~\ref{fig:blobspace}, we presented some of the phases of blob solutions, namely, Myers-Perry black holes and black bars, and their zero modes. Fig.~\ref{fig:endpoints} contains additional details, including other phases found in \cite{Licht:2020odx}. Consistently with the stability argument, collisions with a given initial value of $J/M<2.66$ always end in the stable blobs indicated in the figure: Myers-Perry black holes for $0\leq J/M <2$; black bars for $2\leq J/M < 4/\sqrt{3}$;  dumbbell-shaped blobs \cite{Licht:2020odx} for $4/\sqrt{3}\leq J/M< 2.66$. For larger values of $J/M$, there are no stable stationary blobs, so any black hole merger with $J/M>2.66$ evolves to a horizon-pinching violation of cosmic censorship.\footnote{As discussed in footnote~\ref{foot:odd}, odd-parity modes of bars lead to the formation of singular cusps at their endpoints. These may have interesting implications for the amount of gravitational radiation produced, especially in extrapolating to smaller $D$ where these radiation effects are anticipated to become more important.}

In the context of the simulations here, that pinch off never really happens.  Recall that  we are representing black holes as blobs on a continuous brane.  The neck joining the two blobs just becomes thinner and thinner.  Subleading effects in the $1/D$ expansion, which we have neglected, become more and more important, invalidating our approximation.  However, since cosmic censorship does appear to be violated in the analogous evolution of black strings, it is also expected to be violated in these black hole collisions.

To conclude this section, we would like to discuss how these collision experiments may or may not extrapolate to lower, finite values of $D$.  A big issue in this context is the competition between the dynamical instability associated with large enough $J/M$ and the classical radiation sourced by the non-axisymmetric spinning black objects.  \cite{Andrade:2019edf} perform a detailed estimate where they argue that, very likely down to at least $D\gtrsim 8$, radiation will be a negligible effect, and the dynamical instability will break up the bar into two disks. Their estimates are not precise enough to conclude if the effect will persist for lower $D$, but dedicated numerical simulations indicate that it does, even in $D=6$ \cite{Andrade:2020unpb}. The arguments in \cite{Andrade:2018yqu,Andrade:2019edf} are based on the similarity between the instabilities of black bars and of black strings, and the latter exist down to $D=5$. However, it seems possible that in this dimension the system will lose angular momentum rapidly enough to fall beneath the dynamical instability threshold and will relax back to a spinning black hole. In $D=4$ there are no black strings and all the mergers result in plump horizons, without any reason why they should destabilize and pinch. Therefore cosmic censorship seems very likely to hold in black hole collisions in $D=4$.

\section{Holographic applications}
\label{sec:holography}

One of the grand challenges of theoretical physics is to describe strongly interacting systems.  
In the context of understanding how strong interactions between quarks and gluons leads to the formation of hadrons
in quantum chromodynamics, 
one of the Clay Mathematics Prizes\footnote{%
https://www.claymath.org/millennium-problems
} is offered for proving that the simpler Yang-Mills theory has a mass gap.  
A better understanding of strange metals (or non-Fermi liquids) may unlock the secrets of high temperature superconductivity and yet seems to necessitate accurately modeling strong interactions between electrons.  
Turbulence and shocks in hydrodynamics, which are associated with another Clay Mathematics Prize, involve both the strong interactions between the liquid molecules and also strong time dependence.  Indeed, a second important challenge for theoretical physics is non-equilibrium phenomena.

There are few tools at our disposal to address these types of questions.  Perturbation theory is obviously limited to weakly coupled regimes.  Numerical simulations play an important role in examining turbulence but tend to be expensive and time intensive.  They also, in a time dependent and nonzero density world, are largely limited to classical phenomena.  In the quantum realm, computer simulations in lattice QCD work well where the Monte Carlo simulations do not run into issues with oscillatory path integrals  -- the ``sign problem''.  While these lattice computations can tell us the energy spectrum of QCD and the critical temperature above which QCD deconfines, they cannot in most cases address physics needed to describe fermions, non-zero densities, and time dependence.  

In this context, AdS/CFT correspondence (or holography) presents an additional tool to examine time dependence and strong interactions in quantum field theory.  
The tool works by conjecturing a relationship between a strongly interacting field theory and a classical theory of gravity in one extra dimension.  
Unfortunately, the field theories in question are, so far, not directly related to the field theories we believe describe the world we live in, e.g.\ QCD or QED.  AdS/CFT examples are thus a bit like integrable models -- systems that while not directly related to the physical systems of interest, can be exactly solved and thus may shed light on the original targets.

One uses AdS/CFT to examine QFT by solving 
the equations of motion of the dual gravity theory, effectively performing a saddle point approximation of the field theory in a strongly interacting limit, reducing the computation of a path integral to solving Einstein's equations.
One of the points of this review, however, is that solving Einstein's equations
is still a difficult task.  A large $D$ limit may make this task simpler.  Thus, it becomes natural to ask what additional insight a large $D$ limit may shed on strongly interacting quantum field theories and non-equilibrium phenomena via AdS/CFT.

Many partial answers to this question have already been provided.  We divide our discussion up into three parts: 1) hydrodynamics; 2) applications of AdS/CFT to condensed matter physics, sometimes called AdS/CMT, CMT for condensed matter theory; 3) applications of AdS/CFT to problems in nuclear physics, sometimes called AdS/QCD.

\subsection{Holographic Hydrodynamics}\label{subsec:holohydro}

We review here some applications of the large $D$ formalism to the hydrodynamic limit of strongly interacting field theories with a dual gravity description.  

As AdS/CFT deals with conformal field theories, we must have a traceless stress tensor ${T^\mu}_\mu = 0$, which greatly constrains the form of the constitutive relations, as mentioned briefly at the end of section \ref{sec:hydroreview}.  
 There is no bulk viscosity, only shear viscosity.  For a field theory in $d$ space-time dimensions, 
 the energy density, pressure, and temperature are related in a simple way:
 $\varepsilon = (d-1) P \sim T^d$.  
Furthermore, the speed of sound $c_s^2 = \frac{\partial \varepsilon}{\partial P} = \frac{1}{d-1}$ is suppressed in the large $d$ limit.  Similar to what we did in the discussion of the Gregory-Laflamme instability, we will need to rescale the spatial coordinates by a factor of $\sqrt{d}$ to keep sound modes in the leading order equations of a large $d$ limit.

As was discussed above, black branes in the large $D = d+1$ limit of anti-de Sitter space are described by a hydrodynamic-like set of equations (\ref{conshigherdone}) and (\ref{conshigherdtwo}) with $\epsilon = -1$.  In the context of the discussion of hydrodynamics in section \ref{sec:hydroreview}, these equations have a remarkable property.  They only involve first and second order derivatives.  Viewed as $\partial_\mu T^{\mu\nu} = 0$ in an appropriately chosen hydrodynamic frame, 
the gradient expansion has truncated at this leading order in the $1/D$ expansion \cite{Herzog:2016hob,Emparan:2016sjk}.\footnote{%
See also \cite{Camps:2010br} for the earlier black string analog of this statement.
}
Provided gradients are small compared to $D$, even if the gradients are otherwise large, e.g.\ $1 \ll (\partial_z m)/ m \ll D$, hydrodynamics is exact.
By an appropriate frame redefinition, 
one can put (\ref{conshigherdone}) and (\ref{conshigherdtwo}) in Landau frame with $T^{\mu\nu}$ given by (\ref{hystress}) 
to first order in gradients.  Indeed one can find the map to second order in gradients in
\cite{Emparan:2016sjk,Herzog:2016hob,Rozali:2017bll}.  Such a frame redefinition will introduce in fact an infinite series of higher order gradient corrections.  In other words, the statement that the gradient expansion truncates is not independent of frame choice.  The frame chosen by the large $D$ limit is thus rather special.

These equations  (\ref{conshigherdone}) and (\ref{conshigherdtwo}) constitute a non-relativistic, compressible version of the Navier-Stokes equations.  
The equations must be compressible to allow for sound waves.
The equations are non-relativistic because of the rescaling that has kept the sound modes in the spectrum of the theory.  By zooming in on modes with a speed of order $1/\sqrt{D}$, the speed of light has effectively been pushed off to infinity.  
Indeed, the equations are invariant under the Galilean boost \eqref{galiboost}.   If $m(t,x)$ and $p_i(t,x)$ are a solution, then so are  $m(t, x - v t)$ and $p_i(t, x - v t) - v_i m(t, x- v t)$.

To see the sound modes, similar to what we did in the Gregory-Laflamme case, we can perform a fluctuation analysis $m = m_0 + \delta m \, e^{-i \omega t +i k x}$ and $p = \delta p \, e^{-i \omega t + i k x}$.  We quickly arive at a dispersion relation
\be
\omega = \pm k - i k^2 \ .
\ee
Unlike the Gregory-Laflamme case where an instability appears below a threshold $k_{GL}$, these modes are always stable.
They are sound modes which have speed equal to one, given our rescaling of the spatial coordinates.  They also have a damping term, indicative of the presence of viscosity.  Indeed, we can understand the relation between damping and viscosity by running a similar analysis with the constitutive relation (\ref{hystress}) for $T^{\mu\nu}$ in Landau frame.  In that case, we start with the ansatz $T = T_0 + \delta T e^{-i \omega t + i \tilde k x}$ and $u_x = \delta u \, e^{-i \omega t + i \tilde k x}$ with $u_t = -1$.  
The position coordinate has not been rescaled in (\ref{hystress}) and so we should be careful to keep track that $\tilde k = \sqrt{n} k$.  
We find the dispersion relation, accurate only to terms of $\Or{\tilde k^2}$:
\be
\omega = \pm \frac{\tilde k}{\sqrt{d-1}} - \frac{i (2d-3)}{2d(d-1)} \frac{\eta}{P} \tilde k^2 + \Or{\tilde k^3} \ .
\ee
Taking the large $d$ limit, we identify the shear viscosity with the pressure, $\eta = P$.  In this large $d$ limit, black hole thermodynamics tells us that the entropy density $s = 4 \pi P$, allowing us to recover the well-known holographic result $\frac{\eta}{s} = \frac{1}{4 \pi}$ 
\cite{Kovtun:2004de}.

While equations similar to (\ref{conshigherdone}) and (\ref{conshigherdtwo}) (with $\epsilon = -1$) have been explored in a hydrodynamic context in depth over the years, the connection to black hole physics is relatively new.  It is also surprising that in this context, the description is exact at leading order in a $1/D$ expansion, even when the gradients are large.  We would like to mention two explorations of these equations in more detail.

The first \cite{Herzog:2016hob} explores the physics of shock formation in a 1+1 dimensional setting, keeping only a single $p_i$.  The second \cite{Rozali:2017bll,Andrade:2019rpn} 
explores turbulence in 2+1 and 3+1 dimensions, keeping two or three of the $p_i$.  Through the AdS/CFT correspondence, these explorations correspond to particular fluid-like behaviors of the dual field theories.  In the context of gravity, they have an intriguing interpretation as dynamical motion of black hole horizons.  

\begin{figure}
a) \includegraphics[width=.35\textwidth]{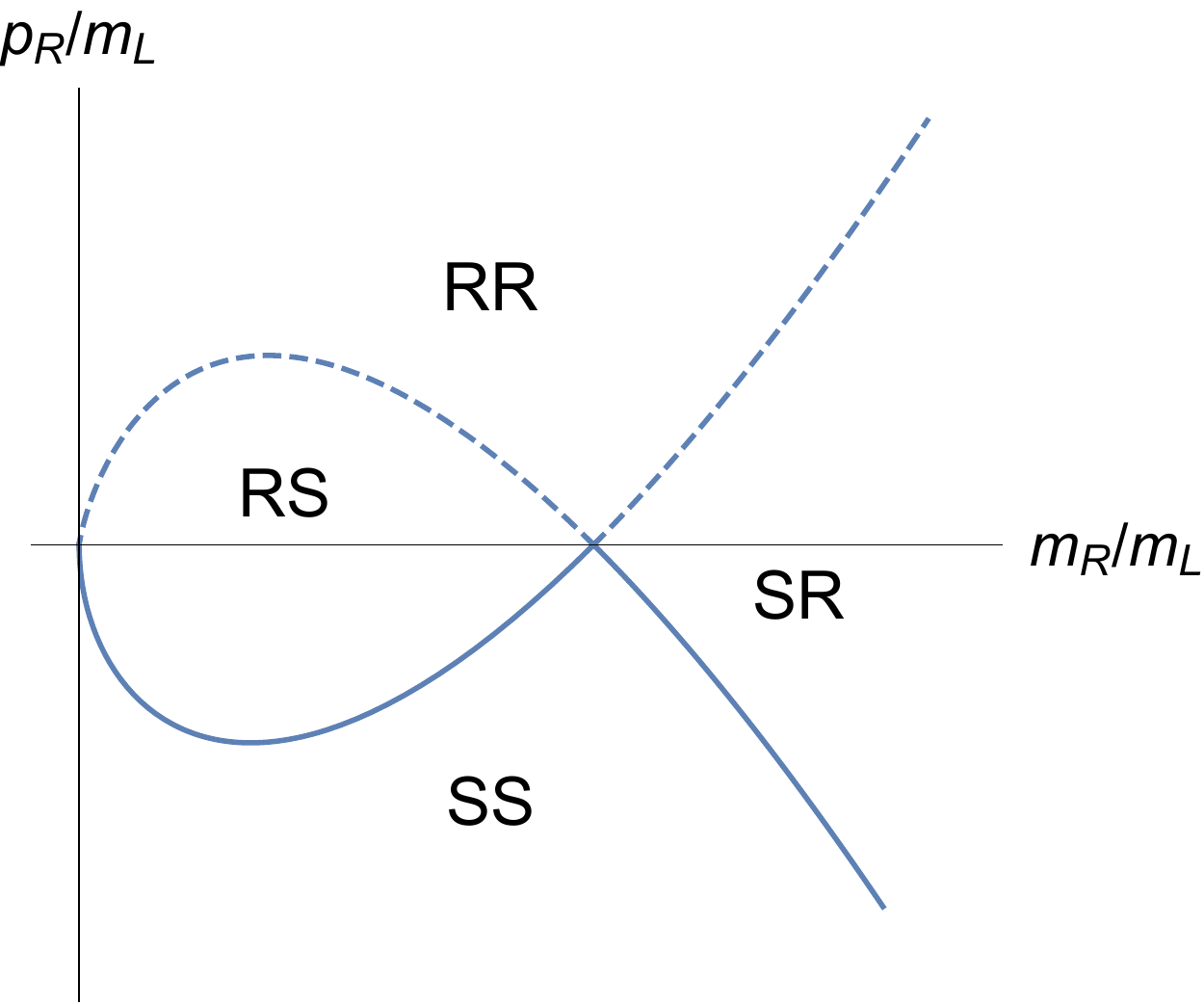}
b) \includegraphics[width=.55\textwidth]{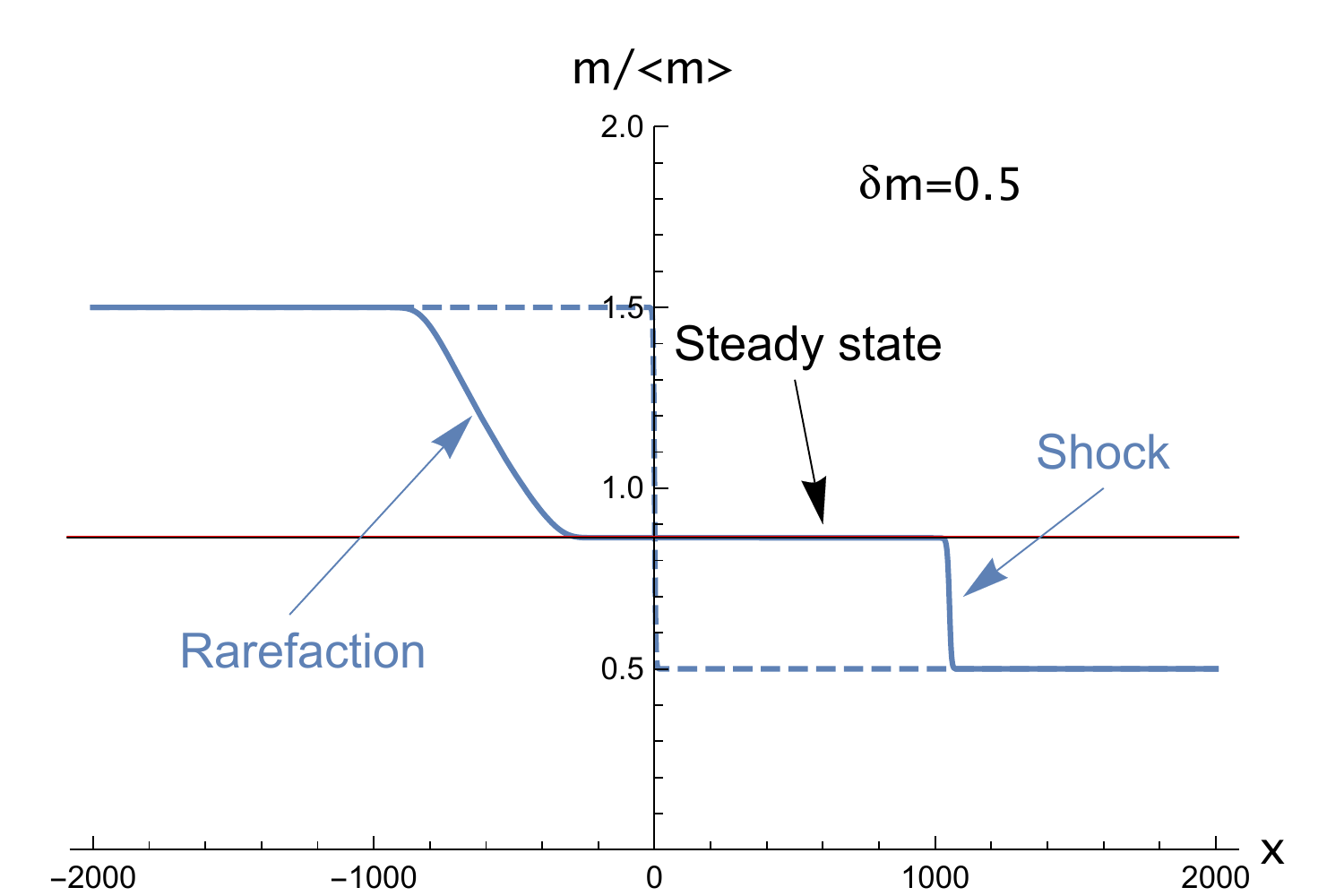}
\caption{
a) A phase diagram for the solution to the Riemann problem in a large $D$ limit.  Given a pair $(m_L, 0)$ and $(m_R, p_R)$, the selection of shock and rarefaction waves is determined by the value of $m_R/m_L$ and $p_R/m_L$.  The dashed and solid lines are ``critical'': The dashed line indicates the values of $(m_R, p_R)$ connected to $(m_L,0)$ by a single rarefaction wave while the solid line indicates the value of $(m_R, p_R)$ connected to $(m_L,0)$ by a single shock wave.  b)
A snapshot of the time evolution of the energy density for a RS case; $x$ is the rescaled position coordinate while 
$\delta m=  (m_L - m_R) / (m_L + m_R)$ and $\langle m \rangle = \frac{1}{2}(m_L + m_R)$. 
[[ \cite{Herzog:2016hob} figs.\ 1 and 5 ]]
\label{fig:fish}
}
\end{figure}

The authors \cite{Herzog:2016hob} consider the case of a Riemann problem where at time zero there is a planar interface in the system.  To the right, the energy and current are $(m_R, p_R)$ while to the left they are $(m_L, p_L)$.  
In fact, by Galilean invariance, we can boost to a frame where $p_L=0$ without loss of generality.
In a typical time evolution, a pair of rarefaction and/or shock waves form and move away from each other, creating in their wake a region with almost constant $m$ and $p$.  
In recent literature, this intermediate region has been called a non-equilibrium steady state (NESS) \cite{Bernard:2016nci}.
Starting with (\ref{conshigherdone}) and (\ref{conshigherdtwo}), \cite{Herzog:2016hob} determine a ``phase diagram'' (see fig.\ \ref{fig:fish}) that describes which pair of waves are formed: rarefaction-shock (RS), shock-shock (SS), shock-rarefaction (SR), or rarefaction-rarefaction (RR).  
Entropy production plays a key role in deciding which type of waves are preferred;
in the cases where a rarefaction wave is preferred, the corresponding shock solution would lead to a
decrease in the entropy of the fluid.
Interestingly, the rules for the phase diagram can be derived 
in a limit where, away from the shock interface,
 the second order derivative contributions
 to 
(\ref{conshigherdone}) and (\ref{conshigherdtwo}), i.e.\ the viscous terms, are dropped and the fluid is treated as ideal. 
Nevertheless, the numerical simulations indicate the viscous effects are small and become less and less important as time increases.

The authors \cite{Rozali:2017bll} explore decaying turbulence described by (\ref{conshigherdone}) and (\ref{conshigherdtwo}) in 2+1 and 3+1 dimensions.\footnote{%
 See \cite{Adams:2013vsa} for earlier tour de force numerical AdS/CFT 
 simulations of decaying turbulence in 2+1 dimensions, not in the large $D$ limit.
}
  They look at counterflow initial conditions, where a sinosoidally varying initial current distribution is allowed to relax.  These initial conditions allow for the initial Mach and Reynolds numbers to be independently varied; low Mach number corresponds to the incompressible limit and high Reynolds number to a situation that should quickly become turbulent.  Their results are largely insensitive to Mach number, and the turbulence they observe largely similar in character to that of incompressible fluids.  
In three spatial dimensions, they see an energy cascade where the energy of the initial counterflow is carried to smaller and smaller scales and eventually dissipated in small eddies.  They observe the standard Kolomogorov scaling law $E(k) \sim r^{2/3} k^{-5/3}$ where $E(k)$ is the energy carried by wave numbers $k$ 
in a shell between $|k|$ and $|k| + \d k$  and $r$ is the energy loss rate.  

In two spatial dimensions, they see instead an inverse cascade where the vortices become larger as time proceeds.  The scaling law they see $E(k) \sim k^{-4}$ is not as expected.  In studies of steady-state turbulence in two dimensions, one typically pumps in energy at some scale $k_{\rm in}$.  There is then an inverse energy cascade with scaling $E(k) \sim r_{\rm IR}^{2/3} k^{-5/3}$ to larger and larger length scales where eventually one must put in some damping term $r_{\rm IR}$ in the simulation to remove energy from the largest vortices.  But there is also a direct enstrophy cascade, with scaling $E(k) \sim r_{\rm UV}^{2/3} k^{-3}$ where $r_{\rm UV}$ is the rate at which enstrophy is removed at small scales.  Enstrophy, roughly speaking, is the square of the vorticity density of the system.  That \cite{Rozali:2017bll}  do not see the standard scaling results $k^{-5/3}$ or $k^{-3}$ in two spatial dimensions may be largely due to the fact that they are looking at decaying turbulence and have not carried out a large enough simulation to see robust scaling results.  
The authors \cite{Chen:2018nbh} later generalized this analysis to include a Gauss-Bonnet term in the Einstein equations.  They find the same qualitative behavior as in \cite{Rozali:2017bll}, including the $k^{-4}$ scaling in two dimensional turbulent flows at large Reynolds number and low Mach number.
More recently, however, \cite{Andrade:2019rpn} considered forced turbulence in the large $D$ limit in an effective 2+1 dimensional system.  
They 
do observe the $k^{-5/3}$ scaling
in the inverse cascade as well as standard, Kolmogorov scaling results for the longitudinal structure functions.
However, they do not discuss the direct enstrophy cascade or the dependence of their results on Mach and/or Reynolds number.

\subsection{AdS/CMT}
\label{subsec:adscmt}

Our survey of the large $D$ AdS/CMT literature will proceed in reverse chronological order.  The hydrodynamic discussion above 
transitions naturally into a discussion of translation symmetry breaking \cite{Andrade:2018zeb}, achieved through fixing
a nontrivial metric on the boundary of the asymptotically AdS space-time.  A drawback of using holography to compute transport coefficients such as charge and heat conductivities is that 
most of the simplest models preserve translation invariance.  In a condensed matter context, this invariance is broken
by a lattice and impurities, leading to qualitative differences in the behavior of transport.  For example, in a system at nonzero charge density, charge currents also carry momentum.  With nothing to dissipate momentum, the DC charge conductivity would be infinite.

Having disposed of the conceptually simplest method of breaking translation invariance, we will then move to an earlier 
work by the same group of authors \cite{Andrade:2015hpa}, where translation symmetry breaking is achieved through adding 
additional scalar fields to the Einstein-Hilbert action.  
Indeed, 
in the context of AdS/CMT applications, Einstein's equations are often solved in the presence of additional fields.  
The holographic dictionary maps the boundary values of bulk scalar fields to scalar operators in the field theory.  The dictionary also posits an equivalence between the boundary values of bulk gauge fields and conserved currents in the dual field theory.  If one wanted to study conductivity, for example, which could be extracted from applying a Kubo formula to a two point current-current correlation function, it is natural to include a bulk gauge field  in the gravity system.  Similarly, if one wanted to model a phase transition with a scalar order parameter, it becomes natural to include a bulk scalar field.  
The idea of adding additional fields to the Einstein-Hilbert action leads us to the last topic, 
the large $D$ holographic superconductor
\cite{Emparan:2013oza,Romero-Bermudez:2015bma}, 
which historically preceded the works on translation symmetry breaking.

\subsubsection*{Breaking Translation Symmetry}

While there exist a variety of techniques to break translation invariance in AdS/CMT models, the reduction in symmetry involved typically greatly increases the difficulty of solving the equations of motion.  
The most obvious method is also technically quite challenging.
%
Beginning with the Einstein-Hilbert action in a background with negative cosmological constant, 
one lets the boundary of this space-time be an arbitrary metric
\be
\label{hAB}
h_{AB} \d x^A \, \d x^B = - \left(1 - \frac{\gamma_{tt}(t,x)}{n} \right) \d t^2 - \frac{2}{n} \zeta_i (t,x)\d t \, \d x^i 
+ \frac{1}{n} \gamma_{ij}(t,x) \d x^i \, \d x^j \ .
\ee
In analogy to the Newtonian limit of general relativity, $\gamma_{tt}$ is a physical potential for the system,
$\zeta_i$ has some similarities to an external gauge field, and 
$\gamma_{ij}$ in a condensed matter context can be thought of as strain disorder.
The reduction in symmetry makes Einstein's equations an enterprise to solve.
The case of $\gamma_{tt}$ was described with relatively heavy duty numerical techniques in
 \cite{Balasubramanian:2013yqa} for the special case of $AdS_4$. 
 If one restricts to boundaries with small temporal and spatial gradients,  a fluid gravity approach \cite{Hubeny:2011hd} can be employed.  The authors \cite{Scopelliti:2017sga} looked at a nonzero $\gamma_{ij}$ through this lens.

Here, we review a large $D$ approach \cite{Andrade:2018zeb}.  One follows the same path 
that led to the hydrodynamic-like equations (\ref{conshigherdone}) and (\ref{conshigherdtwo}).  
In the same way, the gradient expansion
truncates at leading order in the $1/D$ expansion.  
The covariantized version of (\ref{conshigherdone}) and (\ref{conshigherdtwo}) is then
\be
\label{covhydroone}
(\partial_t + K - \nabla_i \nabla^i) m &=& - \nabla_i \tilde p^i \ , \\
(\partial_t + K - \nabla_j \nabla^j) \tilde p_i &=& - \nabla_i m 
- \nabla_j \left( \frac{\tilde p_i \tilde p^j}{m} \right) + \frac{\tilde p_i}{2} {\mathcal R} 
- \frac{\nabla_i (m^2 {\mathcal R})}{2 m} - m \partial_i K + 2 \nabla_j (m {K^j}_i ) \nonumber \\
&& + m \partial_t \zeta_i + 2 (\tilde p^j - \nabla^j m) \nabla_{[j} \zeta_{i]} + \frac{m}{2} \nabla_i \gamma_{tt} \ , 
\label{covhydrotwo}
\ee
where $\tilde p_i = p_i + m \zeta_i$.  The covariant derivatives are computed with respect to the spatial metric $\gamma_{ij}$, 
${\mathcal R}$ is the Ricci scalar of $\gamma_{ij}$, and $K_{ij} = \frac{1}{2} \partial_t \gamma_{ij}$ with $K = \gamma^{ij} K_{ij}$.  

The equations (\ref{covhydroone}) and (\ref{covhydrotwo}) are not the simplest to work with, but yield some interesting results.  Keeping only two spatial directions dynamical, and introducing translation breaking in the form 
\be
\gamma_{ij} = (1 + A_0 \cos (k_L x) \sin (k_L y)) \delta_{ij}
\ee
the authors \cite{Andrade:2018zeb} compute the heat conductivity tensor $\kappa^{ij}$.  They look at nonzero frequency and wave number $e^{-i \omega t + i q x}$ in a perturbative expansion in $A_0$, in the regime where $\omega \sim q \sim (A_0)^2$.  They find
\be
\kappa^{xx} &=& \frac{i \omega }{ \omega( \omega + i \Gamma + 2 i q^2 ) - q^2 } + \Or{A_0} \ , \\
\kappa^{yy} &=& \frac{i}{\omega + i \Gamma + i q^2} + \Or{A_0}
\ee
The mixed components $\kappa^{xy}$ vanish to $\Or{A_0}$.   The damping coefficient has the form
\be
\Gamma = \frac{k_L^2 ( 1 + 2 k_L^2 + 2 k_L^4)}{4 (1 + 2 k_L^2)^2} A_0^2 + \Or{A_0}^4 \ .
\ee
and prevents the conductivities from diverging in a $\omega \to 0$ and $q \to 0$ limit, as they otherwise would without
the explicit translation symmetry breaking introduced through $\gamma_{ij}$.


In attempts to find technically simpler if conceptually less straightforward methods of breaking translation invariance,
people fell upon introducing
a set of scalar fields with a linear spatial profile $\Psi_I = \alpha x^a \delta_{Ia}$ \cite{Donos:2013eha,Andrade:2013gsa}
with $\alpha$ a real number parametrizing the strength of the translation symmetry breaking.  
One is able to keep an ansatz for the metric and fields that requires solving
only ODEs, not PDEs.
Consider the action
\be
I = \int \d^{n+3} x \sqrt{-g} \left( R + 2 \Lambda - \frac{1}{4} F^2 - \frac{1}{2} \sum_{I=1}^{n+1} (\partial \psi_I)^2 \right)
\ee
with negative cosmological constant $\Lambda = -\frac{(n+1)(n+2)}{2 L^2}$ (one can typically set $L=1$ without loss of generality)  and field strength $F = \d A$.  Adding a field strength $F_{\mu\nu}$ allows one to look at charge conductivities in addition to heat conductivities.

A slight generalization of a charged black hole in AdS solves the equations of motion
\be
\d s^2 = - f(r) \d t^2 + \frac{ \d r^2}{f(r)} r^2 \delta_{ab} \d x^a \, \d x^b \ , \; \; \; A = A_t(r) \d t \ ,
\ee
where
\be
f(r) &=& r^2 - \frac{\alpha^2}{2n} - \frac{m_0}{r^n} + \frac{ n \mu^2}{2 (n+1)} \frac{r_0^{2n}}{r^{2n}} \ , \\
A_t(r) &=& \mu \left( 1 - \frac{r_0^n}{r^n} \right) \ .
\ee
From the form of $f(r)$, it is clear that in order to remain a significant effect in the large $D$ limit, one should keep
$\alpha$ of $\Or{\sqrt{n}}$, defining $\hat \alpha \equiv \alpha / \sqrt{n}$.  
The Hawking temperature of the black hole, which is also the temperature of the field theory, depends both on the chemical potential $\mu$ and the inhomogeneity parameter $\alpha$:
\be
T = \frac{f'(r_0)}{4 \pi} = \frac{1}{4\pi} \left( (n+2) r_0 - \frac{\alpha^2}{2 r_0} - \frac{n^2 \mu^2}{2(n+1) r_0} \right) \ .
\ee

We introduce the frequency dependent transport coefficients: charge conductivity $\sigma(\omega)$, thermoelectric coefficient $\beta(\omega)$, and heat conductivity $\kappa(\omega)$.  As expected, the DC conductivities are all finite
\cite{Andrade:2013gsa,Donos:2014cya}:
\be
\sigma(0) = r_0^{n-1} \left( 1 + n^2 \frac{\mu^2}{\alpha^2} \right) \ ,
\ee
\be
\kappa(0) = r_0^{n+1} \frac{(4 \pi)^2 T}{\alpha^2} \ , \; \; \; \beta(0) = r_0^n \frac{4\pi \mu}{\alpha^2} \ ,
\ee
It is also possible to compute $\kappa(\omega)$ exactly when $\mu = 0$, $\alpha = \sqrt{2n}$  and $n$ is odd \cite{Andrade:2015hpa}:
\be
\kappa(\omega) = \frac{2 \pi \cosh \left(\frac{\pi \omega}{2} \right) \Gamma \left( \frac{1}{2}(n - i \omega) \right) \Gamma \left(\frac{1}{2}(n+i \omega) \right)}{\Gamma \left(\frac{n}{2}+1 \right) \Gamma \left(\frac{n}{2} \right) }\ .
\ee
While appearing formidable, the expression is merely a $\frac{n-1}{2}$ order real polynomial in $\omega^2$.

Given this plethora of results and  the initial careful choice of translation breaking to simplify the equations of motion, 
one may legitimately ask why then take a large $D$ limit?  The answer is that it gives one a little bit of extra power 
in looking at the nonzero frequency behavior of these transport coefficients \cite{Andrade:2015hpa}.  Poles in $\sigma$, $\beta$, and $\kappa$, generically at complex values of $\omega$, are given by quasinormal modes of the black hole, which can be analyzed more thoroughly in a large $D$ limit.  It is also possible to compute $\kappa(\omega)$ to the first few orders in $1/n$, at least when $\mu = 0$.  The leading order result is  
\be
\kappa(\omega) = 2 \pi \frac{ 2 - \hat \alpha^2}{\hat \alpha^2 - i \omega} + \ldots
\label{kappa}
\ee
and the somewhat lengthy expressions for the next two orders can be found in \cite{Andrade:2015hpa}.  

The expression (\ref{kappa}) summarizes another key observation in \cite{Andrade:2015hpa} concerning the transition between coherent and incoherent transport.  Coherent transport is governed by a single pole in the transport coefficient, in this a case a purely dissipative mode at $\omega = -i \hat \alpha^2$.  Incoherent transport on the other hand will involve a sum over many incommensurate modes.  \cite{Andrade:2015hpa}  noticed that as $\hat \alpha^2$ grows toward 
$2 - \mu^2$, the transport becomes more incoherent in a large $D$ limit.  In (\ref{kappa}), this changeover is  represented by the vanishing of the residue.

\subsubsection*{Holographic Superconductor}

The holographic superconductor \cite{Hartnoll:2008kx,Hartnoll:2008vx,Gubser:2008px} 
is a system which includes both an abelian gauge $A_\mu$ field and a charged scalar $\Psi$:
\be
\label{holosc}
I = - \int \d^{n+1} x \sqrt{-g} \left(R - 2 \Lambda +  \frac{1}{4} F^2 + | \nabla \psi - i A \Psi |^2 + m^2 |\Psi|^2 \right) \ ,
\ee
where in the absence of the additional fields the negative cosmological constant $\Lambda = -\frac{n(n-1)}{2 L^2}$ would give rise to AdS with radius of curvature $L$.  The interest in this system stems largely from the fact that it provides a proof of principle that the physics of a superconducting phase transition can be added to the strongly interacting, scale invariant field theory described by this action in the absence of $\Psi$.  In other words, there is a dream that this system may shed light on the puzzle of high temperature superconductivity.

The puzzle of high temperature superconductivity is not so much about the superconducting region of the phase diagram, but about the normal phase.  At optimal doping but above the critical temperature, the system behaves like a strange metal or non-Fermi liquid, where the interactions between the electrons are large and a quasiparticle picture does not seem to be valid.  One hypothesis is that this region of the phase diagram is controlled by a scale invariant quantum critical point, i.e.\ by a field theory that may bear some resemblance to conformal field theories dual to gravity systems via AdS/CFT correspondence.  While such AdS/CFT systems do not ordinarily have a superconducting phase, the authors
\cite{Hartnoll:2008kx,Hartnoll:2008vx,Gubser:2008px} demonstrated that it is relatively trivial to add such a feature to the model; just add a charged scalar.\footnote{%
 To be more precise, these systems do not describe superconductivity, where the symmetry is gauged, but superfluidity,
 where the broken symmetry is global.  Depending on the questions asked, the difference in many cases can be ignored.
 }

An issue with the holographic superconductor is that the equations of motion that follow from the action (\ref{holosc}) 
must be solved in most cases numerically.  The large $D$ limit provides the possibility of a paper and pencil approach \cite{Emparan:2013oza,Romero-Bermudez:2015bma}. We will review some of the analysis in \cite{Emparan:2013oza} and provide a brief summary of the results in 
\cite{Romero-Bermudez:2015bma}.  

To further simplify the analysis, \cite{Hartnoll:2008vx} proposed a probe limit, where the background metric is fixed and one looks only at the $(A_\mu, \Psi)$ system.  While \cite{Hartnoll:2008vx} worked in $n=3$, it is straightforward to generalize
\cite{Emparan:2013oza}.  The Schwarzschild metric takes the form
\be
\d s^2 = - r^2 h(r) \d t^2 + \frac{\d r^2 }{r^2 h(r)} + r^2 \d x^2 \ , 
\ee
where
$h(r) = 1 - \left( \frac{r_0}{r}\right)^n$ and the temperature is $\Or{n}$, $T = \frac{n r_0}{4 \pi}$.    
To model the phase transition which preserves translational symmetry on the boundary, it is sufficient to assume 
 $\Psi = \psi(r)$ and $A = \phi(r)$.  Then
\be
\psi'' + \left( \frac{h'}{h} + \frac{n+1}{r} \right) \psi' + \left( \frac{\phi^2}{r^4 h^2} - \frac{m^2}{r^2 h} \right) \psi &=& 0 \ , \\
\phi'' + \frac{n-1}{r} \phi' - \frac{2 \psi^2}{r^2 h} \phi &=& 0 \ .
\ee
The system is solved with the large $r$ boundary conditions
\be
\psi &=& \frac{\psi_+}{r^{\Delta_+}} + \ldots \ , \\
\phi &=& \mu - \frac{\rho}{r^{n-2}} + \ldots \ , 
\ee
where 
\be
\Delta_\pm = \frac{n}{2} \left( 1 \pm \sqrt{1+4 \hat m^2} \right) \ , \; \; \; \hat m = \frac{m}{n} \ .
\ee
One can work in the grand canonical ensemble where the chemical potential $\mu$ is a tuneable parameter.  Regularity of the solution at the black hole horizon then fixes the charge density $\rho$ as a function of $\mu$.  One also tunes the source for the scalar field to zero, where otherwise there would be an additional solution near the boundary that scales as $\psi \sim \psi_-  / r^{\Delta_-}$.  The order parameter for the phase transition, or equivalently expectation value for the scalar, is proportional to $\psi_+$.   

The authors \cite{Emparan:2013oza} were able to analyze this system analytically near criticality using a WKB style approach.  
As in the finite $D$ analysis, 
the phase transition is caused by the scalar becoming tachyonic close to the horizon because of a large value of $\mu$.  
They find an estimate for the critical chemical potential above which the system becomes superconducting
\be
\hat \mu = \sqrt{\frac{1}{4} + \hat m^2} + \left( \frac{\sqrt{\frac{1}{4} + \hat m^2}}{2 n^2} \right)^{1/3} a_1 + o(n^{-2/3}) 
\ee
where $a_1 = 2.33811$ is a zero of the Airy function ${\rm Ai}(x)$ and $\mu = n \hat \mu r_0$.  In this scale invariant setting, the physical quantity should be a dimensionless ratio of the temperature to the charge density (or chemical potential):
\be
\left. \frac{T}{\rho^{1/(n-1)}} \right|_{\rm crit} = \frac{n^2}{4\pi} \left[ \sqrt{\frac{1}{4} + \hat m^2}+ a_1 n^{-2/3} \left( \frac{\sqrt{\frac{1}{4} + \hat m^2}}{2} \right)^{1/3} + \ldots \right]^{-\frac{1}{n-1}} \ .
\ee
This estimate has about a 15\% error for $n=4$ and a 35\% error for $n=3$.  

In addition to this probe analysis, \cite{Emparan:2013oza} contains also a brief discussion of how to move beyond the probe limit and look at the back reaction of the $(A_\mu, \Psi)$ sector on the metric.
While largely numerical in nature, \cite{Romero-Bermudez:2015bma} contains some additional paper and pencil results.  They provide analytic estimates for the optical conductivity at zero temperature, in both the high frequency and low frequency limits.  They also compute the entanglement entropy in a large $D$ expansion.

\subsection{AdS/QCD}

We would like to describe two applications of the large $D$ limit to the AdS/QCD program. 
The first 
\cite{Casalderrey-Solana:2018uag} is part of a larger effort
to shed light on the physics of heavy ion collisions by looking at the hydrodynamics of a strongly interacting field theory with a gravity dual.  
The second \cite{Herzog:2017qwp} attempts to gain insight into the confinement phase transition (or cross-over) in QCD by looking at a similar phase transition in a field theory with a gravity dual.

\subsubsection*{Bjorken Flow}

Boost invariant hydrodynamic solutions are important for modeling heavy-ion collisions studied at RHIC and LHC.   Close 
to the central region of the collision, the particles produced are modeled well by an approximately boost invariant fluid.
In addition to its phenomenological interest, this boost invariant flow, or Bjorken flow, is an interesting laboratory to explore far from equilibrium dynamics.  
 A holographic dual of this flow was first studied in \cite{Janik:2005zt}.  Here, we would like to review how to take a large $D$
 limit of this holographic dual \cite{Casalderrey-Solana:2018uag}.


Bjorken flow is a solution of relativistic hydrodynamic equations that is invariant under Lorentz boosts along one of the spatial directions $x_{\parallel}$ of the system.  If we use Milne-type coordinates,
\be
\tau = \sqrt{t^2 - x_{\parallel}^2} \ , \; \; \; y = \operatorname{arctanh} \left( \frac{x_{\parallel}}{t} \right) \ ,
\ee
with $\tau$ the proper time and $y$ the rapidity, Bjorken flow depends only on $\tau$.  The flow does not depend on the rapidity $y$ or any of the transverse spatial coordinates.  
In the context of the gravity dual, one important difference from what has come before
is that no rescaling of the spatial coordinate has been performed.  Indeed, the flow is relativistic, and 
there is no spatial coordinate, only $\tau$.  

Given the symmetry restriction, the hydrodynamic conservation equations $\nabla_\mu{T^{\mu\nu}}  = 0$ reduce to a single equation
\be
\tau \dot \varepsilon(\tau) + \varepsilon(\tau) + P_L = 0 \ ,
\ee
where $\varepsilon(\tau) \equiv T^{\tau\tau}$ is the energy density and $P_L \equiv T^y_y$ is the longitudinal pressure.  For a conformal fluid at equilibrium, we have $\varepsilon(\tau) = (n-1) P_L$.  We can thus isolate the deviation from equilibrium,
\be
\Delta P_L = P_L - \frac{1}{n-1} \varepsilon(\tau) \ ,
\ee
as coming from gradient corrections in the hydrodynamic expansion.  Given the dependence on a single variable $\tau$, gradients become equivalent to inverse powers of $\tau$.  

To express the hydrodynamic equation at arbitrary order in the gradient expansion in a large $n = D-1$ limit, the 
authors \cite{Casalderrey-Solana:2018uag} find it convenient to introduce two auxiliary quantities, $\tilde \varepsilon = (4 \pi T / n)^n$ and $w_\varepsilon = \tau \tilde \varepsilon^{1/n}$.  With these definitions, the conservation equation takes the form
\be
\tau \frac{\dot {\tilde \varepsilon}}{\tilde \varepsilon} + \frac{n}{n-1} + \sum_{i=1}^\infty \frac{\theta^{(i)}}{w_\varepsilon^j} = 0 \ , \; \; \;
\theta^{(i)} = \sum_j \theta_j^{(i)} \frac{1}{n^j} \ ,
\ee
where the $\theta_j^{(i)}$ are numerical coefficients.  Starting from Einstein's equations, \cite{Casalderrey-Solana:2018uag} provide these coefficients up to and including order $n^{-3}$. 
Consistent with earlier observations that the gradient expansion truncates in the large $D$ limit,
 the sum appears to start with $j = \lfloor \frac{i+1}{2} \rfloor$.  In other words, working to a given order in $1/n$, only a finite number of inverse powers of $\tau$ are required.
 At late times, all the gradient corrections vanish, and the flow is controlled by ideal Bjorken expansion
 \be
 \tilde \varepsilon \sim \frac{\Lambda^n}{(\Lambda \tau)^{n/(n-1)}}  .
 \ee

The information necessary to set-up the initial conditions is mostly lost at late times, reduced to a single constant $\Lambda$.  
The data describing the initial conditions can be recovered by looking at non-perturbative corrections to the flow.  These corrections are governed by non-hydrodynamic (or non-decoupled) black hole quasinormal modes and have a time dependence of the form $e^{-i \Lambda n \tau + \Or{n^{1/3}}}$ \cite{Casalderrey-Solana:2018uag}.  The corrections indicate that the large $D$ expansion is only asymptotic in nature, as discussed in section \ref{subsec:nonconv}.  
In addition to being asymptotic in $1/D$, a similar analysis of the gradient expansion for Bjorken flow shows the gradient series is only an asymptotic series as well \cite{Heller:2013fn}.  
This story has close connections with the
recent interest in resurgence and trans-series.  See \cite{Aniceto:2018bis} for a review.  
In particular, the non-perturbative corrections both in the gradient expansion \cite{Heller:2015dha} and the large $D$ expansion \cite{Casalderrey-Solana:2018uag}
can be cast as trans-series.

\subsubsection*{Small Black Holes}

The deconfinement transition in QCD remains a mysterious affair, and there has been a long standing hope that AdS/CFT might shed some light on its nature.  While we do not have a holographic dual to QCD, one can study an analog
deconfinement phase transition in maximally supersymmetric $SU(N)$ Yang-Mills theory in four dimensions (MSYM).
 MSYM does have a gravity dual, a sector of which is well described by Einstein's equations in five dimensions with a negative cosmological constant.
While in flat space, MSYM is conformal, \cite{Witten:1998zw} realized that this theory does undergo a sort of deconfinement phase transition when placed on a $S^3$.  The sphere introduces a mass gap for the scalars and fermions of MSYM, of order the inverse radius of the sphere.  When the temperature is small compared to the inverse radius, the theory confines.  When the temperature is large, the theory deconfines.  On the gravity side, this first order phase transition is the Hawking-Page phase transition \cite{Hawking:1982dh} for black holes in AdS.

\begin{figure}
\begin{center}
\includegraphics[width=.5\textwidth]{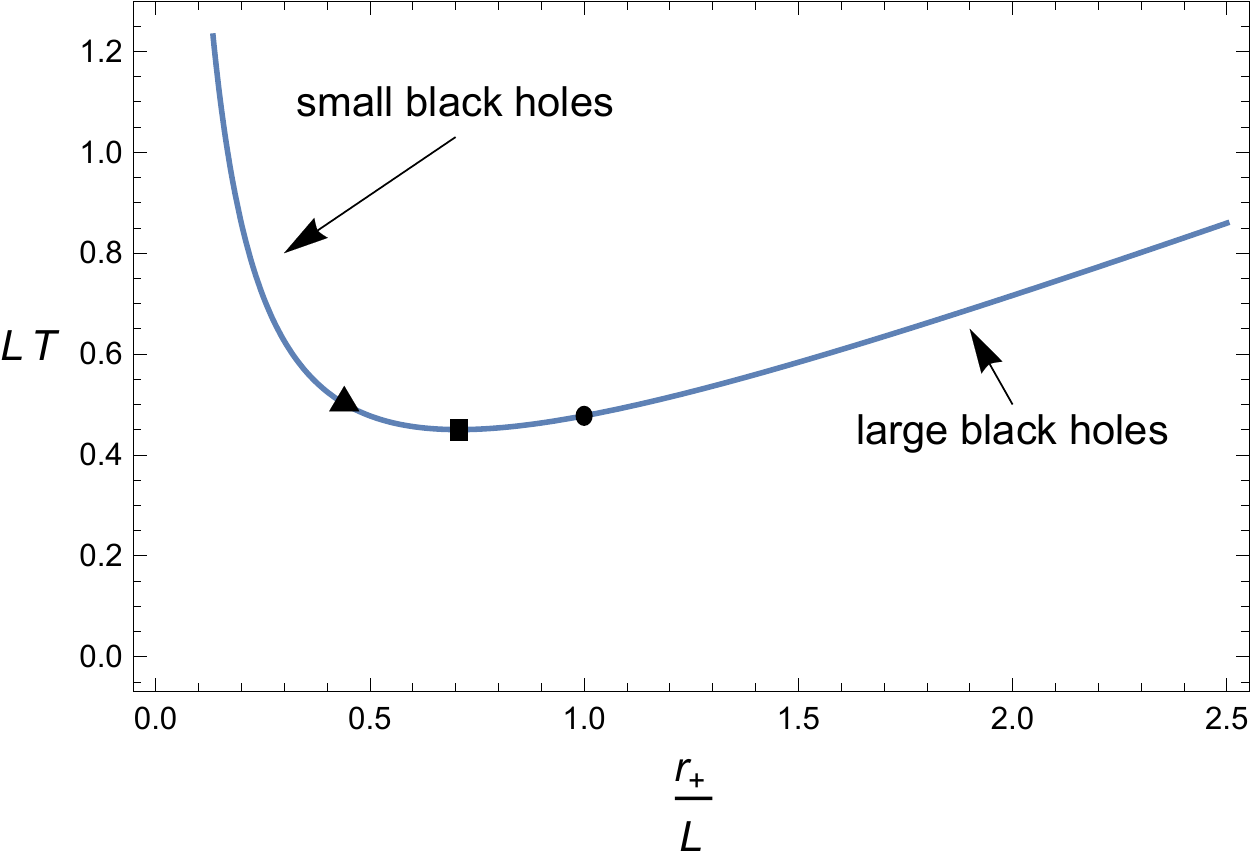}
\end{center}
\caption{
Black hole temperature as a function of radius for $D=5$.  Approaching from the large black hole branch, the circle indicates
the Hawking-Page phase transition to thermal (or empty) AdS in the canonical ensemble, the square is the point at which the heat capacity becomes negative, and the triangle is the Gregory-Laflamme type instability \cite{Hubeny:2002xn}.  As $D$ gets larger, the three special points approach each other.
[[ \cite{Herzog:2017qwp} fig.\ 1 ]]
\label{fig:HP}
}
\end{figure}

While this picture works well in the canonical ensemble, with a heat reservoir and a well-defined temperature, it is less clear what should happen in the micro-canonical ensemble, at fixed energy and entropy.  Black holes in AdS have two branches: a small black hole branch with negative specific heat, similar to black holes in Minkowski space; and a large black hole branch with positive specific heat.  
In the canonical ensemble, one starts on the large black hole branch at high temperature and cools the system down.  
At some point, there is a first order phase transition to empty AdS.  
See fig.\ \ref{fig:HP}.  In the micro-canonical ensemble, one would like to know what happens to 
smaller black holes, beyond the Hawking-Page point.  
There is an old conjecture that it is insufficient to look just at the $AdS_5$ sector of the geometry, that one should consider the full ten dimensional supergravity $AdS_5 \times S^5$ dual to MSYM, that the endpoint of the instability should be a small
black hole in ten dimensional space, with $S^8$ topology \cite{Banks:1998dd,Peet:1998cr}, associated with breaking the $SO(6)$ symmetry of the $S^5$.  (In MSYM, this SO(6) symmetry is the R-symmetry of the superconformal field theory.)
Later numerical studies pinned down the onset of this Gregory-Laflamme like instability \cite{Hubeny:2002xn,Buchel:2015gxa,Dias:2015pda}.  The endpoint was then studied with heavy duty numerics \cite{Dias:2015pda,Dias:2016eto}.
Here we shall review a large $D$ approach \cite{Herzog:2017qwp}, which can in principle be carried out with pen and paper, although more realistically some computer assisted algebra is required.  There are strong similarities to the discussion of the Gregory-Laflamme instability for black strings discussed above.

As the $AdS_5 \times S^5$ type IIB supergravity solution 
involves both the metric and the five-form, we begin with the large $D$ action
\be
S = \int \d^p x \, \d^q x \sqrt{-g} \left(R - \frac{1}{2 q!} |F_q|^2 \right) \ ,
\ee
and focus on the case of interest 
$p=q=d$ where the equations of motion are to be supplemented with a self-duality constraint on $F_d$.
\cite{Herzog:2017qwp} take the
ansatz for the metric and field strength
\be
\label{ansatz}
\d s^2 &=& g_{tt} \d t^2 + 2 \d t \, \d r + 2 g_{t \theta} \d t \, \d \theta + g_{\theta \theta} \d \theta^2 + g_A \d \Omega_{d-2}^2 + g_B \d \Omega_{d-1}^2 \ , \\\
F_d &=& \sqrt{g_A^{d-2} \det (S^{d-2})} (f_{tr} \d t \wedge \d r + f_{t \theta} \d t \wedge \d \theta  + f_{r\theta} \d r \wedge \d \theta ) \wedge \d \Omega_{d-2} \nonumber \\
&& 
+  \sqrt{g_B^{d-1} \det (S^{d-1})} (f_t \d t + f_r \d r + f_\theta \d \theta) \wedge \d \Omega_{d-1} \ 
\ee
where $n = d-3$.
In the $1/n$ expansion, one finds the leading order solution
\be
g_{tt} &=& - \left( 1 + r_c^2 R^{2/n} - \frac{m(t, \theta)}{R} \left(1 + r_c^2 m(t, \theta)^{2/n} \right) \right) + \Or{n^{-1}} \ , \\
g_{t \theta} &=& \frac{1}{n} \frac{p(t,\theta)}{R} + \Or{n^{-2}} \ , \; \; \; g_{\theta \theta} = 1 + \Or{n^{-2}} \ , \nonumber\\
g_A &=& r_c^2 R^{2/n} + \Or{n^{-3}} \ , \; \; \; g_B = \sin^2 \theta + \Or{n^{-3}} \ , \nonumber \\
f_t &=& \Or{n^{-2}} \ , \; \; \; f_r = \Or{n^{-2}} \ , \; \; \; f_\theta = \sqrt{2(n+2)} + \Or{n^{-3/2}} \ .  
\ee
where as usual $R = (r/r_c)^n$.
The functions $m(t,\theta)$ and $p(t,\theta)$ are governed by the hydrodynamic like equations
\be
\label{sfiveone}
\partial_t m - \cot \theta ( r_c \partial_\theta m + p) &=& 0 \ , \\
\label{sfivetwo}
\partial_t p - \cot \theta \left( r_c \partial_\theta p + \frac{p^2}{m} \right) + (1+r_c^2) \partial_\theta m + r_c (2 + \csc^2 \theta) p &=& 0 \ .
\ee
In contrast to what is done in the black string case, here no rescaling of the spatial $\theta$ coordinate has been performed.
It is possible to rescale $\theta$ \cite{Herzog:2017qwp}, zooming in on the equatorial region of the sphere.  The penalty one pays is that it becomes less clear how to deal with the boundary conditions at the poles $\theta = 0$ and $\theta = \pi$.  These boundary conditions are important given that we are looking for 
global solutions on the whole sphere.  Note that $\cot \theta$ in 
(\ref{sfiveone}) and (\ref{sfivetwo}) comes from a connection term in the covariant derivative $\nabla_\theta$. 
The portion $\partial_\theta$ is suppressed relative to the connection term in the large $n$ limit, although could be restored
by rescaling $\theta$.

Interestingly, these hydrodynamic equations can be solved generally, despite their apparent nonlinearity:
\be
p(t,\theta) &=& (- r_c \partial_\theta  + \tan \theta \, \partial_t ) \, m(t,\theta) \ , 
\\
\label{esoln}
m(t,\theta) &=& c \exp \left( \sum_{\ell=1}^\infty (a_{\ell_+} e^{-i \omega_{\ell_+} t} + a_{\ell_-} e^{-i \omega_{\ell_- }t}) \cos^\ell \theta \right) \ , 
\ee
where $c$ and $a_{\ell_\pm}$ are normalization constants and the allowed frequencies are
\be
\omega_{\ell_\pm} = i \left[ - r_c (\ell+1) \pm \sqrt{\ell+r_c^2(\ell+1)} \right] \ ,
\ee
with $\ell = 1, 2, \ldots$ a non-negative integer.  The Gregory-Laflamme like instability identified by \cite{Hubeny:2002xn} occurs where exactly one of these $\omega_{\ell_\pm}$ develops a positive imaginary part, namely $\omega_{1_+}$.  The zero mode first appears when $r_c = 1/\sqrt{2}$.  There are in fact a whole series of such instabilities, for each choice of integer $\ell$, corresponding to higher and higher spherical harmonics on the $S^d$.  These instabilities kick in as the radius of the black hole gets smaller and smaller, at $r_c = 1/\sqrt{\ell+1}$.  

The time dependent solution (\ref{esoln}) does lend some support to the early conjecture \cite{Banks:1998dd,Peet:1998cr} that the endpoint of the instability involves breaking the symmmetry of the $S^d$.  Depending on the initial conditions,
the time dependent solutions lead to black holes which look like spots or belts on the $S^d$.  Similar lumpy black holes were observed in \cite{Dias:2015pda}, for $d=5$, as well.

It would of course be nice to find static lumpy solutions with topology $S^{d-2} \times S^d$, 
or even isolated static solutions where the topology of the black hole changes to $S^{2d-2}$, 
as were found in \cite{Dias:2015pda}  and \cite{Dias:2016eto} respectively in the $d=5$ case.  Indeed, a
more careful consideration of $1/n$ corrections reveals candidate lumpy static solutions and also cousins
of the interpolating solution (\ref{interpolatingsoln}) that takes one from the uniform black hole in the far past to a lumpy black hole in the far future \cite{toappear}.  Finding solutions where the topology changes to $S^{2d-2}$ appears to be beyond the scope of this large $d$ expansion.  There will always be a horizon everywhere on the $S^d$, even though the horizon radius varies.   
%
%

In the study of the black string, we found there was a critical dimension above which the Gregory-Laflamme transition became second order, below which it was first order.  One may ask if a similar situation holds here.  It seems very likely although the jury is still out.  The authors \cite{Dias:2015pda,Dias:2016eto} find that for $AdS_5 \times S^5$, the transition is likely to be first order, while the analysis in \cite{Herzog:2017qwp, toappear} indicates that in the large $d$ limit, the transition is smooth and of second order. 

An important issue is the degree to which the ansatz (\ref{ansatz}) restricts the form of the answer.  The
ansatz allows the $SO(d+1)$ symmetry of the $S^d$ to break only to $SO(d)$ while the actual endpoint could conceivably break more symmetry.  Furthermore, 
there are many  fields in type IIB supergravity, in addition to the metric and five-form; it may be that the endpoint involves these other fields as well, in which case it is also less clear how to extend type IIB appropriately to higher dimensions. 
Hopefully, the situation is as simple as  was conjectured in \cite{Banks:1998dd,Peet:1998cr}, and the existence of the appropriate solutions in \cite{Dias:2015pda,Dias:2016eto,Herzog:2017qwp} lends support to this hope.

\newpage

\part{Other directions}
\label{part:ramblings}

There are several subjects in gravitational theory where a $1/D$ expansion is potentially fruitful, but which nevertheless have received relatively little attention. In the following we discuss a sample of them without any aims of being exhaustive.

\section{Quantum black holes, quantum gravity and strings}
\label{sec:hawkingetal}

The study of quantum effects in black holes seems a natural field where the large $D$ limit may be useful. Even though interacting quantum field theories in $D\geq 4$ have terrible ultraviolet behavior, many effects of quantum fields in curved spacetime arise in free theories and constitute essentially infrared physics, so they might be amenable to study in the $1/D$ expansion. A good example is the study in \cite{Keeler:2019exd} of one-loop determinants in the large $D$ limit. It exploits the decoupling of low frequency fluctuations in order to obtain analytical results, which give indications of the membrane-like nature of black holes in this regime.

Of course the most famous and important of black hole quantum effects is Hawking radiation, so it behooves us to discuss its main features when $D\gg 1$.

\subsection{Large $D$ limit of Hawking radiation}
\label{subsec:hawking}

We begin with some elementary remarks. When $\hbar$ is present, gravitational theory (retaining $G$ but setting $c=1$) acquires a length scale of its own, the Planck scale
\beq
L_\text{Planck}= (G\hbar)^{1/(D-2)}\,.
\eeq
Quantum effects on black holes of radius $r_0$ are then governed by the dimensionless ratio $r_0/L_\text{Planck}$. We may keep it fixed as $D$ grows, or instead change it with $D$ at a specified rate.\footnote{Equivalently, we can redefine the Planck length with suitable $D$-dependent factors.} The choice sets the size in Planck units of the black holes we are considering, and selects which quantum properties remain non-zero and finite in the large $D$ limit. For instance, it turns out to be impossible to take $D\to\infty$ in such a way that both the entropy and the temperature of the black hole are finite. These are\footnote{This discussion refers to Schwarzschild and possibly Myers-Perry black holes. In AdS the analysis can be substantially different.}
\beq
S_{BH}\sim \lp \frac{r_0}{\sqrt{D}L_\text{Planck}}\rp^D
\eeq
and
\beq\label{THEP}
\frac{T_H}{E_\text{Planck}}\sim L_\text{Planck}\frac{D}{r_0}\,,
\eeq
where 
\beq
E_\text{Planck}=\frac{\hbar}{L_\text{Planck}}=\frac{L_\text{Planck}^{D-3}}{G}\,.
\eeq
We see that the entropy stays finite when $D\to\infty$ if the black hole radius is $r_0\sim \sqrt{D}L_\text{Planck}(1+\alpha/D)$ (with constant $\alpha$), while finite temperature requires much larger sizes, $r_0\sim DL_\text{Planck}$. The latter is the condition that the size of the near-horizon region is Planckian parametrically in $D$, i.e., it could still be much larger than $L_\text{Planck}$ but in a $D$-independent manner.

Curiously, keeping $T_H$ finite does not imply finite emission rates, as noted in the first study of Hawking radiation in the large $D$ limit in \cite{Hod:2011zzb}. One might have expected that the typical energy $\hbar\omega$ of Hawking quanta is of the order of $ T_H$. However, the actual energies are much larger, due to the huge increase in the phase space available to high-frequency quanta at large $D$, which grows like $\omega^D$. This shifts the radiation spectrum towards energies much larger than $T_H$, with a peak around
\beq\label{peakom}
\hbar\omega_H\simeq D T_H\sim \hbar\frac{D^2}{r_0}\,.
\eeq
Thus we find a new, ultrashort length scale,
\beq
\lambda_H\sim \frac{r_0}{D^2}
\eeq
for the wavelength of typical Hawking quanta. It means that black holes at large $D$, unlike four-dimensional black holes (but like, say, stars), are very large quantum radiators, whose radius ($\sim r_0$) and typical classical vibrational wavelengths ($\sim r_0/D$) are much longer than the wavelengths that they radiate quantum mechanically. One consequence is that radiated Hawking quanta follow null geodesics in the black hole background, and the geometric optics approximation for graybody factors applies very accurately \cite{Hod:2011zzb}.\footnote{See \cite{Wei:2014bva} for other considerations on Hawking radiation at large $D$.}

Moreover, since the energy per quantum is very large and the time to emit each of them is very short, the black hole evaporates extremely quickly, within a timescale \cite{Holdt-Sorensen:2019tne}
\beq
t_\text{evap}\sim t_\text{Planck}\lp\frac{4\pi}{D}\rp^{D+1/2}S_{BH}^{\frac{D-1}{D-2}}\,.
\eeq
The factorial rate $D^{-D}$ makes this time potentially much shorter than the scrambling time
\beq
t_\text{scr}\sim \frac{t_\text{Planck}}{\sqrt{D}}S_{BH}^{\frac{1}{D-2}}\ln S_{BH}\,,
\eeq
and when this happens, the assumption of semiclassicality of the evaporation becomes very questionable. Conversely, this result puts constraints on the size of black holes that, in a given dimension, admit a semiclassical description \cite{Holdt-Sorensen:2019tne}.

More elucidation of the import of all these observations is desirable, which should be relevant for the further use of the large $D$ expansion in this context, as a conceptual guide and also as a calculational method.

\subsection{Large $D$ matrix models}
\label{subsec:frank}

In sec.~\ref{sec:crazy} we saw intriguing hints of a relation between string theory and the 2D black hole that appears near the horizon of large $D$ Schwarzschild. It is not clear, though, how to better ground these observations, since the microscopic description of that black hole in string theory is not very well understood. Nevertheless, there are other classes of near-horizon limits of large $D$ black holes, e.g., if one includes charge, which arise as solutions of different 2D dilatonic gravities. Some of these hold better promise for associating large $D$ black holes to a class of microscopic quantum theories.

In order to motivate this connection, we begin by recalling that our best quantum theories of gravity so far -- those based on holographic dualities -- take the form of theories of large $N$ matrices. Four-dimensional large $N$ gauge theories dual to AdS$_5$ gravity are a well known example with full fledged gravitational dynamics, but they are extremely difficult to solve. Matrix models in zero spacetime dimensions are much more tractable, and in the limit $N\to\infty$ they reproduce features of strings, gravity and black holes, but the absence of temporal dynamics limits their usefulness. Models of quantum mechanical matrices should provide a more realistic set up, but unfortunately it is in general very difficult to do explicit calculations with them. For this reason, the advent of the SYK model -- a quantum mechanical model of matrices where a `melonic' subclass of the planar diagrams dominates at large $N$ \cite{Kitaev:2015} -- triggered great advances in the microscopic understanding of black holes \cite{Maldacena:2016hyu}. The main properties that make it tractable have also been found in certain models of tensors \cite{Witten:2016iux}. Of interest to us here is a class of mixed matrix-vector models \cite{Ferrari:2017ryl} which, as we shall see, have reasonable hope of connecting to large $D$ black holes.

Matrix quantum mechanics can be motivated by the study of D0-branes in string theory: these are pointlike objects (i.e., not extended in space like $p$-branes) which are described using bosonic and fermionic $SU(N)$ matrices $X^{ij}_\mu(t)$, $\psi^{ij}_\mu(t)$, where the matrix indices $i,j$ stand for open string degrees of freedom, and the index $\mu$ indicates directions in the target space where the 0-brane lives, so we can regard it as an $O(D-2)$ vector index. String theory determines uniquely the Hamiltonian for these matrices, providing a quantum theory of gravity in which the regime of classical gravity emerges when $N\to\infty$ \cite{Banks:1996vh}.

We may consider other models of this kind with different Hamiltonians, e.g., with interactions not restricted by supersymmetry, in the expectation that the large $N$ limit also leads to a dual gravitational theory. Such models are in general as hard to solve as any matrix quantum mechanics, but \cite{Ferrari:2017ryl} made a remarkable observation: there exists models such that, if one expands them in $1/N$, and then takes the limit $D\to\infty$ in a specific way, the same melonic class of diagrams as in SYK becomes dominant.\footnote{For more work on these models, see \cite{Azeyanagi:2017drg,Ferrari:2017jgw,Azeyanagi:2017mre,Ferrari:2019ogc,Carrozza:2020eaz}.}
The idea is highly suggestive: taking the limit $N\to\infty$ results in a regime of classical gravity which is difficult to solve; but if we then send $D\to\infty$, substantial simplifications occur that make the theory much more manageable. The similarity to the large $D$ limit of classical black holes is tantalizing.

Since SYK models at low energies are dual to the dynamics of the near-AdS$_2$ throats that appear in Reissner-Nordstr{\"o}m black holes close to extremality -- not solutions of the dilaton gravity theory of \eqref{eq:larged3}, but of Jackiw-Teitelboim gravity \cite{Almheiri:2014cka} -- the natural expectation is that a model of the type of \cite{Ferrari:2017ryl} gives a quantum mechanical description of the large $D$ limit of near-extremal charged black holes. There seem to be the right ingredients, but more work is needed to support this connection.

\subsection{Large $D$ entanglement}
\label{subsec:entangle}

Quantum entanglement in a many-body system or a quantum field theory naturally presents a marked dependence on the dimensionality of space. The larger the number of dimensions is, the more neighbouring degrees of freedom are available to be entangled with. Similarly to our introductory remarks on $D\to\infty$ as a mean field theory limit, one expects that entanglement becomes strongly localized as the number of dimensions grows large, eventually resulting in the decoupling (in average) of correlations between points in the system.

The correlations between two subsystems $A$ and $B$ separated by a common spatial boundary can be quantified in terms of the entanglement entropies $S(A)$ and $S(B)$ of the density matrices $\rho_A$ and $\rho_B$ obtained by tracing out the degress of freedom of the complementary subsystem, respectively $B$ and $A$. Of special interest to us is that, for strongly coupled conformal field theories, there exists a prescription to compute these entropies by means of a dual gravitational bulk spacetime. This is the celebrated Ryu-Takayanagi (RT) entanglement entropy formula \cite{Ryu:2006bv}, covariantly extended in \cite{Hubeny:2007xt}. The entanglement entropy associated to a region $A$ is given by the area (divided by $4G\hbar$) of an extremal surface in an AdS bulk that is homologous to $A$, minimized over all such possible surfaces.

Within this circle of ideas, \cite{Colin-Ellerin:2019vst} chose to study the mutual information $I(A:B)$ as an appropriate measure of the correlations between two distant regions. This is defined as the amount by which the entropies of the separate regions differ from the entropy of their union, i.e.,
\beq
I(A:B)=S(A)+S(B)-S(A\cup B)\,.
\eeq
It vanishes if the systems are uncorrelated, and it reaches a maximum $I(A:B)=2S(A)=2S(B)$ when the joined system is pure, $S(A\cup B)=0$. 

Two convenient regions for the holographic study of their mutual information are two `caps' centered around opposite poles of the $S^{D-2}$ at the boundary of AdS$_D$. A simple argument suggests that the behavior of $I(A:B)$ must change discontinuously as the caps grow in size: when they are small, the RT surface for $S(A\cup B)$ will consist of two disjoint surfaces anchored at the boundaries of $A$ and $B$, so $I(A:B)=0$ (in holography, this vanishes to leading order in $1/N$, or in $G\hbar$). As the caps become larger and get closer to each other, the RT minimal surface for $S(A\cup B)$ will jump to a cylinder stretched between the boundaries of $A$ and $B$, thus making $I(A:B)>0$.

The calculations that explicitly show this behavior cannot be done analytically in any finite $D$ except in $D=3$. \cite{Colin-Ellerin:2019vst} took a large $D$ limit of these equations, and found enough simplification to compute analytically the separation at which the phase transition occurs. Their result shows that, when $D$ is very large, the (averaged) correlations between distant regions are small (zero, at leading large $N$ order) unless the two regions become infinitesimally close and occupy the entire volume of the boundary theory. This is naturally interpreted as a manifestation of the spatial localization of correlations when $D\to \infty$.

Given this confirmation that the structure of holographic entanglement simplifies in the large $D$ limit, there is hope that the emergence of spacetime from quantum entanglement may be more easily understood in a $1/D$ expansion. 

\subsection{Large $D$ strings}

The study of strings in the limit of large $D$ is an old subject, in particular in the context of the effective string model for confinement in QCD. \cite{Alvarez:1981kc} exploited the mean-field character of the large $D$ limit in order to study the corrections from quantum fluctuations to the linear potential between two quarks joined by a Nambu-Goto string.

With the advent of AdS/CFT, where the effective string is a fundamental string in AdS$_5$, \cite{Vyas:2012dg} investigated a holographic bulk dual of that approach in a manner that may relate to the large $D$ limit in gravity. As a very simple illustration of the expected effects, consider the holographic representation of a quark as a string hanging from the boundary into the bulk. The gluonic cloud around the quark at the boundary corresponds to the gravitational field that the string in the bulk creates around itself. When $D$ gets large, this field becomes more and more strongly localized near the string and vanishes a short distance away from it. On the boundary, this means that the quantum gluonic cloud is becoming averaged out beyond distances $\Or{1/D}$, an effect related to the spatial decoupling of correlations that we have discussed in sec.~\ref{subsec:entangle}. So the large $D$ localization of classical gravity is indeed dual to a mean field limit in a quantum theory.\footnote{More recently, the study of large $D$ strings in \cite{Ambjorn:2016hao} may also contain other lessons for gravity.}

Although strings in theories at large $D$ might appear to necessarily be effective strings, valid only at low energies, an intriguing construction in \cite{Friess:2005be} suggests that critical strings might exist as conformally invariant 2D sigma models in AdS$_D$ when $D$ is large. The uses of these theories and their possible connections to large $D$ gravity need to be better understood.

\section{Higher-derivative gravity}
\label{sec:hider}

In dimensions $D\geq 4$, Einstein's theory can be extended to a class of theories with higher-derivative terms in the action, whose field equations nevertheless are of second derivative order. These are the Lovelock theories, the simplest of which consists of the addition of a Gauss-Bonnet term in $D\geq 5$, but as $D$ grows higher an increasing number of terms are allowed in the action \cite{Lovelock:1971yv}. Even though their consistency as classical theories has not been completely established (see e.g., \cite{Reall:2014pwa,Papallo:2017qvl,Kovacs:2020ywu}), they are a natural field of study for the large $D$ program.

The same elementary remark can be made about these theories as in sec.~\ref{subsec:hawking}. Since each additional Lovelock term comes with a length scale of its own, in the form of an undetermined coefficient (coupling) in the action, there is a choice to be made about how these lengths change with $D$. That is, there is not a `correct' way of taking the large $D$ limit for these theories, but rather there are different trajectories in the space of $D$-dependent couplings, leading to different limits as $D\to\infty$. Which limit one chooses may be dictated by the kind of physics that one intends to select, or more pragmatically, by the simplifications that result from a specific choice.\footnote{In principle, string theory determines the values of these couplings in $D=10$ or lower.}

Remarkably, there are exact solutions for static black holes with arbitrary numbers of Lovelock couplings \cite{Boulware:1985wk,Wheeler:1985nh,Wheeler:1985qd,Cai:1998vy}. A first study of the large $D$ limit in these theories analyzed elementary properties of these black holes \cite{Giribet:2013wia}. In a similar context, but with additional terms in the action motivated by string theory, \cite{Prester:2013gxa} took a large $D$ limit in order to explicitly solve the equations for a class of `small black holes' of string-scale size.

A more comprehensive study was begun in \cite{Chen:2015fuf} with the investigation of the quasinormal spectrum of large $D$ black holes in Gauss-Bonnet theory. An important feature of the Lovelock black holes is that, if the large $D$ limit is taken in such a way that the effects of the higher couplings remain finite near the horizon, then the geometry loses most of the simplifying properties of the pure Einstein theory. In particular, it is not possible in general to solve for the quasinormal modes in closed analytic form. One may, however, find solutions in a perturbative expansion for small values of those couplings, as was done in  \cite{Chen:2015fuf}. This same limitation extends to all subsequent studies of higher-derivative black holes at large $D$. Even though it may appear to be a serious restriction, one must bear in mind that when gravitational theory is regarded as an effective field theory, the higher derivative operators must consistently be treated as perturbatively small corrections.

Within this framework of ideas, the different effective theories of black holes that we have discussed in sec.~\ref{sec:eft}, as well as many of their applications, have been extended to higher-derivative theories (mostly Gauss-Bonnet) in the Beijing variety of the Japan-Barcelona style in \cite{Chen:2016fuy,Chen:2017wpf,Chen:2017hwm,Chen:2017rxa,Chen:2018nbh,Chen:2018vbv,Li:2019bqc,Guo:2019pte}, and in the Indian style in \cite{Saha:2018elg,Kar:2019kyz}.

\section{Beyond decoupling}
\label{sec:beyond}

\subsection{Non-decoupled waves and hair}
\label{subsec:ndwaves}

Much interesting black hole physics takes place outside the regime of decoupled dynamics that is so elegantly captured by the large $D$ membrane theories. 
We lack a general conceptual framework for non-decoupled black hole dynamics of the kind that underlies the physics of the decoupled sector. 
Nevertheless, there are several examples of phenomena outside the remit of decoupled physics for which the large $D$ expansion has proved efficient. Perhaps they contain clues for more systematic approaches.

An early instance is the study of the scattering of scalar waves off a black hole, which was fully solved analytically in \cite{Emparan:2013moa}. The result was cast into the form of effective boundary conditions at the `hole' for waves that propagate in the far zone.

We have also discussed in sec.~\ref{subsec:nondecqnm} how the calculation of the universal spectrum of non-decoupled quasinormal modes easily yields the real part of frequencies with an accuracy that compares well with other analytical methods. The very broad universality of the spectrum, consisting of almost normal, non-dissipative modes, may allow further analysis, including possibly the quantization of black hole oscillations.

A problem that is technically similar to the non-decoupled quasinormal oscillations -- solved using a WKB approximation near a peak of a radial potential -- is the study of scalar field condensation in a holographic superconductor, which we reviewed in sec.~\ref{subsec:adscmt}. Actually, this is among a class of phenomena that have attracted considerable attention in recent years: the formation of hair as a scalar condensate around black holes. Hair loss can be prevented, in AdS and in other situations, when the black hole is inside some sort of `box' that does not allow the scalar to disperse away. When $D$ is large, in known models the hair condenses just outside the near-horizon zone. With the hair in this region, the gains from considering large $D$ are meager: the complete description of the condensate generally cannot be reduced to a fully analytical solution, and must be solved numerically like when $D$ is finite.\footnote{There are analytic holographic superconductors \cite{Herzog:2010vz}.}
We do not know of any argument why the scalar field could not condense within the near-horizon zone, so it would be interesting if a well-motivated and workable model of it could be constructed.

Finally, a related topical problem that is hard to solve with numerical approaches (since it involves widely separate scales) is the evolution of `black hole bomb' superradiant instabilities of rotating black holes, in AdS and in other `box' spacetimes.\footnote{Among many references to this subject, \cite{Press:1972zz,Cardoso:2004nk,Cardoso:2006wa} are relevant early ones.} Studying it in a large $D$ approach requires handling the non-perturbative (in $1/D$) coupling of the near and far zones in a dynamical situation, which is insufficiently understood yet.

\subsection{Short-scale structure, singularities, and topology change}

There are many situations where structure on very short distances appears in a gravitational system, often involving the presence of singularities. 
While there are many types of singularities and our understanding of them is incomplete, it seems likely that the strength of a singularity often will depend on the number of dimensions.\footnote{Even the nature of the singularity may change with $D$, as is known to occur for BKL singularities \cite{Damour:2002et}.} Given the increased localization of gravity, we may expect that as $D$ grows, the divergences become stronger near a curvature singularity. This suggests to look for an appropriate scaling with $D$ that magnifies the singular region and, by isolating it, possibly simplifies its investigation. 

Opportunities of this kind arise in the deep non-linear regime of the GL instability, which we have discussed in sec.~\ref{sec:GL}.
Singularities play a role both in the phase space of static solutions, and in the time evolution of the instability.
The former were successfully dealt with in \cite{Emparan:2019obu}, who studied the topology-changing merger transition (in solution space) between black strings and black holes localized in a Kaluza-Klein circle. A new kind of large $D$ scaling of the black string was found that blows up the merger region and reduces the problem not only to a tractable form, but also to a mathematically appealing one: the near-horizon geometry varies along the Kaluza-Klein circle direction following the equations of Ricci flow.\footnote{Unknown to the authors of \cite{Emparan:2019obu} at the time, the same large $D$ limit had been studied earlier in \cite{Perelman:2006un}.} A singular pinch on the black string horizon corresponds to the endpoint of a certain Ricci flow. \cite{Emparan:2019obu} could then obtain a complete understanding of the singularity and its deformation and resolution across the topology-changing horizon.

In the time dependent setting, a naked singularity forms dynamically in the evolution of the GL instability, as we mentioned in sec.~\ref{sec:GL}. This has been observed only numerically \cite{Lehner:2010pn}. An analytic understanding of it using the large $D$ limit is a challenging problem, still open.

\subsection{Critical collapse and Choptuik scaling}
\label{subsec:choptuik}

Another important instance where the dynamics of gravity leads to very short distance physics, indeed naked singularities, is the phenomenon of critical collapse, discovered in a numerical study by Choptuik in \cite{Choptuik:1992jv} (see \cite{Gundlach:2007gc} for an excellent review).

Consider a one-parameter family of initial data for a spherical scalar field cloud. For definiteness we take the parameter, call it $a$, to be the amplitude of the cloud, but it could also be, for instance, its radial velocity or its width. Left to evolve under the action of gravity, the cloud will collapse, and the outcome will depend on its initial amplitude. For small amplitudes, the field contracts and then bounces back and disperses away to infinity. For large amplitudes, instead, the collapse forms a black hole (with part of the scalar field dispersing to infinity) -- a bigger black hole for bigger initial amplitude. By decreasing the parameter $p$, we can tune to a regime where the mass of the black hole scales as
\beq
M\propto (a-a_*)^\gamma\,,
\eeq
where $a_*$ is the critical parameter that separates the two outcomes, and the critical exponent is numerically determined to be $\gamma\approx 0.374$. This value is universal for all one-parameter families of scalar field initial data. Moreover, all near-critical data approach the same geometry in a finite region of space for a finite time, from either side of the threshold. The critical solution $\phi_*(t,r)$ repeats itself with discrete self-similarity, or scale-echoing, such that
\beq
\phi_*(t,r)=\phi_*(e^\Delta t,e^\Delta r)\,,
\eeq
where the logarithmic period is found to be $\Delta\approx 3.44$.

This phenomenon is as intriguing as it is exciting. By fine-tuning the initial data, 
we can use the gravitational attraction to create regions of increasingly large curvature, eventually reaching a naked singularity, or more physically, a region of Planck-scale curvature where quantum gravity effects are made visible to observers afar.\footnote{Due to the fine-tuning required, critical collapse is often dismissed as an unphysical violation of cosmic censorship. We disagree with this view: the distinction is that of experimental physics vs.\ observational physics. Although very hard to observe in the skies by astronomers, critical gravitational collapse to form a Planck-size ``black hole'' may well be engineered by the experimentalists of a sufficiently advanced civilization. Observe that the fine-tuning is a moderate power law (not an exponential) in the initial data. }

What is the meaning of the exponents $\gamma$ and $\Delta$? In statistical mechanics, when critical exponents are irrational, it is due to large quantum fluctuations near criticality. In gravitational physics, critical phenomena typically arise associated to physics on a pre-existing horizon, and the exponents take simple, rational values of mean-field theory type. 
Choptuik's criticality is remarkable in this respect since $\gamma$ seems to be irrational, coming out of a fully naked singularity. Its mystery might start to unravel if we could find a way of computing it analytically, possibly in a perturbative expansion where the leading order value were rational, from some sort of limiting solvable mean field theory which we could then correct order by order. 

The large $D$ limit appears to have the right ingredients for this problem. Numerical investigations of scalar field critical collapse at increasing values of $D$ reveal an interesting pattern \cite{Sorkin:2005vz,Bland:2005vu}: as $D$ grows large, $\gamma$ approaches the value $1/2$. In fact, the data are well approximated by
\beq
\gamma\approx \frac12 \lp 1-\frac1{D}\rp
\eeq
(curiously, for $D=4$ this yields $\gamma=0.375$, although a better fit to the data at larger $D$ is $\gamma\approx \frac12 \lp 1-\frac1{D+1}\rp$; more accurate fits are given in \cite{Rozali:2018yrv}). The behavior with $D$ of the echoing period is less clear, since this quantity is in general hard to extract numerically, even more so as the number of dimensions increases. 
But, in any case, these observations provide a good motivation to attack the problem of critical collapse using a large $D$ approach, going beyond the scope of the effective membrane theory.

The first order of business would be identify the appropriate scalings with $D$ of field variables and coordinates. To approach this problem, \cite{Rozali:2018yrv} first examined the wave equation for a spherically symmetric 
scalar field in flat spacetime,\footnote{A mass term may be added, but one expects it will become irrelevant at the short scales near criticality.}
\beq\label{scawave}
\partial_t^2 \varphi=\partial_r^2\varphi +\frac{D-2}{r}\partial_r\varphi\,.
\eeq
In the study of non-decoupled waves in a black hole background (both quasinormal modes and scattering), the appropriate scalings were $t= \bar t/D$ and $r\to 1+\ln\sR/D$, i.e. $\partial_t,\,\partial_r \sim D$, which focus on the region near the horizon while preserving the hyperbolic (wave-like) character of the equation. Instead,  \cite{Rozali:2018yrv} noticed that if $\partial_t\sim \sqrt{D}$ and $\partial_r \sim \Or{1}$, that is, scaling
\beq\label{sqrtD}
t=\frac{\tau}{\sqrt{D}} 
\eeq
and then taking $D\to\infty$, the wave equation \eqref{scawave} turns into a parabolic equation
\beq\label{parab}
\partial_\tau^2 \varphi=\frac1{r}\partial_r\varphi\,.
\eeq
This is not obviously giving what we are after, but there are not that many non-trivial large $D$ scalings of the equation, and moreover it is the same scaling (with $t$ and $r$ swapped) as was taken in \cite{Emparan:2019obu} for the study of the black hole/black string transition -- a problem linked by double analytic continuation to that of critical collapse \cite{Kol:2004ww}. Eq.~\ref{parab} admits simple exact solutions, which, as \cite{Rozali:2018yrv} found, can be extended with a similar scaling in $D$ to the case when the field $\varphi$ is coupled to gravity. These gravitating solutions do not describe a collapsing field, but instead a class of oscillating `soliton stars', which have an intriguing feature: when the amplitude of the oscillations reaches a critical value, a divergence appears. If the divergence is regarded as a signal of the formation of the horizon of a black hole, one finds a critical exponent $\gamma=1/2$.

So, even if the gravitating scalar solution in \cite{Rozali:2018yrv} is not the critical collapse that we seek -- it is not self-similar, and it is not a collapse -- these findings suggest that the scaling \eqref{sqrtD} may zoom in on the right regime of physics.

If a large $D$ expansion manages to yield a simple enough solution to collapse near criticality, then one may consider embedding it in AdS with the aim of improving the understanding of the non-linear instability of AdS \cite{Bizon:2011gg}.

\section{Gravitational radiation}
\label{sec:gwaves}

\subsection{Shockwave collisions}

The collision of gravitational shockwaves in $D$ dimensions was studied in \cite{Herdeiro:2011ck,Coelho:2012sy}. It was found that when $D$ is very large the result for the radiated energy admits a very simple fit \cite{Coelho:2012sya}. This suggests that, rather than first solving the problem for arbitrary $D$ and then taking $D\gg 1$, it may be fruitful to set up the problem from the start in terms of a $1/D$ expansion, hoping that the equations become simple enough to allow a more complete study.

\subsection{An effective theory of gravitational waves?}

Before we conclude this review, we would like to speculate whether, in addition to effective theories of black holes, a large $D$ effective theory of gravitational waves might be possible. Since the vacuum theory $R_{\mu\nu}=0$ is essentially a theory of black holes and of the gravitational waves that interact with them, we would achieve a complete reformulation of General Relativity around the limit $D\to\infty$. As we have seen, the coupling between the effective membrane and the gravitational waves is non-perturbative in $1/D$. Therefore the two sectors of the theory -- black holes and gravitational waves -- can be studied independently before coupling them.

The studies of perturbative quantum General Relativity in \cite{Strominger:1981jg,BjerrumBohr:2003zd} may be regarded as attempts to exploit the large number $\sim D^2$ of graviton polarizations in order to tame the gravitational fluctuations of the geometry away from black holes. However, we already mentioned in the introduction that this route does not appear to lead very far, so other ideas must be tried.

In order to formulate an effective theory of gravitational waves, it would seem necessary to have a separation of length scales. There are two natural ones: their amplitude and their wavelength. Perhaps an effective theory is possible in regimes where these lengths are parametrically separated in $D$. It is not clear whether or how this may employ any of the two aspects of the large $D$ limit that we discussed in the introduction -- the large number of graviton polarizations, and the localization of the gravitational interaction -- or instead, and more interestingly, other new concepts will be required.

\acknowledgments

We are greatly indebted to our collaborators Tom{\'a}s Andrade, 
Jorge Casalderrey-Solana,
Daniel Grumiller, Keisuke Izumi, Young-Shin Kim, David Licht, Raimon Luna, Marina Mart$\acute{\imath}$nez, Ben Meiring,  Tetsuya Shiromizu, Michael Spillane, Takahiro Tanaka, and Amos Yarom. RE owes special gratitude to Ryotaku Suzuki and Kentaro Tanabe for their innumerable essential contributions to launching and developing our joint work on the large $D$ program. We also acknowledge discussions with Joan Camps, \'Oscar Dias, Pau Figueras, Gavin Hartnett, Christiana Pantelidou, Moshe Rozali, Jorge Santos, Arunabha Saha, Benson Way, and Ben Withers, over several topics covered in this review.
We are grateful to Sayantani Bhattacharyya and Shiraz Minwalla for sharing with us their deep knowledge of the subject.

RE is supported by ERC Advanced Grant GravBHs-692951 and MEC grant FPA2016-76005-C2-2-P.
CH is supported by U.K.\ Science \& Technology Facilities Council Grant ST/P000258/1 and 
a Wolfson Fellowship from the Royal Society.

Finally, we must mention that -- much beyond the usual support by tax-paying -- finishing this review under the dire conditions in the world in March 2020 would have been impossible without the generosity, work and dedication of our fellow citizens -- health workers and many others who keep our societies running in the face of unprecedented adversity in our time.

\bibliography{LargeDReview}

\end{document}